\documentclass[twocolumn,tighten,times]{aastex631}

\usepackage{xstring}
\usepackage{savesym}
\savesymbol{tablenum}
\usepackage[print-unity-mantissa=false]{siunitx}
\restoresymbol{SIX}{tablenum}
\usepackage{amsmath}
\usepackage{multirow}
\usepackage{newtxmath}
\usepackage{definitions}
\usepackage{enumitem}

\graphicspath{{./}{figures/}}

\newcommand{\nov}{AT\,2020nov}
\newcommand{\msun}{\text{M\textsubscript{\(\odot\)}}}

\newcommand{\uqtyexp}[5]{#1 \genfrac{}{}{0pt}{}{\scriptstyle+#2}{\scriptstyle-#3} \times 10^{#4} \, \mathrm{#5}}
\newcommand{\uqty}[4]{%
  \ensuremath{#1{\raisebox{0.4ex}{\scriptsize $^{+#2}_{-#3}$}}}\,\si{#4}%
}\newcommand{\ueqty}[5]{%
  \ensuremath{#1{\raisebox{0.3ex}{\scriptsize $^{+#2}_{-#3}$}} \times 10^{#4}}\,\si{#5}%
}
\newcommand{\unum}[3]{%
  \ensuremath{#1{\raisebox{0.4ex}{\scriptsize $^{+#2}_{-#3}$}}}%
}

\newcommand{\eline}[3][a]{%
    \IfStrEqCase{#1}{%
        {a}{#2${\,}\textsc{#3}$}%
        {f}{[#2${\,}\textsc{#3}$]}%
    }[failed]%
}
\newcommand{\elinesgl}[4][a]{%
    \IfStrEqCase{#1}{%
        {a}{#2${\,}\textsc{#3}{\;} \lambda #4$}%
        {f}{[#2${\,}\textsc{#3}]{\;} \lambda #4$}%
    }[failed]%
}
\newcommand{\elinedbl}[5][a]{%
    \IfStrEqCase{#1}{%
        {a}{#2${\,}\textsc{#3}{\;} \lambda\lambda #4, #5$}%
        {f}{[#2${\,}\textsc{#3}]{\;} \lambda\lambda #4, #5$}%
    }[failed]%
}

\usepackage{booktabs}
\newcolumntype{P}[1]{>{\centering\arraybackslash}p{#1}}
\newcolumntype{L}[1]{>{\arraybackslash}p{#1}}

\DeclareSIUnit\angstrom{\text{Å}}
\begin{document}

\title{\nov{}: Evidence for Disk Reprocessing in a Rare Tidal Disruption Event}

\author[0000-0003-1714-7415]{Nicholas Earl}
\affiliation{Department of Astronomy, University of Illinois, 1002 W. Green St., Urbana, IL 61801, USA} 
\affiliation{Center for Astrophysical Surveys, National Center for Supercomputing Applications, Urbana, IL, 61801, USA}

\author[0000-0002-4235-7337]{K.~Decker~French}
\affiliation{Department of Astronomy, University of Illinois, 1002 W. Green St., Urbana, IL 61801, USA} 
\affiliation{Center for Astrophysical Surveys, National Center for Supercomputing Applications, Urbana, IL, 61801, USA}

\author[0000-0003-2558-3102]{Enrico~Ramirez-Ruiz}
\affiliation{Department of Astronomy and Astrophysics, University of California, Santa Cruz, CA 95064, USA}

\author[0000-0002-4449-9152]{Katie~Auchettl}
\affiliation{Department of Astronomy and Astrophysics, University of California, Santa Cruz, CA 95064, USA}
\affiliation{School of Physics, The University of Melbourne, VIC 3010, Australia} 

\author[0000-0002-6248-398X]{Sandra~I.~Raimundo}
\affiliation{DARK, Niels Bohr Institute, University of Copenhagen, Jagtvej 128, 2200 Copenhagen, Denmark}
\affiliation{Department of Physics and Astronomy, University of Southampton, Highfield, Southampton SO17 1BJ, UK}

\author[0000-0002-5680-4660]{Kyle~W.~Davis}
\affiliation{Department of Astronomy and Astrophysics, University of California, Santa Cruz, CA 95064, USA}

\author[0000-0003-4127-0739]{Megan~Masterson}
\affiliation{MIT Kavli Institute for Astrophysics and Space Research, Massachusetts Institute of Technology, Cambridge, MA 02139, USA}

\author[0000-0001-7090-4898]{Iair~Arcavi}
\affiliation{The School of Physics and Astronomy, Tel Aviv University, Tel Aviv 69978, Israel}

\author[0000-0002-1568-7461]{Wenbin~Lu}
\affiliation{Department of Astronomy, University of California, Berkeley, CA 94720, USA}

\author[0000-0003-4703-7276]{Vivienne~F.~Baldassare}
\affiliation{Department of Physics \& Astronomy, Washington State University, Pullman, Washington 99164, USA}

\author[0000-0003-4263-2228]{David~A.~Coulter}
\affiliation{Space Telescope Science Institute, Baltimore, MD 21218, USA}

\author[0000-0001-5486-2747]{Thomas de Boer}
\affiliation{Institute for Astronomy, University of Hawaii, 2680 Woodlawn Drive, Honolulu, HI 96822, USA}

\author[0000-0001-7081-0082]{Maria~R.~Drout}
\affiliation{David A. Dunlap Department of Astronomy and Astrophysics, University of Toronto, 50 St. George Street, Toronto, Ontario, M5S 3H4, Canada}

\author[0009-0008-6396-0849]{Hannah~Dykaar}
\affiliation{David A. Dunlap Department of Astronomy and Astrophysics, University of Toronto, 50 St. George Street, Toronto, Ontario, M5S 3H4, Canada}

\author[0000-0002-2445-5275]{Ryan~J.~Foley}
\affiliation{Department of Astronomy and Astrophysics, University of California, Santa Cruz, CA 95064, USA}

\author[0000-0002-8526-3963]{Christa~Gall}
\affiliation{DARK, Niels Bohr Institute, University of Copenhagen, Jagtvej 128, 2200 Copenhagen, Denmark}

\author[0000-0003-1015-5367]{Hua~Gao}
\affiliation{Institute for Astronomy, University of Hawaii, 2680 Woodlawn Drive, Honolulu, HI 96822, USA}

\author[0000-0003-1059-9603]{Mark~E.~Huber}
\affiliation{Institute for Astronomy, University of Hawaii, 2680 Woodlawn Drive, Honolulu, HI 96822, USA}

\author[0000-0002-6230-0151]{David~O.~Jones}
\affiliation{Institute for Astronomy, University of Hawai'i, 640 N A'ohoku Pl, Hilo, HI 96720, USA}

\author[0000-0001-5710-8395]{Danial~Langeroodi}
\affiliation{DARK, Niels Bohr Institute, University of Copenhagen, Jagtvej 128, 2200 Copenhagen, Denmark}

\author[0000-0002-7272-5129]{Chien-Cheng~Lin}
\affiliation{Institute for Astronomy, University of Hawaii, 2680 Woodlawn Drive, Honolulu, HI 96822, USA}

\author[0000-0002-7965-2815]{Eugene~A.~Magnier}
\affiliation{Institute for Astronomy, University of Hawaii, 2680 Woodlawn Drive, Honolulu, HI 96822, USA}

\author[0000-0001-6350-8168]{Brenna~Mockler}
\affiliation{The Observatories of the Carnegie Institute for Science, 813 Santa Barbara St., Pasadena, CA 91101, USA}

\author[0009-0005-1158-1896]{Margaret Shepherd}
\affiliation{Department of Astronomy, University of Illinois, 1002 W. Green St., Urbana, IL 61801, USA} 

\author[0000-0003-1535-4277]{Margaret E. Verrico}
\affiliation{Department of Astronomy, University of Illinois, 1002 W. Green St., Urbana, IL 61801, USA} 
\affiliation{Center for Astrophysical Surveys, National Center for Supercomputing Applications, Urbana, IL, 61801, USA}

\begin{abstract}

We present a detailed analysis of \nov{}, a tidal disruption event (TDE) in the center of its host galaxy, located at a redshift of $z = 0.083$. \nov{} exhibits unique features, including double-peaked Balmer emission lines, a broad UV/optical flare, and a peak log luminosity in the extreme ultra-violet (EUV) estimated at $\sim$\LpeakTDE. A late-time X-ray flare was also observed, reaching an absorbed luminosity of \qty{1.67e43}{erg.s^{-1}} approximately \qty{300}{days} after the UV/optical peak. Multi-wavelength coverage, spanning optical, UV, X-ray, and mid-infrared (MIR) bands, reveals a complex spectral energy distribution (SED) that includes MIR flaring indicative of dust echoes, suggesting a dust covering fraction consistent with typical TDEs. Spectral modeling indicates the presence of an extended, quiescent disk around the central supermassive black hole (SMBH) with a radius of $\sim$\SpecDiskOuterRadius{}. The multi-component SED model, which includes a significant EUV component, suggests that the primary emission from the TDE is reprocessed by this extended disk, producing the observed optical and MIR features. The lack of strong AGN signatures in the host galaxy, combined with the quiescent disk structure, highlights \nov{} as a rare example of a TDE occurring in a galaxy with a dormant but extended pre-existing accretion structure.

% We present a detailed analysis of AT~2020nov, a tidal disruption event (TDE) in the center of its host galaxy, located at a redshift of $z = 0.083$. AT~2020nov exhibits unique features, including double-peaked Balmer emission lines, a broad UV/optical flare, and a peak log luminosity in the extreme ultra-violet (EUV) estimated at $\sim$$45.66^{+0.10}_{-0.33} \; \mathrm{erg} \, \mathrm{s^{-1}}$. A late-time X-ray flare was also observed, reaching an absorbed luminosity of $1.67 \times 10^{43} \; \mathrm{erg} \, \mathrm{s^{-1}}$ approximately 300 days after the UV/optical peak. Multi-wavelength coverage, spanning optical, UV, X-ray, and mid-infrared (MIR) bands, reveals a complex spectral energy distribution (SED) that includes MIR flaring indicative of dust echoes, suggesting a dust covering fraction consistent with typical TDEs. Spectral modeling indicates the presence of an extended, quiescent disk around the central supermassive black hole (SMBH) with a radius of $\sim$$5.06^{+0.59}_{-0.77} \times 10^4 \; \mathrm{R_g}$. The multi-component SED model, which includes a significant EUV component, suggests that the primary emission from the TDE is reprocessed by this extended disk, producing the observed optical and MIR features. The lack of strong AGN signatures in the host galaxy, combined with the quiescent disk structure, highlights AT~2020nov as a rare example of a TDE occurring in a galaxy with a dormant but extended pre-existing accretion structure.

\end{abstract}


\section{Introduction} \label{sec:intro}

Tidal disruption events (TDEs) are unique astrophysical phenomena that occur when the orbit of stars brings them close enough to super-massive black holes (SMBHs) for their self-gravity to be overpowered by tidal forces \citep{hills1975,rees1988}. After disruption, approximately half of the stellar material remains bound to the SMBH and returns to the pericenter of the star's orbit. The relativistic apsidal precession of the orbits leads to self-intersection and subsequent circularization of the bound stellar material into an accretion disk on timescales of a few times the return time of the most bound stellar debris \citep{kochanek1994,ramirez-ruiz2009,hayasaki2013a,guillochon2014,piran2015,hayasaki2016a,bonnerot2020}. While the luminous flare produced in these events can reach or even exceed the Eddington luminosity \citep{hung2017,wevers2019b,yao2023}, the precise mechanism powering them is not fully understood. TDEs therefore act as an important probe of the accretion physics around SMBH and a means by which to study black hole demographics by illuminating otherwise dormant black holes.

Canonically, the formation of a compact accretion disk from the material of the disrupted star is thought to be the source of the electromagnetic radiation, and the luminosity of this disk is expected to follow the rate of debris fallback ($\sim$$t^{-5/3}$) due to the uniform distribution of mass per binding energy of the stellar debris \citep{phinney1989c}. Recent studies show that the light curve shape depends on stellar structure and composition \citep{lodato2009,ramirez-ruiz2009,law-smith2020}, with the rise and peak of the light curve influenced by the star's internal density profile \citep{lodato2009,ramirez-ruiz2009,guillochon2013,law-smith2017,golightly2019,coughlin2022}. According to some models, the disk is formed quickly due to general relativistic apsidal precession causing the returning stellar debris streams to intersect and lose their orbital energy, leading to circularization \citep{rees1988,dai2015,guillochon2015,bonnerot2017}. If most of the debris rapidly forms an accretion disk, it would predict highly super-Eddington accretion power, potentially reaching up to \qty{1e46}{erg.s^{-1}}. However, a significant portion of the debris may be expelled by radiation pressure during disk formation \citep{strubbe2009,metzger2016}. Additionally, this classical model suggests an accretion disk radius roughly on the scale of the tidal radius ($10$ and $10^2 R_{\rm g}$, where $R_{\rm g}$ denotes the gravitational radius). But for optically observed TDEs, the inferred black body radii are one to two orders of magnitude larger than the expected extent of the disk \citep{gezari2014,arcavi2014,hinkle2021b,charalampopoulos2022}. This requires that the X-ray and extreme ultraviolet (EUV) photons produced near the black hole are reprocessed to optical/UV wavelengths by disk outflows or optically thick intermediate material \citep{guillochon2014,metzger2016,roth2016,dai2018,lu2020,thomsen2022,bu2022}, or that the released energy comes from shocks caused by colliding debris streams which release kinetic energy at large self-intersection radii \citep{lodato2012, piran2015}. Alternatively, the cooling envelope model posits that after a star is disrupted by a SMBH, the stellar debris rapidly forms a quasi-spherical, pressure-supported envelope rather than a compact accretion disk. This envelope undergoes gradual Kelvin-Helmholtz contraction, radiating energy primarily in the optical/UV bands \citep{metzger2022,sarin2024}. The model predicts that the envelope's effective temperature rises over time while its photosphere radius contracts, leading to an optical light curve that decays roughly as $t^{-3/2}$. The SMBH accretion rate peaks only after the envelope contracts to the circularization radius, which can take months to years, explaining the delayed rise in X-ray and radio emissions observed in some TDEs. This model could account for the large photosphere radii and high optical luminosities observed in TDEs, which are not easily explained by traditional accretion disk models.

The discrepancy between the observed optical/UV energy ($\sim$\qty{1e51}{erg}) and the bolometric energy predicted by theoretical models leads to a ``missing energy'' problem, suggesting that a significant portion of the emission must be released in the extreme ultraviolet (EUV), carried away by outflows, or absorbed by the black hole \citep{lu2018}. Additionally, the cooling envelope model for TDEs proposes that much of the energy may be stored in a hot, extended atmosphere around the disrupted star, which radiates primarily in the EUV and soft X-rays, thereby explaining the discrepancy. \citet{mockler2021} found that the total radiated energy in TDEs is potentially higher than early estimates based on optical light curves alone, which could help resolve this issue. Furthermore, recent studies of infrared dust echoes from TDE candidates \citep[e.g.,][]{jiang2021a,panagiotou2023,dodd2023,masterson2024} report radiated energies of $\gtrsim10^{52}\rm\, erg$, suggesting that in these cases, the ``missing energy'' is radiated predominantly in the EUV. We discuss the dust echo behavior of \nov{} in Section \ref{sec:euv-primary}.

Observations of nuclear transients with rise times of weeks to months have seen a significant increase in recent years due in part to all-sky surveys and their ability to probe the nuclear environments of galaxies at high cadences. Recent population studies of TDEs have identified distinct spectroscopic classes based on their optical spectra: TDE-H (hydrogen lines only), TDE-He (helium lines only), TDE-H+He (hydrogen with \eline{He}{ii}) \citep{arcavi2014,velzen2020,nicholl2022}, and TDE-featureless (lack of emission features) \citep{hammerstein2023}. Some TDEs also show Bowen fluorescence lines, particularly \eline{N}{iii} and \eline{O}{III}, indicating the presence of dense material and extreme UV/X-ray radiation \citep{blagorodnova2019,leloudas2019,velzen2020}.

Double-peaked or asymmetric emission lines have been observed in TDEs and are often associated with the presence of a disk structure. For instance, the TDE candidate PTF09djl exhibited a double-peaked broad H$\alpha$ line, which was well modeled with a relativistic elliptical accretion disk \citep{arcavi2014,liu2017}. The peculiar substructures, with one peak at the line rest wavelength and the other redshifted to about \qty{3.5e4}{km.s^{-1}}, were attributed to the orbital motion of the emitting matter within the disk plane. Similarly, the TDE ASASSN-14li displayed single-peaked asymmetric line profiles, which were also modeled with a relativistic elliptical disk \citep{cao2018}. Double-peaked emission lines have also been observed in several other cases, including SDSS J0159 \citep{merloni2015,zhuang2021}, PS18kh \citep{holoien2019a}, AT~2018hyz \citep{short2020,hung2020}, AT~2019qiz \citep{nicholl2020,short2023}, AT~2020zso \citep{wevers2022}. 

Double-peaked emission lines are similarly observed in Active Galactic Nuclei (AGN) in the permitted H and He lines and are proposed to originate in the outer parts of a disk, typically at $\gtrsim 10^3R_{\rm g}$ from the SMBH \citep{storchi-bergmann2017}. For example, the broad double-peaked profiles in the spectra of Arp 102B and NGC 1097 have been attributed to a disk origin, with the line-emitting region taken to be the outer regions of the accretion disk \citep{storchi-bergmann2017}. The variability in the profiles, such as changes in the relative strengths of the blue and red peaks, has been linked to the rotation of spiral arms or hotspots in the disk \citep{ward2024}. While TDEs and AGN exhibit some similarities in their spectra and light curves, important differences exist. Spectroscopically, TDEs often display broad emission lines with velocities of \qtyrange{3e4}{13e4}{km.s^{-1}} and may show Bowen fluorescence lines of \eline{O}{iii} and \eline{N}{iii}, which are rare in AGN spectra \citep{gezari2021,trakhtenbrot2019}. Photometrically, TDEs demonstrate a more coherent decay in their light curves, following a power-law decline that roughly scales with $t^{-5/3}$. In contrast, AGN light curves exhibit ``red noise'' with power spectra declining from timescales of months or years to hours, and individual modes have little or no phase coherence \citep{lawrence1987,mchardy1987,lawrence1993,aranzana2018}. X-ray properties also differ, with TDEs often showing extremely soft spectra compared to AGN \citep{auchettl2017}. Mapping double-peaked emission features in spectra can provide insights into the distinct physical processes in these systems.

Distinguishing TDEs from impostor nuclear transients is challenging due to the overlapping properties of TDEs with other transient phenomena such as AGN flares and Ambigous Nuclear Transients (ANTs). The classification is further complicated by the variability in observed TDE signatures and the non-uniform sensitivity of detection methods across different candidates \citep{zabludoff2021}. To accurately identify TDEs, a multi-wavelength observational approach is essential, capturing data in X-ray, UV, and optical bands to construct well-sampled light curves and spectral profiles \citep{gezari2012,mattila2017,auchettl2018}. For example, ANTs, which may be exotic TDEs, AGN flares, or Bowen fluorescence flares \citep{kankare2017,trakhtenbrot2019,neustadt2020,hinkle2021a,holoien2022,wiseman2024}, exhibit high dust covering fractions similar to AGN, suggesting the presence of dusty tori \citep{hinkle2024}, and show variability on timescales intermediate between AGN and TDEs. The optical spectra of ANTs typically show broad Balmer and \eline{He}{ii} line emission alongside narrow emission, including \eline[f]{O}{iii}, indicative of ionization processes occurring farther from the SMBH, similar to AGN. 

TDEs occurring in AGN pose additional challenges \citep{dodd2021}, as the interaction between the debris stream and the existing accretion disk can produce complex radiative outcomes that deviate from typical TDE or AGN spectra \citep{chan2019}. The bright nature of AGN, combined with potential flare-like features, makes confirming a TDE particularly difficult. The challenge of distinguishing traditional AGN emission from TDE emission has led to TDE searches being predominantly conducted in quiescent galaxies. However, it has been suggested that the occurrence rate of TDEs may be higher in a gaseous disk compared to a quiescent galaxy, particularly for SMBHs with masses less than \qty{1e7}{\msun} \citep{karas2007,wang2024,kaur2024}. Despite this, TDE candidates associated with AGN constitute only a small fraction of the overall observed TDE sample \citep{zabludoff2021}. For instance, \citet{liu2020} reported the discovery of a TDE candidate in the AGN SDSS J0227-0420, which exhibited distinctive power-law behavior in its long-term IR/optical/UV light curves. Similarly, \citet{yan2018} proposed that the decades-long decay of the X-ray light curve in the low-luminosity AGN NGC 7213 is best explained by the disruption of a main-sequence star. The soft X-ray spectrum and characteristic light curve of GSN 069 have also been attributed to a TDE \citep{shu2018}. Furthermore, \citet{blanchard2017} identified a TDE (PS16dtm) in a narrow-line Seyfert 1 galaxy, confirmed by the detection of a mid-infrared (MIR) echo from the AGN's dusty torus \citep{jiang2017,yang2019}. Another potential candidate is the energetic transient event PS1-10adi, discovered in an active galaxy and analyzed by \citet{jiang2019a}, which was possibly caused by a TDE \citep{kankare2017}. Some recurring nuclear transients have been suggested to be linked to TDEs occurring within AGN disks. \citep{payne2021}. Even the extensively studied ASASSN-14li exhibits signs of AGN activity, including the presence of a pre-TDE nuclear radio source \citep{alexander2016}. Likewise, a significant number of TDE host galaxies have been identified as LINERs or Seyfert 2 galaxies \citep{french2020b}.

Recent simulations have shown that TDEs are significantly influenced by the presence of an accretion disk surrounding the black hole. When the debris stream from a disrupted star returns to pericenter, it collides with the AGN disk, exciting shocks that lead to rapid inflow and energy dissipation at super-Eddington rates \citep{chan2019,chan2020}. This interaction modifies the circularization process by dissipating the kinetic energy of the debris and expediting its inflow toward the black hole. If the debris stream is sufficiently dense, it can penetrate the disk and cause a second collision, further dissipating energy and producing high-energy emission. These processes can result in distinct observational signatures, including smoother light curves and harder X-ray spectra, which differentiate TDEs in AGNs from ordinary AGN flares \citep{chan2021}.

In this paper, we report the discovery and identification of the TDE \nov{}. Spectroscopic observations reveal double-peaked emission line profiles while the photometry exhibits strong optical excess, with a transient flare seen in the optical, UV, and X-ray. A MIR flare is also detected at the time of the disruption, indicating the presence of dust near the nucleus. In Section \ref{sec:observations} we present the multi-band observational data. Modeling of the multi-band SED and long-term light curve, along with black hole mass estimates, are presented in Section \ref{sec:photo-analysis}. In Section \ref{sec:spec-analysis}, we analyze the spectroscopic features and model the double-peaked emission line profiles arising from the elliptical disk. We contextualize and discuss our results in Section \ref{sec:discussion} and summarize our conclusions in Section \ref{sec:conclusion}. Throughout this work, we adopt a flat $\Lambda$CDM cosmology with $H_0 = \qty{67.4}{km.s^{-1}.Mpc^{-1}}$, $\Omega_m = \num{0.32}$, and $\Omega_\Lambda = \num{0.68}$ \citep{planck2020}.

\section{Observations \& Data Reduction} \label{sec:observations}

\nov{} was discovered by the Zwicky Transient Facility \citep[ZTF;][]{bellm2019,masci2019,patterson2019} (internal designation ZTF20abisysx) at right ascension $16 \mathrm{h}\,58 \mathrm{m}\,12.99 \mathrm{s}$ and declination $2^{\circ}\,07'\,03.18''$. The transient position is consistent with the center of galaxy WISEA J165812.98+020703.0 (recorded as LEDA 1216501 in SIMBAD) at a redshift of 0.0826 and luminosity distance of \qty{389}{Mpc} as derived from narrow post-transient emission lines. \nov{} was initially identified by the ALeRCE broker and subsequently reported to the Transient Name Server (TNS). Additional reports came from the Young Supernova Experiment \citep[YSE;][]{jones2021} via the Panoramic Survey Telescope and Rapid Response System \citep[Pan-STARRS;][]{kaiser2002}, the Asteroid Terrestrial impact Last Alert System \citep[ATLAS;][]{tonry2018}, and Gaia Science Alerts \citep{hodgkin2013}.

Follow-up observations using the SED Machine IFU Spectrograph on the Palomar 60-inch \citep{blagorodnova2018} was taken and posted to the TNS \citep[Classification Report No\. 8221,][]{dahiwale2020} 20 days after discovery on 2020-07-16, revealing a  blue continuum, broad H$\alpha$+\eline{N}{ii} line complex, and a broad \elinesgl{He}{ii}{4868} line profile characteristic of optical/UV TDEs \citep{dahiwale2020}.

\subsection{Ground-based Imaging}

\begin{figure*}
    \centering 
    \includegraphics[width=\textwidth]{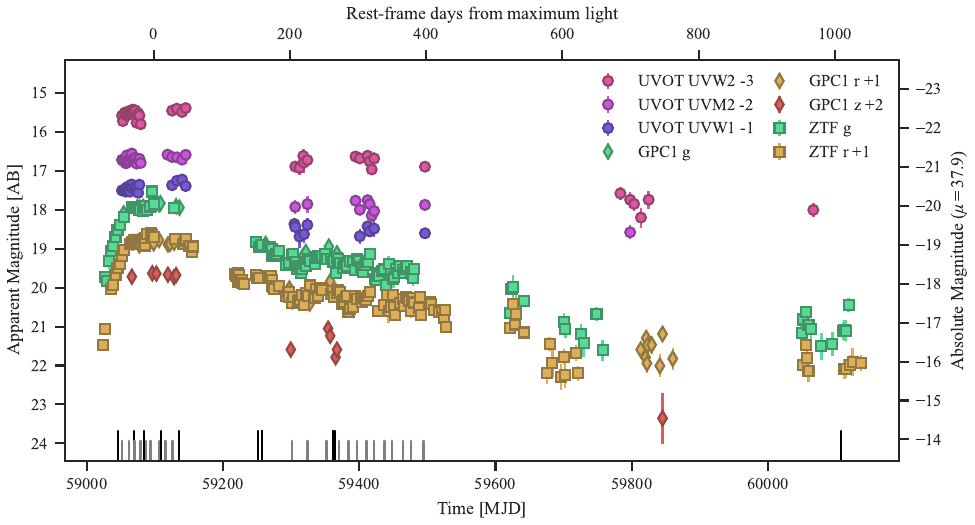}
    \caption{Optical and UV light curves of \nov{} from ZTF, PS1 (GPC1), and Swift (UVOT). The $g$-band, $r$-band, and $z$-band used difference imaging to remove host contribution, while estimates of the host flux from the population synthesis analysis were used for the Swift data (see Section \ref{sec:observations}). Spectral observations taken by the YSE collaboration are represented by thick black vertical lines, while shorter grey vertical lines show the lower resolution FLOYDS spectra.}
    \label{fig:uvopt_phot}
\end{figure*}

High-cadence ($\sim$\qty{3}{day}) imaging data for \nov{} was obtained through the YSE \citep{jones2019} in $g$, $r$, $i$, and $z$ filters using the Pan-STARRS telescope (PS1). The photometric reduction pipeline for YSE utilizes \texttt{photpipe} \citep{rest2014} and each image template was taken from stacked PS1 exposures, mostly from the PS1 3$\pi$ survey. To ensure accuracy, all images and templates were resampled and astrometrically aligned to match a skycell in the PS1 sky tessellation. By comparing the PSF photometry of stars to updated stellar catalogues of PS1 observations \citep{chambers2017}, an image zero-point was determined. To match the PSF of the nightly images, the PS1 templates were convolved with a three-Gaussian kernel, subtracted from the nightly images using the High Order Transform of PSF ANd Template Subtraction \citep[HOTPANTS;][]{becker2015} code, and a flux-weighted centroid was found for each supernova (SN) position. Finally, Point Spread Function (PSF) photometry was performed using forced photometry, and the brightness of the SN for that epoch was determined by applying the nightly zero-point to the photometry.

Complimentary data was obtained in the $g$ and $r$ bands through the ZTF public survey \citep{bellm2019}. To generate the host-subtracted light curves, PSF photometry was performed using the \texttt{photutils} python package on the ZTF difference images using the provided estimates for the point spread function. The resulting host-subtracted photometry is shown in Figure \ref{fig:uvopt_phot}.

\subsection{UVOT Observations}

UV and optical photometry from the UV-Optical Telescope \citep[UVOT;][]{gehrels2004, roming2005} onboard the Neil Gehrels Swift Observatory (Swift) were obtained over the course of 37 epochs using target-of-opportunity observations. Aperture photometry was performed on the UVOT sky images using the \texttt{uvotsource} routine as available in the 6.29 release of the HEASoft software. We use the recommended \qty{5}{''} aperture, or approximately twice the instrument point-spread function, in order to avoid aperture corrections and to keep the results consistent with the magnitude calibrations. The count rate from our source-free \qty{20}{''} background aperture positioned at right ascension $16 \mathrm{h}\,58 \mathrm{m}\,15.65 \mathrm{s}$ and declination $2^{\circ}\,05'\,53.74''$ is subtracted from the calculated source aperture count rate and the magnitude is derived using the coincidence-loss corrected net count rate. 

Due to the lack of UV host galaxy observations prior to the transient, we calculate the contribution of the host using pre-event archival data of the galaxy. Section \ref{sec:host-sed-model} details our approach to deriving host magnitudes by fitting a spectral energy distribution (SED) to the archival data. The host galaxy flux in the UVOT bands were estimated from the results of our population synthesis modeling of the host galaxy (see Section \ref{sec:host-sed-model}). To correct for the enclosed energy in the photometric reduction of the UVOT sky images, we scaled the host flux by the fraction expected to fall within the source aperture of each UVOT band. The value is determined by taking the ratio of archival Galaxy Evolution Explorer \citep[GALEX;][]{martin2005} far-ultraviolet (FUV), near-ultraviolet (NUV), and Pan-STARRS $g$, $r$, $i$, $z$, $y$ flux within a \qty{5}{''} radius to the amount within \qty{14}{''}, the estimated radius which contains 100\% of the host flux as extrapolated from the reported GALEX GCAT \citep{seibert2012} values. The ratios are interpolated to the UVOT bands and results in scaling fractions of 0.56, 0.60, and 0.66 for the UVW2, UVM2, and UVW1 bands, respectively. Applying this correction to the host flux in each UVOT band and subtracting from the extracted photometry yields the host-subtracted transient flux.

\subsubsection{XRT Observations} \label{sec:obs-x-ray}

In addition to Swift UVOT observations, we obtained X-ray data with Swift's X-Ray Telescope \citep[XRT;][]{burrows2005} through target-of-opportunity requests. An initial analysis revealed a notable X-ray source at the position of \nov{}, clearly visible in stacked images generated using the UK Swift Science Data Centre's online analysis tools \citep{evans2009}. We further supplemented these observations with a late-time XMM-Newton observation taken \qty{786}{days} post-discovery, as well as archival ROSAT data from the host approximately \qty{30}{years} prior to the detection of \nov{}. To analyze the evolution of the X-ray emission in detail, we used \texttt{PyXspec}\footnote{https://heasarc.gsfc.nasa.gov/docs/xanadu/xspec/python/html/index.html}, the Python interface to the \texttt{XSPEC} package \citep{arnaud1996}.

We measure a 3$\sigma$ upper limit of \qty{1.61e43}{erg.s^{-1}} on the absorbed luminosity in the \qtyrange{0.3}{10}{keV} energy range from the ROSAT observation.
While not highly constraining, this limit suggests that the host galaxy was not exhibiting quasar-like activity during the 1990 observations \citep[e.g.][]{yuan1993,green1995}. However, if the actual X-ray emission was below the derived upper limit, it could still be consistent with levels typical of Seyfert galaxies or Low-Ionization Nuclear Emission-line Regions (LINERs) \citep{roberts2000,hernandez-garcia2016}. This aligns with the BPT diagnostics discussed in Section \ref{sec:emission-line-profiles}.
The Swift X-ray observations span from \qty{36}{days} before to \qty{737.5}{days} after the optical/UV peak, with a single late-time XMM-Newton observation taken as the X-ray emission was fading. To enhance the signal-to-noise ratio in the extracted count rates, we grouped the data into time bins of variable size. All count rates and hardness ratios have been checked for whether they are consistent with a random fluctuation or exceed the background.

\begin{figure}
    \centering 
    \includegraphics[width=\columnwidth]{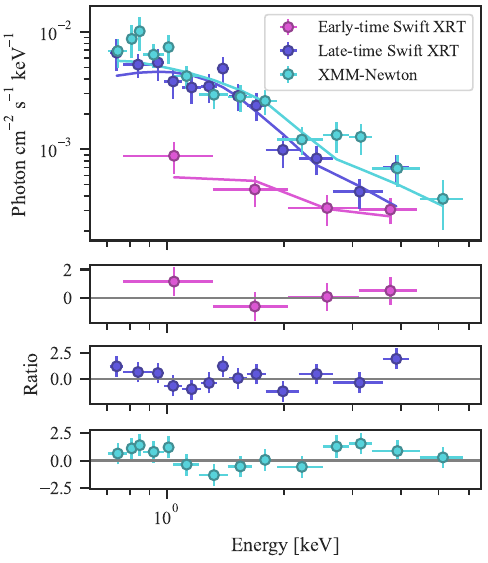}
    \caption{X-ray spectra from the merged early- (magenta) and late-time (purple) Swift XRT and XMM-Newton observations (cyan). Fitted absorbed power law models are shown as solid lines. Residuals are shown in the bottom panels.}
    \label{fig:xrt_spec}
\end{figure}

Swift X-ray emission is detected in the period leading up to the optical/UV peak, and again $\sim$\qty{300}{days} after the peak. Although the coverage is relatively sparse, these observations sample distinct regimes in the evolution of \nov{}, and we therefore divide the X-ray spectral reduction into two epochs: during the rise to the optical/UV flare ($\lesssim \qty{59078}{MJD}$) and during the X-ray brightening (\qtyrange[range-units=single]{59078}{59670}{MJD}). We additionally extract the spectrum for the late-time XMM-Newton observation. We fit the spectra with a model composed of an absorbed power law component redshifted to the host. A \texttt{TBabs} \citep{wilms2000} Galactic absorption component is also included, fixed to a neutral column density along the line of sight of $N_\mathrm{H} = \qty{6.38e20}{cm^{-2}}$ \citep{hi4pi2016}. The extracted spectra and power law model are shown in Figure \ref{fig:xrt_spec}.

The best-fit power law models yield power law indices of $\Gamma = \unum{0.93}{0.42}{0.41}$, \unum{2.17}{0.32}{0.31}, and \unum{1.75}{0.36}{0.34} for the early-time Swift spectra, late-time Swift spectra, and the late-time XMM-Newton spectrum, respectively. These values are much harder than typical optical TDEs which often have steeper power law indices ($\Gamma \gtrsim 2$) likely due to thermal emission components or softer X-ray spectra as their accretion disks stabilize over time \citep{auchettl2017,jonker2020,guolo2024}. This evolution in the spectral slope of \nov{} could be indicative of changes in the emission mechanisms and physical conditions within the accretion flow. 

\begin{figure}
    \centering 
    \includegraphics[width=\columnwidth]{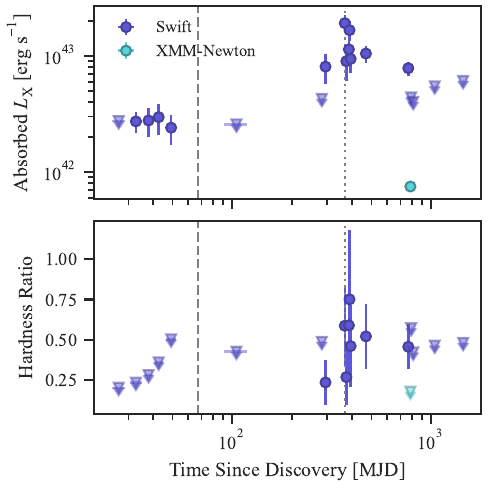}
    \caption{\textit{Top:} Absorbed luminosity of the Swift XRT light curve (purple) and the single XMM-Newton (cyan) observation in the \qtyrange{0.3}{10}{keV} energy band. The X-ray flare occurs $\Delta t = +\qty{300}{days}$ (vertical dotted line) after the optical/UV peak (vertical dashed line). Derived 3$\sigma$ upper limits are shown as inverted triangles. \textit{Bottom:} X-ray hardness ratio computed as $(\mathrm{hard} - \mathrm{soft})/(\mathrm{hard} + \mathrm{soft})$. The hardness ratio peaks during the flare, reaching approximately \num{0.75(0.43)}.
    }
    \label{fig:xrt_lc}
\end{figure}

To convert count rates into flux, flux upper limits, and luminosity, we assumed an absorbed power law with a photon index averaged from the best-fit indices of the extracted spectra. We show the X-ray light curve for our Swift XRT and XMM-Newon observations in Figure \ref{fig:xrt_lc}. We follow the formalism of \citet{auchettl2017} and calculate the hardness ratio evolution using $(H-S)/(H+S)$, where $S$ is the count rate in the \qtyrange{0.3}{2}{keV} band and $H$ the count rate in the \qtyrange{2}{10}{keV} band. The hardness ratio reaches a maximum of $\mathrm{HR} \approx \num{0.75(0.43)}$ during the period of X-ray brightening, but in general appears harder than typical optically selected X-ray TDEs with HR values consistent with type I and type II AGN \citep{auchettl2018, guolo2024}. Given that the emission remains relatively hard over time, it is plausible that the observed X-ray flare involves early interactions or is influenced by a high-temperature outflow or jet. Alternatively, the hard emission could result from lower-energy disk photons being scattered by a hot corona \citep{hajela2024}.

A late-time X-ray brightening event occurs $\sim$\qty{300}{days} after the optical/UV peak, with the peak reaching an absorbed luminosity of \qty{1.67(0.31)e43}{erg.s^{-1}}, and flux of \qty{8.5(1.5)e-13}{erg.cm^2.s^{-1}}, almost an order of magnitude above the pre-flare observations.

\subsection{IR Data}

\begin{figure}
    \centering 
    \includegraphics[width=\columnwidth]{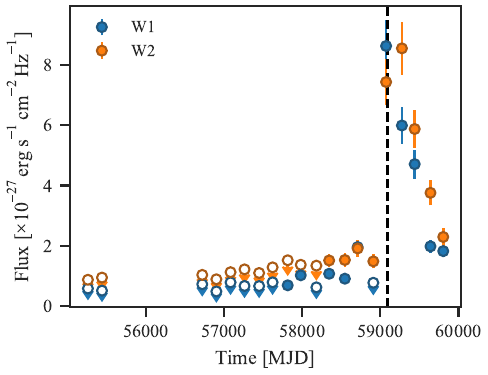}
    \caption{Mid-IR light curves of WISE \qty{3.4}{\micron} (blue) and \qty{4.6}{\micron} (orange) observations. The 7$\sigma$ upper limits are shown as open circles with downward arrows. The time of the optical/UV peak is shown as the vertical dashed line, coincident with the dramatic increase in the IR flux. Evidence for activity begins $\sim$\qty{3.5}{years} before the flare associated with \nov{} (See Section \ref{sec:ir-discussion}).}
    \label{fig:ir_lc}
\end{figure}

\nov{} was observed in mid-infrared (MIR) W1 (\qty{3.4}{\micron}) and W2 (\qty{4.6}{\micron}) bands over the course of $\sim$\qty{12}{yrs} as part of the NASA WISE and NEOWISE missions \citep{mainzer2014}. Data was reduced as part of a systematic search of the survey data for transients conducted by \citet{masterson2024}. The procedure involved the use of a modified code by \citet{de2020} based on the ZOGY algorithm \citep{zackay2016} to perform image subtraction directly on the time-resolved, coadded W1 and W2 data released as part of the unWISE project \citep{lang2014,meisner2018}. In total, 20 epochs of observations were obtained over $\sim$\qty{12.5}{years} with each epoch composed of an unWISE stack of NEOWISE data containing 12 exposures. Each exposure is 7.7 seconds and was acquired over approximately one day at the same sky position. To avoid time-varying artifacts as a consequence of scattered light, a 7$\sigma$ detection threshold was used for the WISE data. We show the MIR light curve in Figure \ref{fig:ir_lc}.

\subsection{Spectroscopy} \label{sec:obs-spec}

Follow-up spectra of \nov{} were primarily obtained over a period of $\sim$15 months, with a final follow-up spectrum taken $\sim$\qty{3}{years} after discovery. Observations were performed with the Kast Double Spectrograph on the Lick Shane \qty{120}{in} telescope \citep{miller1994}, the Low-Resolution Imaging Spectrometer \citep[LRIS]{oke1995} on the \qty{10}{m} Keck~I telescope, the Wide-Field Spectrograph \citep[WiFeS]{dopita2007,dopita2010} on the ANU \qty{2.3}{m} telescope, and the Goodman High Throughput Spectrograph \citep[GHTS]{clemens2004} on the \qty{4.1}{m} Southern Astrophysical Research Telescope (SOAR). Additional spectroscopic data were obtained from the FLOYDS spectrograph on the Faulkes Telescope North (FTN) and South (FTS) instruments of the Las Cumbres Observatory \citep{brown2013}. Including the classification spectrum taken by the SED Machine IFU spectrograph, nine of the 32 spectra were obtained before the light curve peak. A foreground extinction correction of $E(B-V)=\num{0.0960(0.0019)}$ was applied to all spectra using the dust maps of \citet{schlafly2011} and the extinction curve from \citet{fitzpatrick1999}. The spectra are additionally corrected for redshift.

\begin{figure*}
    \centering
    \includegraphics[width=\textwidth]{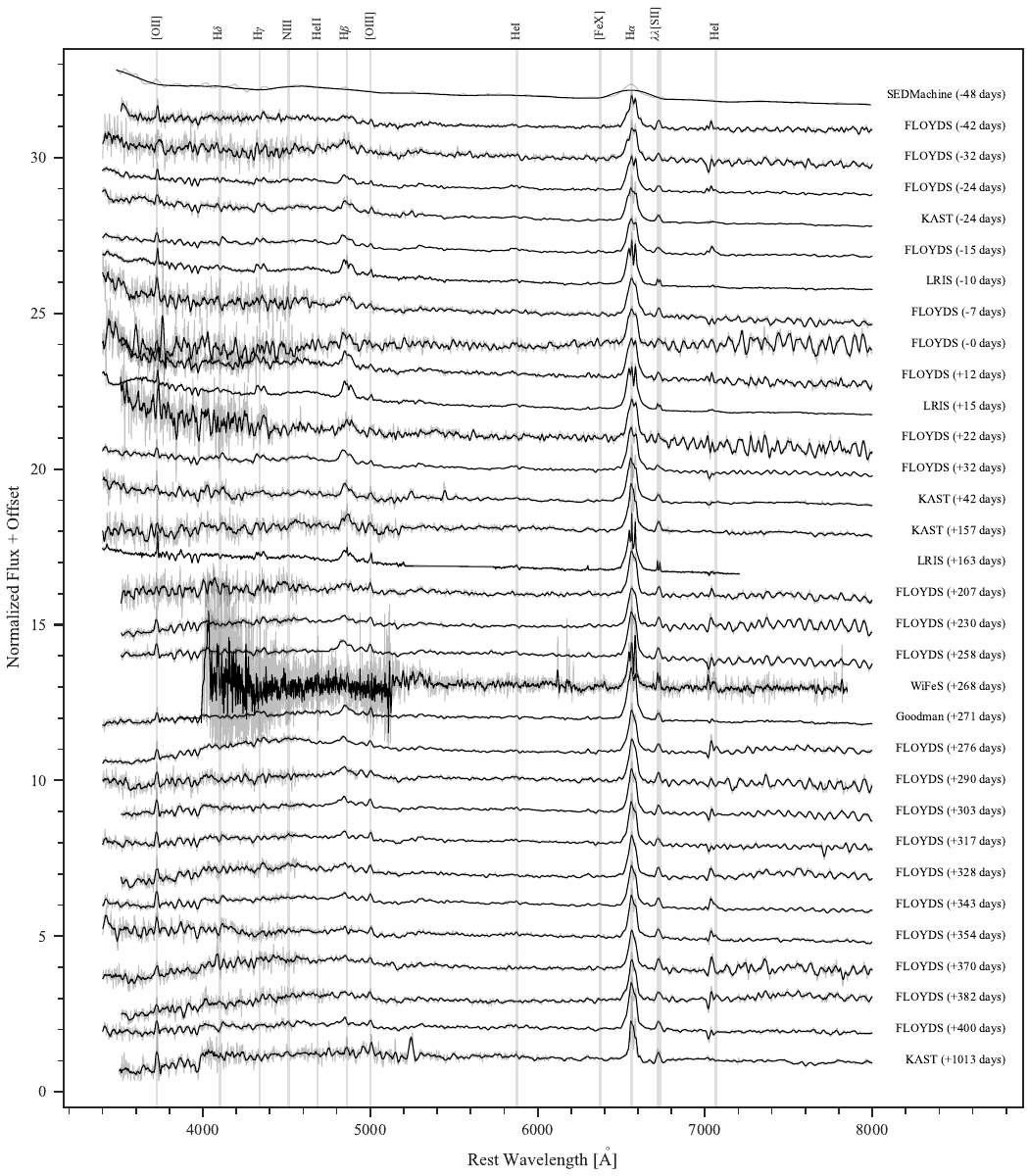}
    \caption{Spectroscopic evolution of \nov{}, covering \qty{-48}{days} before peak (2020 August 26) to \qty{1013}{days} after peak. Section \ref{sec:obs-spec} describes the flux calibration to the photometry. Common emission features of TDEs are indicated with light grey lines. The double-peaked features are not well resolved in the FLOYDS spectra, but integrated values of their Balmer features follow the trend of the higher resolution spectra that indicate the double-peaked features disappear by \qty{1013}{days} after peak. Smoothed spectra using a Savitsky-Golay filter are shown in black over top the normalized flux (grey).}
    \label{fig:all_spectra_with_floyds}
\end{figure*}

All spectra include the host contribution as no pre-flare spectra of the host is available. The spectra are scaled to ZTF $r$-band photometry and are shown with corrections in Figure \ref{fig:all_spectra_with_floyds}. The evolution of the spectra show an initial hot, blue continuum before becoming host-dominated at late times. Vertical grey lines highlight the location of typical TDE emission lines whose behavior evolves over the span of spectroscopic observations.

The spectral data from the Kast, Goodman, and LRIS instruments were reduced using the {\tt UCSC Spectral Pipeline}\footnote{\url{https://github.com/msiebert1/UCSC\_spectral\_pipeline}} \citep{siebert2019b,foley2003,silverman2012b}. The optical algorithm of \citet{horne1986} was used to extract the one-dimensional spectra from the combined two-dimensional spectral frames after bias, flat-field, and gain variation corrections were applied. The two-dimensional spectral frames were also trimmed, and cosmic-ray rejection performed using the \texttt{pzapspec} algorithm prior to combining. Internal comparison-lamp spectra were used for the wavelength calibration with linear shifts determined by cross-correlation of the sky lines with a master night-sky spectrum. Flux calibration involved utilizing standard stars with similar airmass as that of the scientific exposures, choosing both ``blue'' (hot subdwarfs or sdO) and ``red'' (low-metallicity G/F) standard stars in the selection. Atmospheric extinction corrections were derived using the telluric absorption in the standard stars which was determined by fitting their flux-calibrated continuum, with the strength of the absorption estimated using the relative airmass between the standard star and the scientific image. Cross-correlation was used to identify and account for minor shifts in the telluric A and B bands. In the case of dual-beam spectrographs, the two sides were combined by scaling one spectrum to align with the flux of the other in the overlap region and using their error spectra for correct spectral weighting.
oto
WiFeS spectra were obtained using the Wide Field Spectrograph (WiFeS) mounted on the \qty{2.3}{meter} telescope at the Australian National University (ANU2.3m) located at Siding Spring Observatory (SSO) \citep{dopita2007,dopita2010}. The spectra were captured in `Classical Equal' mode, utilizing the $R = 7000$ grating to cover the wavelength range of \qtyrange{4180}{7060}{\text{\AA}}. Data reduction was performed using \texttt{PyWiFeS} \citep[version 0.7.4]{childress2014}, with sky subtraction based on a 2D sky spectrum obtained during the observations. For further details, see \citet{carr2024}.

FLOYDS optical spectra cover \qtyrange{3500}{10000}{\text{\AA}} at a resolution $R \approx \numrange[range-phrase=\textup{--}]{300}{600}$ and were reduced using the \texttt{floydsspec} pipeline\footnote{\url{https://github.com/svalenti/FLOYDS_pipeline/}}, which performs flux and wavelength calibration, cosmic-ray removal, and final spectrum extraction and is described in \cite{valenti2014}.

\section{Photometric Analysis} \label{sec:photo-analysis}

\subsection{Host SED Model} \label{sec:host-sed-model}

To estimate the host galaxy's contribution to the UV bands and to measure properties such as stellar mass, metallicity, and dust extinction, we utilize archival optical and UV observations of LEDA 1216501. Although the host galaxy lies outside the field of spectroscopic surveys like the Sloan Digital Sky Survey (SDSS), data are available from the Pan-STARRS and GALEX surveys. The Pan-STARRS DR1 dataset provides pre-transient photometry of the host in the $g$, $r$, $i$, $z$, and $y$ filters. Additionally, we obtained FUV and NUX data from the archival GALEX All-Sky Imaging Survey (AIS). Near-infrared data in the $J$, $H$, and $K_s$ bands are available from the Two Micron All Sky Survey \citep[2MASS;][]{cutri2003}, and photometry in the $W1$ to $W4$ bands comes from the Wide-field Infrared Survey Explorer \citep[WISE;][]{wright2010}. The archival host magnitudes are listed in Table \ref{tab:host_data}.

\begin{table}
\centering
\caption{Archival Host Photometry of LEDA 1216501}
\label{tab:host_data}
\begin{tabular}{lccc}
\toprule
Filter & Magnitude & Uncertainty & Source \\ 
\midrule
FUV & 20.82 & 0.23 & GALEX \\
NUV & 20.09 & 0.07 & GALEX \\
$g$ & 17.56 & 0.01 & Pan-STARRS \\ 
$r$ & 16.82 & 0.01 & Pan-STARRS \\
$i$ & 16.49 & 0.01 & Pan-STARRS \\
$z$ & 16.29 & 0.01 & Pan-STARRS \\ 
$y$ & 16.22 & 0.02 & Pan-STARRS \\
$J$ & 15.70 & 0.10 & 2MASS \\
$H$ & 15.33 & 0.09 & 2MASS \\
$K_s$ & 15.49 & 0.11 & 2MASS \\
$W1$ & 16.05 & 0.01 & WISE \\
$W2$ & 16.48 & 0.02 & WISE \\
$W3$ & 14.92 & 0.05 & WISE \\
$W4$ & 14.29 & 0.25 & WISE \\
\bottomrule
\end{tabular}
\tablecomments{Magnitudes use the AB photometric system.}
\end{table}

The physical properties of the galaxy were extracted using stellar population synthesis models fit to the (SED) composed of the archival photometry. The fitting was performed using the \texttt{Bagpipes} \citep{carnall2018} code with a non-parametric star formation history (SFH). The chosen parameter set allows for the metallicity, stellar mass, and current star-formation rate to vary. We use the provided \citet{calzetti2000} attenuation curve to model contribution from dust, and therefore include the overall normalization as a free parameter in the fitting. The outputs of SED fitting are strongly dependent on the parameter priors, as such we make use of Gaussian- and $\log_{10}$-based priors where appropriate to deal with the complications of the age-dust-metallicity degeneracy \citep{bell2001}, and rely on the ability of \texttt{Bagpipes} to fully sample the posterior probability density.

\begin{figure}[t]
 \centering 
 \includegraphics[width=\columnwidth]{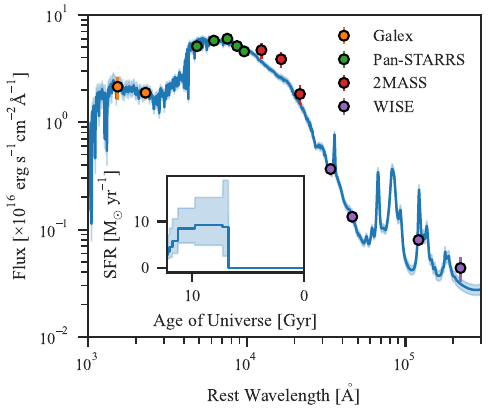}
\caption{Model of the host galaxy SED fit using \texttt{Bagpipes}. The best-fit model and 1-$\sigma$ dispersion of the realizations are shown in blue, while the archival photometry are plotted as colored circles. The star-formation history derived from \texttt{Bagpipes} is shown in the inset, indicating a relatively constant SFR before abruptly ending $\sim$\qty{7}{Gyr} ago.}
\label{fig:prosp_res}
\end{figure}

The best-fit model to the host SED is shown plotted with a 1-$\sigma$ dispersion in Figure \ref{fig:prosp_res}. The fit results in a value of $\log_{10}(\mathrm{M_*}/M_\odot) = \StellarMass$ for the stellar mass and a metallicity of $\mathrm{Z}/Z_\odot = \Metallicity$. We note a modest dust extinction value of $A_\mathrm{V} = \Av$. A specific star formation rate of $\log_{10} \mathrm{sSFR} = \sSFR$ is derived using the fit star formation rate and stellar mass. The star formation history indicates that strong star formation ended in the galaxy approximately \qty{7}{Gyr} ago.

We leveraged the method outlined by \citet{stern2012} to assess the presence of AGN dust heating using WISE photometry, focusing specifically on the mid-infrared color indicators. The $W1 - W2$ color, which we measured to be \qty{0.22(0.03)}{Vega.mag}, does not strongly support the presence of a robust, pre-existing AGN, as this value is below the typical threshold associated with AGN-dominated systems \citep{stern2012}. However, it's important to note that this does not entirely rule out the presence of a weaker AGN, as some low-luminosity or obscured AGNs can exhibit similarly modest $W1 - W2$ colors.

To further refine our analysis, we applied the two-dimensional color cut method from \citet{wright2010}, which incorporates the $W2 - W3$ color. Our findings show that $W2 - W3 = \qty{3.38(0.06)}{Vega.mag}$, placing the object in a region of the color-color diagram typically occupied by bluer, star-forming spiral galaxies and Luminous Infrared Galaxies (LIRGs). This position in the color space is more indicative of galaxy-dominated emission rather than AGN activity, suggesting that the mid-infrared emission is likely driven by star formation processes within the host galaxy. Nonetheless, the possibility of a weak AGN contributing to the observed emission cannot be completely excluded, particularly in cases where the AGN is heavily obscured or intrinsically faint. Thus, while the WISE color cut does not suggest a strong pre-existing AGN, the data still leave room for the presence of a weaker AGN component. 

\subsection{SED Modeling} \label{sec:sed-modeling}

% Previous studies have demonstrated that the optical and UV spectral energy distributions (SEDs) of TDEs can often be well modeled using a black body \citep{gezari2014,arcavi2014,velzen2021}, which provides estimates of the bolometric luminosity, temperature, and radius of the thermalized emission. However, for \nov{}, we find that a single black body model is insufficient to capture the full shape of the SED at any epoch (see Figure \ref{fig:sed-fits}). While the peak emission may lie blueward of the optical/UV bands, the observed data consistently display a significant red excess, which is not well explained by the red tail of a single black body spectrum. Instead, the optical/UV data approximately follow a $\nu^{4/3}$ power law in $\nu L_\nu$, suggesting that this excess arises from reprocessed emission within a relatively quiescent, flat disk structure surrounding the black hole \citep{chiang1997}.

Previous studies have shown that the optical and UV spectral energy distributions (SEDs) of TDEs are often well described by a black body model \citep{gezari2014,arcavi2014,velzen2021}. This enables estimates of the bolometric luminosity, temperature, and the thermalized radius, which is inferred from the Stefan-Boltzmann law, $R_\text{th} = \sqrt{{L_\text{bol}}/{(4\pi \sigma T^4)}}$, where $\sigma$ is the Stefan-Boltzmann constant. This thermalized radius represents the spatial scale from which the majority of the black body emission originates, often interpreted as the radius of an optically thick reprocessing layer \citep{loeb1997, guillochon2014}.

\begin{figure*}[t]
    \centering 
    \includegraphics[width=\textwidth]{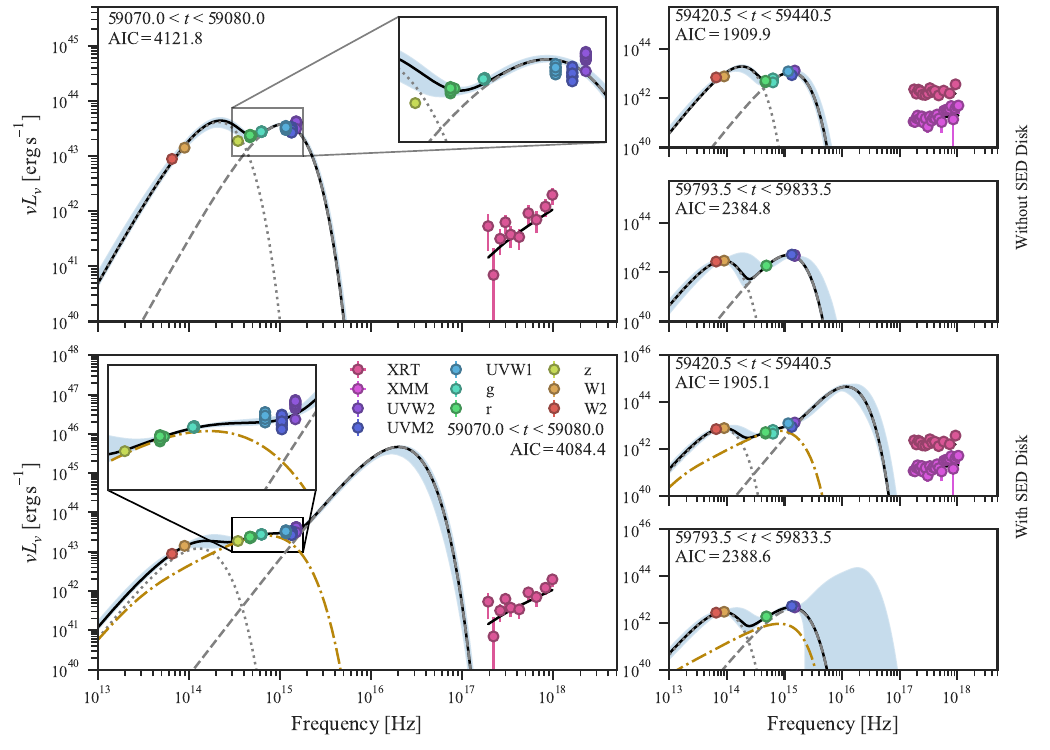}
    \caption{Fitting results for two approaches to modeling the SED of \nov{}. The top row uses a traditional approach of representing the optical/UV emission with a single black body (grey dashed line), while the bottom row additionally includes the passive SED disk model (gold dash-dotted line) proposed to reprocess the EUV emission from the TDE as described in Section \ref{sec:sed-modeling}. We simultaneously model the MIR emission (grey dotted line) in both cases in order to capture the behavior of the dust distribution around the SMBH. The relative fit quality is quantified by the AIC which strongly favors the addition of the disk to capture the optical excess in the redder optical bands. Swift XRT data (pink), XMM-Newton data (magenta), and their associated power law fit are also shown but not used in the fitting.}
    \label{fig:sed-fits}
\end{figure*}

For \nov{}, however, a single black body model fails to reproduce the full shape of the SED at any observed epoch (see the top row of Figure \ref{fig:sed-fits}). Although the peak of the emission likely lies blueward of the optical/UV bands, the observed SED consistently shows a shallower slope than can be fit by a single black body. The optical/UV SED is therefore inconsistent with emission from a simple thermal source. Instead, the optical and UV data approximately follow a $\nu^{4/3}$ power-law dependence in $\nu L_\nu$, which suggests that some of the optical/UV flux emission originates from reprocessed radiation within a geometrically thin, optically thick disk structure surrounding the black hole \citep{chiang1997}. Thus, the primary emission of the TDE likely originates in the EUV, while the observed optical/UV power-law emission is better attributed to the disk's thermal continuum.

To more accurately model the SED, we employ a three-component approach: (1) a black body to capture the TDE emission peaking in the EUV, (2) a passive (i.e. lacking intrinsic luminosity) disk model that reprocesses EUV radiation and contributes to the observed optical/UV emission, and (3) an additional black body to capture the MIR emission observed by WISE. This MIR component is modeled simultaneously since it is affected by the redder portions of the disk model and helps account for any extended thermal emission from cooler regions within the TDE environment. Although a black body is used to model the MIR component, we acknowledge that its shape may be difficult to constrain given the limited MIR data available, and the emission region may not be perfectly represented by a simple black body.

We also include our X-ray observations from Swift on Figure \ref{fig:sed-fits} and show the model fits derived in the previous section using the \texttt{XSPEC} software package. Although these X-ray data provide insight into the high-energy processes near the black hole, they are not used directly in the optical/UV SED fitting. In TDEs, the X-ray emission is thought to arise from a distinct component, such as an inner accretion disk or corona, which is separate from the thermal emission that produces the optical/UV SED. The X-ray model serves to complement our understanding of the multi-wavelength behavior of \nov{} without influencing the parameter estimation for the optical/UV SED. %It is clear that the energy emitted in the optical/UVcannot originate in the X-ray, supporting the view that the primary emission comes from the EUV. 

By incorporating these three components--the EUV black body, the passive disk, and the MIR black body--our SED model aims to capture the complexities of \nov{}'s emission. The reprocessed EUV emission from the passive disk, in particular, appears to be the primary driver of the optical/UV excess. We compute the SED of the disk structure following the approach of \citet{chiang1997} as
\begin{align}
    \nu L_\nu = 4 \pi d^2 \nu F_\nu = 8 \pi^2 \nu \int_{a_\text{in}}^{a_\text{out}} da \; a B_\nu (T_e),
\end{align}
where $B_\nu(T_e)$ is the Planck function and $d$ is the distance to the source with the integration done from the inner ($a_\text{in}$) to the outer ($a_\text{out}$) radius of the disk. For a thin disk with an aspect ratio that does not scale with radius, the temperature, $T_e$, takes the form
\begin{align}
    T_e \approx \left( \frac{2}{3\pi} \right)^{1/4} \left( \frac{R_\circledast}{a} \right)^{3/4} T_\circledast,
\end{align}
where $R_\circledast$ and $T_\circledast$ are the radius and temperature of the TDE black body respectively, and are tied to corresponding values in the TDE black body model during fitting. This multi-component modeling offers a more thorough representation of the various emission regions, providing a clearer understanding of the TDE environment and the impact of an extended disk structure on the observed SED.

\begin{table}
\centering
\caption{Comparison of SED Fitting Results}
\label{tab:bb-params}
\begin{tabular}{lrrrrr}
\toprule
\# & \multicolumn{2}{c}{IR Black Body} & \multicolumn{2}{c}{TDE Black Body} & Disk \\ 
\midrule
 & $\log_{10} T$ & $\log_{10} R$ & $\log_{10} T_\circledast$ & $\log_{10} R_\circledast$ & $\log_{10} a_\text{in}$ \\ 
 & [K] & [cm] & [K] & [cm] & [cm] \\
\midrule
\multicolumn{6}{c}{With SED Disk} \\ 
\midrule
1 & $3.19_{-0.16}^{+0.11}$ & $16.6_{-0.14}^{+0.22}$ & $5.38_{-0.18}^{+0.03}$ & $13.6_{-0.06}^{+0.17}$ & $14.9_{-0.11}^{+0.07}$ \\
2 & $3.02_{-0.09}^{+0.14}$ & $16.9_{-0.25}^{+0.17}$ & $5.15_{-0.14}^{+0.14}$ & $13.5_{-0.26}^{+0.15}$ & $14.5_{-0.37}^{+0.23}$ \\
3 & $3.03_{-0.10}^{+0.18}$ & $16.7_{-0.31}^{+0.20}$ & $4.54_{-0.39}^{+0.79}$ & $13.9_{-0.94}^{+0.61}$ & $14.2_{-0.52}^{+1.08}$ \\ 
\midrule
\multicolumn{6}{c}{Without SED Disk} \\ 
\midrule
1 & $3.41_{-0.16}^{+0.02}$ & $16.5_{-0.05}^{+0.16}$ & $4.13_{-0.02}^{+0.04}$ & $15.0_{-0.05}^{+0.04}$ & -- \\
2 & $3.34_{-0.17}^{+0.02}$ & $16.5_{-0.05}^{+0.22}$ & $4.20_{-0.04}^{+0.09}$ & $14.6_{-0.10}^{+0.06}$ & -- \\
3 & $3.04_{-0.10}^{+0.24}$ & $16.6_{-0.35}^{+0.20}$ & $4.16_{-0.02}^{+0.34}$ & $14.5_{-0.51}^{+0.03}$ & -- \\ 
\bottomrule
\end{tabular}
\tablecomments{Numbers correspond to the epochs defined in Section \ref{sec:sed-modeling}.}

\end{table}

To analyze the SED of \nov{}, we divide the photometry into three distinct epochs that satisfy specific observational requirements: each epoch must contain UV-band and optical-band observations as well as infrared data from the WISE observations. These criteria allow us to examine the transient event at key stages, resulting in epochs that roughly correspond to the following phases: the initial peak of the transient, a possible plateau phase marked by X-ray brightening, and a late-time phase where the transient fades into a remnant stage. 

For each epoch, we fit the SED using the three-component model composed of a black body to capture the EUV emission of the TDE, a passive disk model to address the optical/UV excess of the reprocessed disk emission, and a second black body model for the MIR emission. The SED fits for each epoch are shown in Figure \ref{fig:sed-fits}, where the corresponding X-ray data are also included for reference. As mentioned previously, while the X-ray observations are shown, they are not incorporated into the fitting process as X-ray emission often probe different parts of the disk and is generally decoupled from the optical/UV emission. To assess the model fits, we also show a comparison between a single-component black body fit to the optical/UV photometry (top row) and the more complex model including the passive disk component (bottom row).

Model selection is evaluated using the Akaike Information Criterion \citep[AIC;][]{akaike1974}, which balances fit quality with model complexity, thereby reducing the likelihood of overfitting. Unlike the chi-squared metric, which purely minimizes discrepancies between observed and predicted values, AIC penalizes models with excessive parameters, making it more appropriate for comparing models of varying complexity. For the first epoch, the AIC analysis strongly favors the inclusion of the passive disk component, with $\Delta \mathrm{AIC} > 10$, indicating a significantly improved fit. In the second epoch, the inclusion of the disk is again favored, with $3 < \Delta \mathrm{AIC} < 7$. By contrast, the third epoch favors a simpler single-component black body model, as the disk contribution diminishes and becomes difficult to constrain due to the lack of multi-band optical/UV data at late times. This lack of data increases uncertainties around the peak of the TDE black body, which we represent as a shaded blue region. The fitted parameters for each epoch are summarized in Table \ref{tab:bb-params}.

The fits to the SEDs provide strong evidence for a substantial passive disk-like structure, which reprocesses high-energy EUV photons emitted by the TDE accretion flow, re-emitting them at lower energies and producing the observed optical/near-UV emission. This results in an optical/near-UV emission that is substantially lower than the total energy output from the TDE. Simulations of TDEs, such as those by \citet{dai2018}, predict a black body that is EUV-bright, consistent with the component we require in our model to reproduce the observed SED through disk reprocessing. We estimate the total energy output for each of the emission components by integrating $L_\nu$ over frequency, then integrating again over the three epochs. We infer a total energy output from the TDE black body component of $\EradTDE$, with $\EradDisk$ being reprocessed by the extended passive disk and observed in the optical/UV bands. These fitting results suggest that the black body component in \nov{} peaks at a higher temperature and lower radius than typically seen in optical/UV TDEs.

\subsection{Estimating Black Hole Mass}\label{sec:BH_mass}

It is challenging to constrain the black hole mass for \nov{} due in part to the flattening caused by excess flux in the redder bands of the SED which disallows traditional bolometric luminosity calculations derived from single black body fits (e.g. \texttt{MOSFiT} \citep{mockler2019} or \texttt{TDEMass} \citep{ryu2020b}). We instead utilize scaling relations for the black hole mass with the stellar velocity dispersion ($M_\text{BH}$--$\sigma_*$) and galaxy stellar mass ($M_\text{BH}$--$M_\text{stellar}$) to provide estimates of the black hole mass.

Using the $M_\text{BH}$ versus $M_\text{stellar}$ scaling relation described in \citet{reines2015} for their primary sample of local broad-line AGN, and using our host stellar mass of $\log_\text{10} M_\mathrm{stellar}/ \msun = \StellarMass$ derived from the population synthesis analysis (see Section \ref{sec:host-sed-model}), we find a black hole mass of $\log_{10} M_\text{BH} / \msun = \MbMstRV$. \citet{greene2020} similarly adopt the dynamical sample from \citet{kormendy2013} and supplement with recently published galaxies to derive their relation, which provides us an estimate of $\log_{10} M_\text{BH} / \msun = \MbMstGr$. Finally, \citet{yao2023} find a relation for TDEs by fitting a linear model to the inferred $M_\text{BH}$ of the TDE hosts and the stellar mass derived from galaxy SED fitting. Their results yield us an estimate of $\log_{10} M_\text{BH} / \msun = \MbMstYao$.

Additionally, we measure the stellar velocity dispersion from the optical spectrum taken with Kast in 2021. We use \texttt{pPXF} \citep{cappellari2017} to fit the observed spectrum with a set of stellar templates \citep{verro2022} which are shifted and broadened to match the observed stellar absorption features, while masking the emission lines. We run a Monte Carlo simulation (500 realizations) to estimate the 1$\sigma$ uncertainty on the measured stellar velocity dispersion. We obtain $\sigma = \qty{127(28)}{km.s^{-1}}$. Using the $M_\text{BH}$--$\sigma_*$ scaling relation from \citet{kormendy2013}, we obtain an estimate for the black hole mass of $\log_{10} M_\text{BH} / \msun = \MbsigKH$ based on the stellar velocity dispersion. \citet{greene2020} also provide an updated scaling relation which yields an estimate of $\log_{10} M_\text{BH} / \msun = \MbsigGr$.

Given the consistency between the $M_\text{BH}-\sigma_*$ and $M_\text{BH}-M_\text{stellar}$ relationships from \citet{greene2020}, we adopt the black hole mass derived from our measured velocity dispersion as the nominal value for \nov{}, obtaining $\log_{10} M_\text{BH} / M_\odot = \MbsigGr$. We infer the bolometric luminosity of $\log_{10} L_\text{peak} = \LpeakTDE$ by measuring the maximum luminosity of the TDE EUV component in our SED fit during the first epoch (see Figure \ref{fig:sed-fits}), as the majority of the bolometric luminosity originates from the EUV emission and is subsequently reprocessed by surrounding material to lower energy wavelengths (see Section \ref{sec:euv-primary}).

With this peak luminosity, we calculate an Eddington ratio ($\lambda_\text{Edd} = L_\text{BB} / L_\text{Edd}$) of approximately 1.25, where $L_\text{Edd}$ is given by $L_\text{Edd} = 4 \pi G c M_\text{BH} / \kappa$, and we adopt $\kappa \approx \qty{0.34}{cm^2.g^{-1}}$ as the opacity for electron scattering. This Eddington ratio indicates that \nov{} is in the super-Eddington regime, where the luminosity exceeds the limit at which radiation pressure would theoretically halt accretion under isotropic conditions. Such super-Eddington accretion is consistent with theoretical predictions for TDEs involving black holes with masses less than $\sim$\qty{3e7}{\msun} \citep{strubbe2009,lodato2011,decolle2012,metzger2016}. Optical/UV TDEs have been shown to exhibit a wide range of Eddington ratios, spanning from sub-Eddington to super-Eddington regimes \citep{auchettl2017,yao2023}.

\subsection{Fallback Rate}

We attempted to fit the $g$- and $r$-band luminosities of \nov{} using the light curve model proposed by \citet{velzen2021}. However, this model imposes a hard transition between the rising and declining phases of the light curve--a feature that is not evident in \nov{} due to its broad peak. The absence of a sharp transition makes it challenging to reliably capture the fallback behavior and to determine when the rise ends and the fallback begins. This is particularly important when testing the theoretical $t^{-5/3}$ decline expected for TDEs, as the determination of the fallback rate is highly sensitive to the choice of starting time and the modeling of the rise and fall phases.

To address this issue, we adopt a smoothly broken power-law (SBPL) model, which is often used in light curve fitting of other transients \citep{ryde1999, schulze2011}. The SBPL model introduces a smoothness parameter ($\Delta$) that allows for a gradual transition between the rise and decline phases, ensuring proper treatment of the peak in the light curve without imposing a hard break. The SBPL function is defined as:
\begin{align}
    f(t) = A \left( \frac{t}{t_b} \right)^{-\alpha_1} \left\{ \frac{1}{2} \left[ 1 + \left( \frac{t}{t_b} \right)^{1/\Delta} \right] \right\}^{(\alpha_1 - \alpha_2)\Delta},
\end{align}
where $\alpha_1$ and $\alpha_2$ are the power-law indices for the rise and decline phases respectively, $t_b$ is the break time corresponding to the peak of the light curve, and $A$ is a normalization constant.

\begin{figure}[t]
    \centering 
    \includegraphics[width=\columnwidth]{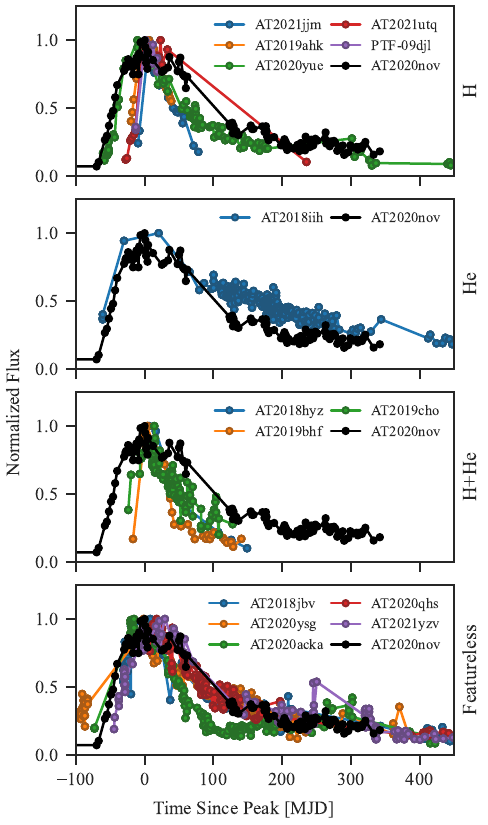}
    \caption{Comparison of the $r$-band light curves of \nov{} (black) and a selection of TDEs whose estimated $\log M_\mathrm{BH}$ are within $\pm \qty{0.5}{dex}$ of \nov{}, categorized by their spectral type. The sharp rise, broad peak, and shallow fallback are characteristic of TDEs with larger black holes, and does not appear to strongly favor any spectral category.
    }
    \label{fig:lc-compare}
\end{figure}

By fitting this SBPL model to the monochromatic luminosities in the $g$ and $r$ bands, we obtain decline power-law indices of \PLExpG{} and \PLExpR{}, respectively. The inclusion of the smoothness parameter allows us to model the gradual transition observed in \nov{}, accommodating the broad peak and providing a better fit to the data. Figure \ref{fig:lc-compare} illustrates the $r$-band light curve of \nov{}, highlighting its broad-peak nature in contrast to other TDEs categorized by their spectral types. We limit the comparison to TDEs with black holes within $\pm 0.5$ dex of \nov{} within each category, finding that the broad-peak behavior does not appear to be unique to \nov{}, but may instead be a consequence of a large black hole mass.

\begin{figure}[t]
    \centering 
    \includegraphics[width=\columnwidth]{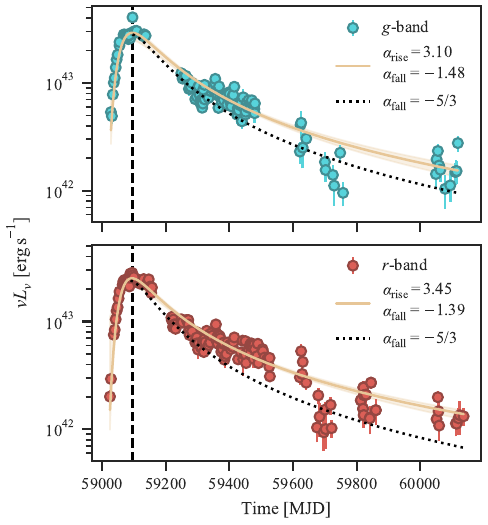}
    \caption{Luminosity rise and decay rates of the $g$- and $r$- band photometry with fits performed using a smoothly broken power law model. The decline from peak follows a $\sim$$t^{-4/3}$ power-law, consistent with super-Eddington disk accretion. The expected fallback rate of $t^{-5/3}$ is shown in the black dotted line. The vertical dashed line shows the time of optical/UV peak.}
    \label{fig:bolo_fits}
\end{figure}

Figure \ref{fig:bolo_fits} presents the fitted SBPL models alongside the theoretical expectation of a $t^{-5/3}$ decline. We find that the $t^{-5/3}$ model provides a reasonable fit to the initial fallback rate, with significant deviations occurring only around \qty{200}{days} after the optical/UV peak. These deviations suggest that the disrupted material from the TDE may have transitioned from being fallback-dominated to disk-dominated, producing late-time disk emission. Such behavior has been observed in a growing number of TDEs \citep{velzen2019b}, indicating that the standard $t^{-5/3}$ decline may not always adequately describe the late-time light curve evolution.

Likewise, the width of the peak in the light curve could indicate differing physical processes depending on the assumed model underlying the tidal disruption. If the primary source of emission comes from an accretion flow, \citet{metzger2016} have shown that the optical radiation will be advected through the subsequent accretion-rate-powered outflow where it then adiabatically transfers a significant amount of energy to the outflow wind. In such a case, for black holes with mass $M_\text{BH} \lesssim \qty{7e6}{\msun}$, the light curve peak can be suppressed and delayed due to the adiabatic losses, both diminishing and potentially extending the peak duration. On the other hand, if the TDE emission is caused by the shocks from debris stream collisions, the escape time of photons from the shock-heated debris may be diffusion-limited, and in sufficiently low mass black holes, the diffusion time will be the dominate timescale of the light curve at peak \citep{auchettl2017,velzen2021}. From our SMBH estimates, it does not appear that either case is applicable to \nov{}.

\subsection{Dust-covering Factor}

The dust covering factor is an important parameter for understanding the environment around supermassive black holes. These factors are typically calculated as the ratio of infrared to optical/UV luminosity, representing the proportion of high-energy emission reprocessed by surrounding dust \citep{stalevski2016}. MIR echoes of TDEs are characterized by significant luminosity increases, typically ranging from \qtyrange{1e41}{1e42}{erg.s^{-1}}. These echoes exhibit a dust-covering factor ($f_c$) of around 1\% at subparsec scales, suggesting a sparse or geometrically thin and flat dust distribution around the SMBH \citep{jiang2021b}. 

Recently, a new population of MIR TDEs has been discovered through their significant MIR flares \citep{masterson2024}. Characterized by peak IR luminosities of $L_{W2} \gtrsim \qty{1e42}{erg.s^{-1}}$ and light curves that show a fast rise followed by a slow, monotonic decline, these MIR-selected TDEs are often missed by optical surveys due to heavy dust obscuration, which reprocesses the optical, UV, and X-ray emission into the IR band. Additionally, they exhibit higher dust-covering factors ($f_c \gtrsim 10\%$) compared to the approximately 1\% seen in optically selected TDEs, indicating that they occur in more dust-rich environments \citep{masterson2024}. In contrast, active galactic nuclei (AGN) show long-term variability and AGN-like WISE colors, while MIR TDEs show no significant prior variability and lack AGN-like colors. Furthermore, AGN typically have much higher dust-covering factors, often close to one-half, as inferred from spectral energy distribution (SED) decomposition of the primary and reprocessed emission \citep{jiang2021b}. This higher $f_c$ in AGN is consistent with the presence of a substantial and geometrically thick torus of dust surrounding the accretion disk.

In addition to mid-infrared TDEs, a class of events known as ambiguous nuclear transients (ANTs) has been observed. ANTs are optical transients that also exhibit MIR flares, and they may represent exotic TDEs or smooth flares originating in AGN. These transients display a wide range of dust-covering factors. For instance, the mean dust-covering factor for detected MIR flares in ANTs is \num{0.38(0.04)}, which is significantly higher than those of optically selected TDEs and similar to those found in AGN \citep{hinkle2024}. This suggests that ANTs often occur in environments with substantial dust.

For \nov{}, we calculate a dust covering factor of approximately 1.1\% when incorporating both the mid-infrared black body emission and the passive disk-like structure from our SED model fits. The passive disk contributes to the obscuration of the primary EUV emission, so the bolometric luminosity of the obscuring material includes contributions from both the IR and disk components, resulting in a calculated luminosity of $\log_\mathrm{10} L_\text{IR+Disk} = \LradIRDisk$. Alternatively, when considering only the mid-infrared black body without the disk contribution, the dust covering factor is reduced to 0.5\%. 

It is important to note that this approach differs from other studies of TDE dust covering factors, which mostly compare optical and IR emissions due to the absence of a disk component to infer the EUV luminosity. Our inclusion of the disk component allows for a more comprehensive assessment of the obscuring material's impact on the EUV emission. The dust covering factor of 1.1\% aligns well with expectations for TDEs, placing \nov{} within the regime of typical optically selected TDEs rather than dust-rich AGN or ambiguous nuclear transients (ANTs).

\subsection{Pre-flare Variability} \label{sec:ir-discussion}

At the WISE mid-IR wavelengths of \qty{3.4}{\micron} (W1) and \qty{4.6}{\micron} (W2), the average variability associated with \nov{} reaches \qty{0.39}{mag} and \qty{0.19}{mag}, respectively, based on photometric observations above the 7-sigma threshold, as shown in the difference imaging in Figure \ref{fig:ir_lc}. This variability was measured with respect to neighboring data points using the structure function, which provides a quantitative assessment of changes over time. A measurable increase in mid-IR flux begins approximately 1,112 days before the mid-IR flare associated with \nov{}, with no significant variability observed between 7 and 4 years prior to the flare. The flare itself corresponds to a sharp rise in mid-IR emission, occurring with a short delay after the initial optical/UV flare. At the time of this mid-IR flare, the variability intensifies, reaching \qty{1.64}{mag} in the W1 band and \qty{1.75}{mag} in W2. For comparison, \citet{kozlowski2016} constrain typical mid-IR variability in dust-obscured AGNs to less than \qty{0.3}{mag} over a 7-year period. In contrast, the mid-IR variability of \nov{}, observed over a shorter span of about 3 years prior to the flare, is notably more pronounced, suggesting that this behavior may reflect either normal AGN variability or variability associated with the TDE in the lead-up to the event.

\begin{figure*}[t]
    \centering 
    \includegraphics[width=\textwidth]{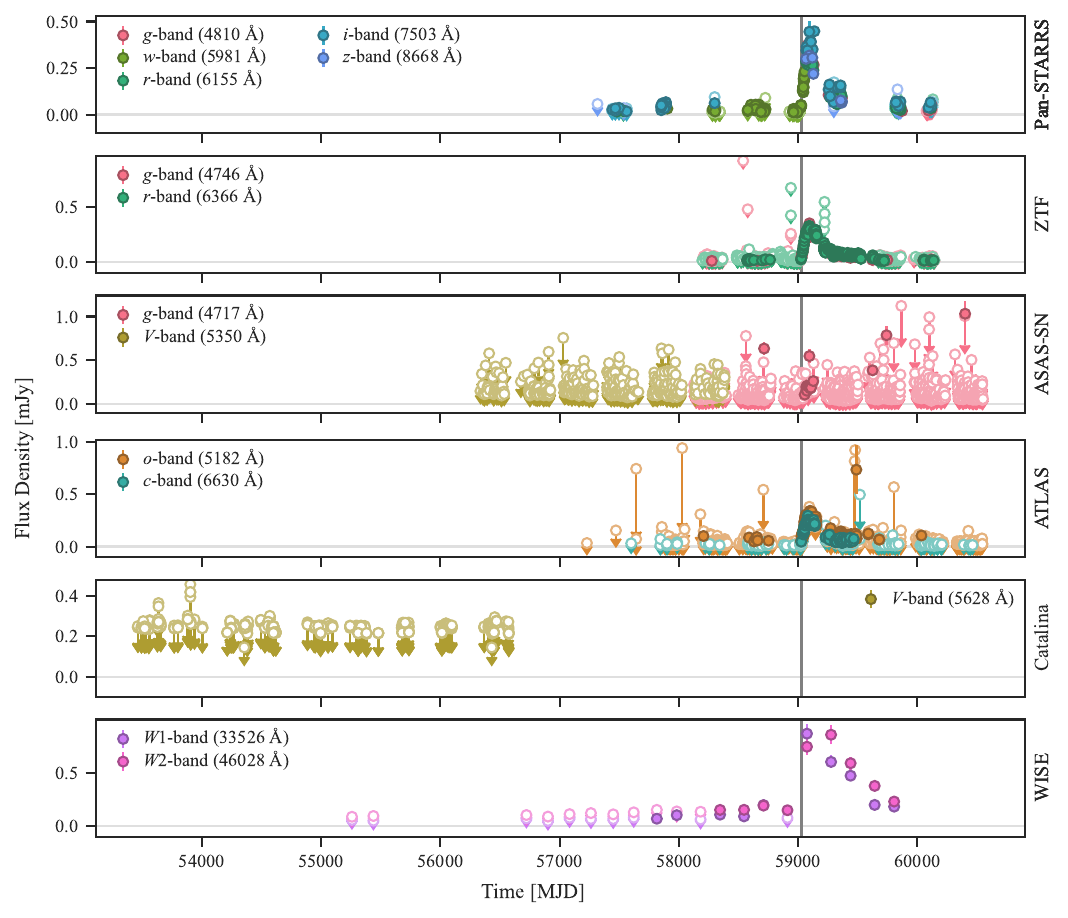}
    \caption{Pre-flare optical (ASAS-SN, ZTF, ATLAS, Catalina, Pan-STARRS) and mid-infrared (WISE) photometry of \nov{}. Derived 3$\sigma$ (7$\sigma$) upper limits are shown for the optical (mid-infrared) as open circles with downward arrows.}
    \label{fig:pre-flare-lc}
\end{figure*}

The analysis of pre-flare optical variability for the \nov{} utilizes archival photometry from several wide-field astronomical surveys: ASAS-SN, ATLAS, Catalina, ZTF, and Pan-STARRS. The All-Sky Automated Survey for Supernovae (ASAS-SN) provides high-cadence, all-sky monitoring, capturing transient events across multiple years. The Asteroid Terrestrial-impact Last Alert System (ATLAS) is designed primarily for near-Earth object detection, but its frequent observations also make it a valuable resource for identifying variability in extragalactic sources. The Catalina Real-Time Transient Survey (Catalina) is another survey aimed at detecting transient and variable phenomena, with a particular focus on identifying changes over longer time spans. The Zwicky Transient Facility (ZTF) is a high-cadence, wide-field survey that offers extensive temporal coverage in multiple optical bands, making it ideal for tracking rapid variability. Finally, the Panoramic Survey Telescope and Rapid Response System (Pan-STARRS) conducts deep, high-resolution imaging of the sky, allowing for detailed measurements of both faint and bright sources over extended periods. Together, these datasets provide a comprehensive picture of the optical variability preceding the TDE, enabling an analysis of potential precursor activity and variability patterns that may inform the nature of the disruption event. We plot the photometric behavior of the archival optical data along with the mid-IR observations from WISE in Figure \ref{fig:pre-flare-lc}.

We used the Python fitting software \texttt{qso\_fit}\footnote{http://butler.lab.asu.edu/qso\_selection/index.html} to analyze the pre-flare optical photometry of \nov{} and assess whether the light curves exhibited variability characteristic of an AGN  \citep{butler2011}. This software is designed to model quasar light curves by fitting them to a damped random walk (DRW) model, which is a common approach for characterizing the stochastic variability typical of AGN. We apply a 3$\sigma$ cut on the observations for the ASAS-SN, ZTF, ATLAS, Catalina, and Pan-STARRS data to determine detections. Only the ATLAS $o$-band, ZTF $g$- and $r$-bands, and Pan-STARRS $w$-, $r$-, and $i$-bands had detectable variability. Analysis of the pre-flare ZTF data revealed a low significance of the variability being DRW-like, with $\sigma_{\text{QSO}} = 0.92$ in the g-band and $0.71$ in the r-band. Typically, variable AGN exhibit a dispersion from the random walk model greater than 3, but we found $\sigma_{\text{var}} = 1.2$ and $2.4$ in these bands, respectively. Likewise, we measured $\sigma_{\text{QSO}} = 0.67$ in the ATLAS $o$-band with $\sigma_{\text{var}} = \num{3.1e-8}$. For the Pan-STARRS $w$-, $r$-, and $i$-bands, variability fitting returned values of $\sigma_{\text{var}} = 17$ ($\sigma_{\text{QSO}} = 0.89$), $\sigma_{\text{var}} = 1.2$ ($\sigma_{\text{QSO}} = 1.3$), and $\sigma_{\text{var}} = 4.7$ ($\sigma_{\text{QSO}} = 1.7$), respectively. Only the Pan-STARRS $w$- and $i$-bands demonstrate significant variability. In the former case, the significance of non-QSO variability is $> 3$, ruling out QSO-like variability; while in the latter case, $\sigma_\text{not QSO} < 3$, making the classification ambiguous. This suggests that it is unlikely the host of \nov{} experienced AGN activity in the recent past or that the transient is due to AGN activity. This conclusion is consistent with both the host galaxy's location on the BPT diagram and the archival WISE color cuts, further reinforcing the idea that AGN-related processes are not driving \nov{}.

\section{Spectroscopic Analysis} \label{sec:spec-analysis}

\begin{figure*}[t]
    \centering
    \includegraphics[width=\textwidth]{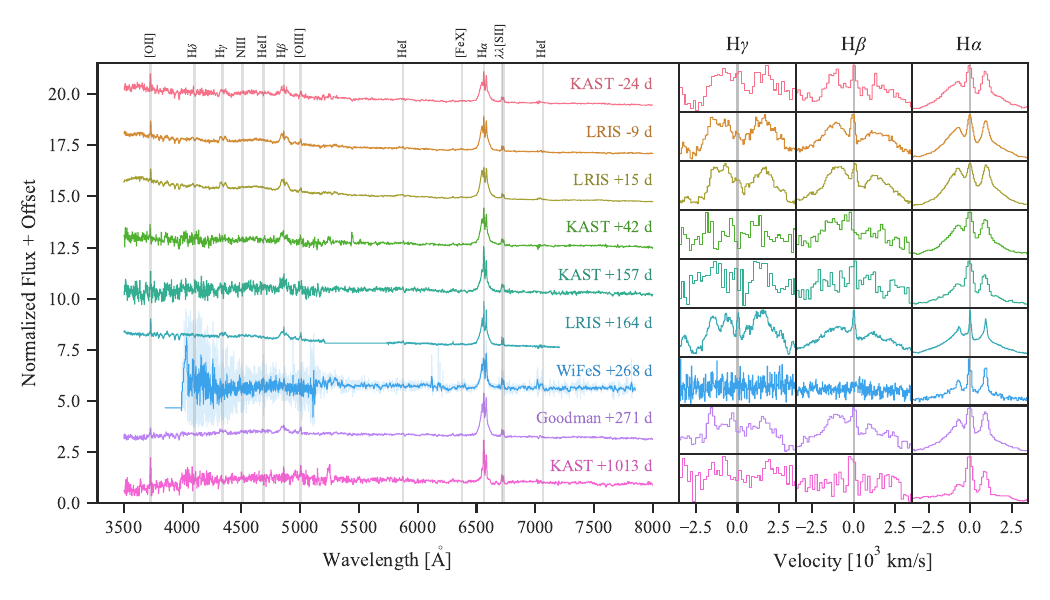}
    \caption{\textit{Left:} Evolution of the high resolution spectra of \nov{} covering phases from \qty{17}{d} before to \qty{1020}{d} after the optical/UV peak. Each spectrum is labeled with the phase and telescope/instrument used for the observation. Common TDE emission features are marked with vertical lines. The continuum transitions over time from hot and blue (TDE-like) to host-dominated. \textit{Right:} Zoom-ins of the double-peak Balmer features seen in the H$\alpha$, H$\beta$, and H$\gamma$ emission.}
    \label{fig:all_spectra}
\end{figure*}

The spectral evolution of \nov{} is characterized by a prominent blue continuum that diminishes approximately \qty{270}{days} after the optical/UV peak, along with broad emission lines of hydrogen and helium that are typical of TDEs \citep{leloudas2019,charalampopoulos2022}. The lack of a pre-transient host spectrum introduces challenges due to contamination from host galaxy emission in our analyses. Nevertheless, the broad features of the transient emission generally allow us to distinguish them from the narrow emission lines originating from the host, enabling modeling of their behavior despite some uncertainties regarding the exact source of the emission.

The high resolution spectra reveal a distinctive double-peaked structure, particularly pronounced in the Balmer emission lines (see the right panels of Figure \ref{fig:all_spectra}). In TDE spectra, double-peaked Balmer emission profiles can arise from multiple mechanisms, including the formation of an elliptical accretion disk around the SMBH. For instance, in AT~2018hyz, the double-peaked Balmer lines suggest the presence of such a disk formed through the efficient circularization of infalling stellar debris, redistributing angular momentum to create an extended H$\alpha$-emitting disk \citep{hung2020}. Similarly, the TDE candidate PTF09djl exhibits unique double-peaked H$\alpha$ profiles attributed to emission from a relativistic elliptical accretion disk, where the orbital dynamics of matter within a highly inclined and eccentric disk shape the line profiles \citep{liu2017}.

Simulations of tidal disruptions also indicate that a non-axisymmetric disk coupled with a debris tail can produce variable double-peaked profiles, resulting from the uneven distribution of debris and the presence of a ``tidal tail'' \citep{bogdanovic2004}. Additionally, outflows and optically thick winds may contribute to double-peaked emission lines, with the wind's kinematics and density significantly influencing the observed spectral features \citep{roth2018,parkinson2022}.

In the case of \nov{}, it is plausible that the double-peaked features originate from the same disk component required to explain the optical/UV spectral energy distribution in the previous section. This disk could be a pre-existing structure around the SMBH rather than one formed solely from tidally disrupted material. The rotational motion within such a disk would naturally produce double-peaked emission lines. The observation of double-peaked emission lines 24 days before the peak of the optical/UV flare supports this interpretation, suggesting that the disk may already be present prior to the TDE. If the double-peaked emission lines indeed arise from the same disk structure used in the SED modeling, we would expect similarities in their geometries, particularly in the fitted radii. Therefore, consistency between the radii derived from modeling the double-peaked profiles and those from the SED analysis would support the hypothesis that a single, coherent disk structure--either pre-existing or formed from the disrupted material--is responsible for both the continuum and the emission line features observed in \nov{}.

In this section, we model these double-peaked Balmer lines to test these possibilities and to determine whether the emission arises from a newly formed accretion disk of stellar debris or from a pre-existing disk structure.

\subsection{Spectral Features}

\begin{figure*}
    \centering 
    \includegraphics[width=\textwidth]{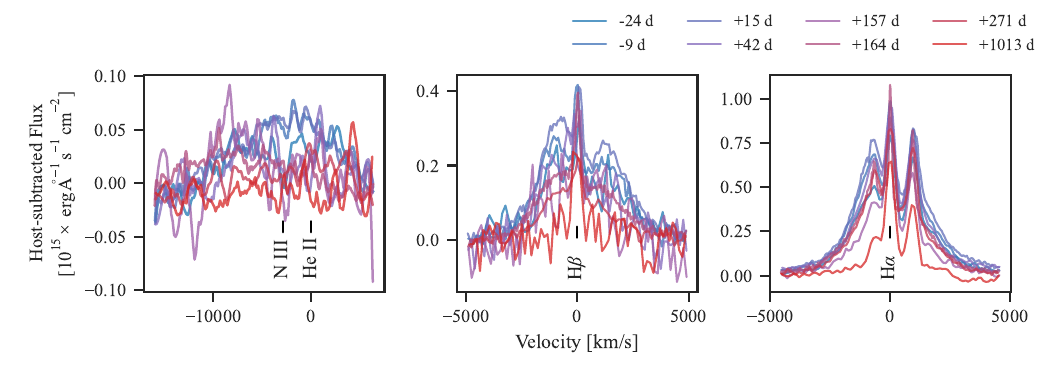}
    \caption{Stacked spectra of \nov{} after scaling and fitting of the \texttt{Bagpipes} continuum with a 3rd degree polynomial has been removed. \textit{Left:} Zoom-in of the \eline{N}{iii}+\eline{He}{ii} complex, demonstrating the disappearance of the \eline{He}{ii} at late-times and a decrease in the amplitude of double-peaked Balmer features. The data has been smoothed using a Savitsky-Golay filter to better highlight the evolution. \textit{Middle, Right:} Zoom-ins of the H$\alpha$ and H$\beta$ Balmer features. The colors indicate the phase at which the spectrum was observed, from early (blue) to late (red). In general, the plot shows only a small decrease in the FWHM of the emission lines.\label{fig:stacked_spec}}    
\end{figure*}

The double-peaked Balmer profiles of \nov{} are well-formed as early as \qty{24}{days} before the optical/UV peak, exhibiting less temporal variation compared to other double-peaked tidal disruption events (TDEs) like AT~2020hyz \citep{short2020}. The early emergence of these double-peaked profiles in \nov{} is relatively unusual. In contrast, the distinct double-peaked features in AT~2020hyz only appeared \qty{51}{days} after discovery, which was after the optical/UV peak in its light curve \citep{hung2020}. Similarly, AT~2018zso did not show double-peaked profiles until around the time of its optical peak \citep{wevers2022}. PTF09djl displayed double-peaked H$\alpha$ emission with peaks observed in spectra taken \qtyrange{19}{79}{days} after the peak in the optical/UV \citep{liu2017}. We show the stacked evolution of the spectra for \nov{} from early (blue) to late (red) times in Figure \ref{fig:stacked_spec}. The zoom-ins of H$\alpha$ and H$\beta$ in the bottom row demonstrate the transient nature of the double-peaked profiles. Measurable Balmer emission is still present in the spectra following the observational gap for a period of time similar to other optical/UV TDEs \citep[e.g.,][]{holoien2020, hung2020}. The double-peaked profiles are asymmetric, with the blue side higher and less extended than the red side. On average, the blue and red peaks show offsets of $\sim$\qty{-1150}{km.s^{-1}} and $\sim$\qty{1400}{km.s^{-1}}, respectively. Notably, the profiles are narrower than those seen in other TDEs with similar emission line profiles.

In addition to the strong double-peaked Balmer lines of H$\alpha$, H$\beta$, and H$\gamma$, we also note the possible detection of broad \elinedbl{N}{iii}{4097}{4104} (hereafter referred to collectively as \elinesgl{N}{iii}{4100}) and \eline{He}{ii}. We note though that this latter emission may be due to \elinedbl{N}{iii}{4634}{4641} (hereafter \elinesgl{N}{iii}{4640}). The recent evidence for metal lines in the optical spectra of TDEs has been attributed to Bowen fluorescence \citep{blagorodnova2019,leloudas2019} wherein the photon emission from the recombination of \eline{He}{ii} causes a cascade of resonant transitions in \eline{O}{iii} and \eline{N}{iii} \citep{bowen1935}. Inspection of our higher resolution LRIS spectra shows that the emission near $\lambda 4100$ is double peaked, suggesting that we are detecting weak H$\delta$ emission and not \elinesgl{N}{iii}{4100}. Similarly, it is difficult to determine if the broad emission around \elinesgl{N}{iii}{4640} is truly due to Bowen fluorescence, as it may be blended with the nearby \elinesgl{He}{ii}{4686} emission. In studying the spectroscopic properties of several TDEs, \citet{charalampopoulos2022} determined that \elinesgl{N}{iii}{4100} and \elinesgl{N}{iii}{4640} have a flux ratio of 1:1 when they are present. Further, the physical mechanism behind Bowen fluorescence predicts the simultaneous detection of \eline{N}{iii} and \eline{O}{iii} \citep{blagorodnova2018,trakhtenbrot2019}, but the absence of O$\,\textsc{iii} < \qty{4000}{\angstrom}$ emission and inconsistent flux ratios of \eline{N}{iii} suggest that Bowen fluorescence is not responsible for the observed features.

If we take into consideration the weak emission from the \elinesgl{N}{iii}{4100} line, especially in comparison to TDEs with strong Bowen features such as AT~2018dyb \citep{leloudas2019}, the blended \eline{N}{iii}+\eline{He}{ii} may instead be the consequence of a blue-shifted \elinesgl{He}{ii}{4686}, with a velocity offset of \qty{3237.4(425.3)}{km.s^{-1}}. Alternatively, we consider that this feature may be contribution from narrow low-ionization \eline{Fe}{ii} lines as seen in AT~2018fyk \citep{wevers2019b}. \citet{wevers2019b} suggest that the emergence of \eline{Fe}{ii} lines in TDE spectra originating from dense gas near the accretion disk could be accompanied by a broad component as \eline{Fe}{ii} requires a dense and optically thick obscuring medium, along with a strong soft X-ray source. While we do note tentative evidence for a \eline{Fe}{ii} complex appearing at \qty{4923}{\angstrom} and between \qtyrange{5169}{5317}{\angstrom}, the measured flux is minimal when the emission is available at all and shows little temporal evolution. \citet{blanchard2017} also identify \eline{Fe}{ii} emission in the spectra of PS16dtm, attributing its appearance to broad-line regions surrounding the black hole, but the lack of evolution in the profile of the \eline{N}{iii}+\eline{He}{ii} blended complex in \nov{} does not support this interpretation. Nonetheless, the \eline{N}{iii}+\eline{He}{ii} feature does appear to be present before and after the optical peak and only fades at late times. Low-ionization \elinesgl{He}{ii}{5876} emission is also detected, appearing strongest near the optical peak and subsequently fading quickly. As such, we primarily attribute the emission to \elinesgl{He}{ii}{4686}, proposing \nov{} to be an H+He TDE in the categorization scheme of \citet{gezari2012,velzen2021,hammerstein2023}.

\subsection{Constraints on AGN Contribution}

\begin{figure*}[t]
    \centering 
    \includegraphics[width=\textwidth]{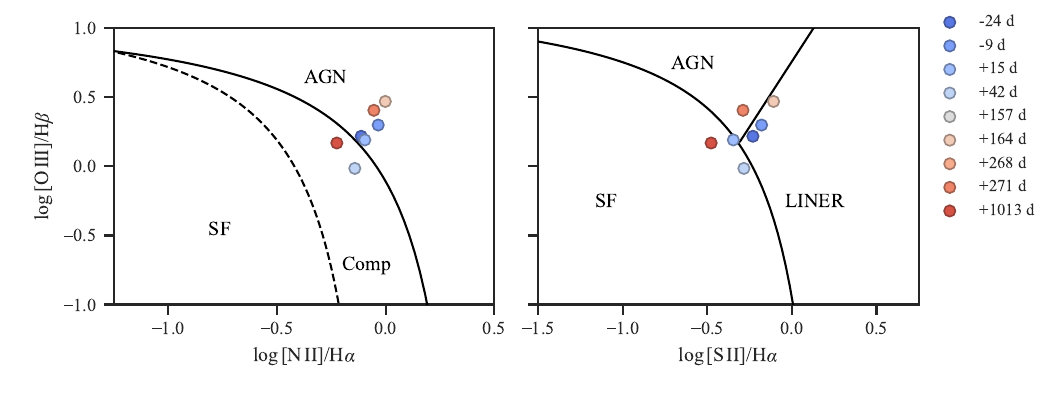}
    \caption{Ratios of the narrow lines measured at each spectral epoch, from early- (dark blue) to late-time (dark red). The results straddle the cutoffs between both composite and AGN, as well as between star-forming, AGN, and LINER galaxies, indicating that the emission line behavior is not strongly AGN-like.}
    \label{fig:bpt_diagram}
\end{figure*}

The weak \elinedbl{O}{iii}{4959}{5007} and broad, ambiguous \eline{N}{iii}+\eline{He}{ii} complex suggests minimal AGN contribution. To estimate the upper limit of the contribution from any pre-existing AGN, we fit a Gaussian model to \elinesgl[f]{O}{iii}{5007} in our late-time Kast spectrum. This yields a luminosity of $L_{5007} = \qty{5.40e40}{erg.s^{-1}}$, which is comparable to the least luminous type-1 and type-2 AGN in the sample of 47 local AGN analyzed by \citet{heckman2004}. Using the correction in \citet{heckman2004}, we derive a total bolometric luminosity upper limit of $L_\text{bol,AGN} = \qty{1.89e44}{erg.s^{-1}}$, more than an order of magnitude below the measured EUV peak. Additionally, we take the measured narrow line emission in each spectrum and plot their ratios on a Baldwin-Philips-Terlevic (BPT) diagram \citep{baldwin1981} in Figure \ref{fig:bpt_diagram} with classification regions marked as described in \citet{kewley2006}. The host galaxy is near the cutoff between composite and AGN in the \eline{N}{ii}/H$\alpha$ BPT diagram, with most of the spectra lying either in the star-forming or LINER regions of the \eline{S}{ii}/H$\alpha$ diagram. This is consistent with the relatively strong and persistent \elinesgl[f]{O}{ii}{3727} line seen in all spectra which, while commonly observed in AGN, can also arise from star formation. However, other such lines indicative of AGN activity, including the forbidden \elinesgl[f]{Ne}{v}{3426}, are not seen. We find that by the time of our late-time Kast spectrum at $+1013$ days, the galaxy has transitioned into the composition/star-forming region of the BPT diagram, suggesting that any AGN activity is not persistent.

\subsection{Emission Line Profiles} \label{sec:emission-line-profiles}

\begin{figure*}[t]
    \centering 
    \includegraphics[width=\textwidth]{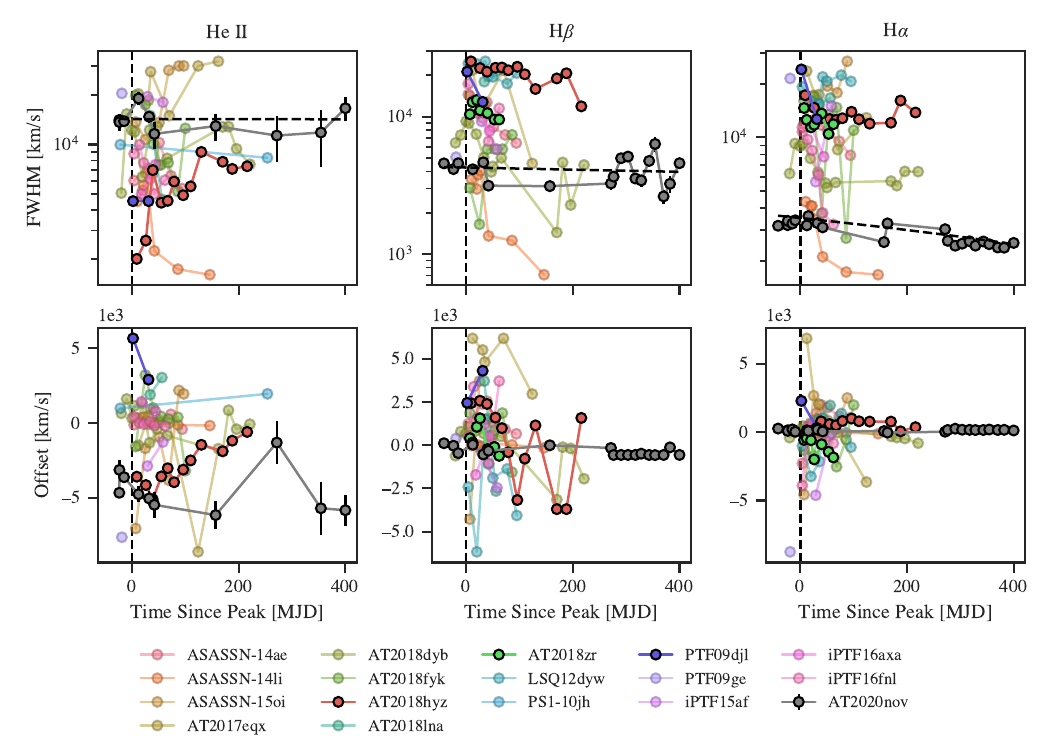}
    \caption{Evolution of emission line FWHM and offsets in \nov{} (shown as grey circles) compared to TDEs in the sample of \citet{charalampopoulos2022}. Vertical dashed line shows the time of the optical/UV peak. Other TDEs with double-peaked emission lines are shown with black marker edges.
    \textit{Left:} \eline{He}{ii} shows a decline in the FWHM, and an offset that declines initially after the peak. \textit{Middle:} The FWHM evolution of H$\beta$ appears flatter than other TDEs, but still declines with time as expected for the velocity width behavior of TDEs. The offset remains relatively constant. \textit{Right:} The H$\alpha$ FWHM is narrower than other TDEs, especially those that show double-peaked emission lines. The offset stays constant.}
    \label{fig:emission-line-evo}
\end{figure*}

To investigate the evolution of the emission line features observed in the spectra of \nov{}, we construct a compound model composed of Gaussian profiles representing the emission line complexes. Due to the lack of a pre-flare host spectrum, we fit a host template to the late-time Kast spectrum using \texttt{Bagpipes}, with the resolution set to match that of each observed spectrum. As part of the fitted model, we scale the template using a low-degree polynomial to approximate the continuum in each spectrum. Finally, we fit this compound model using a Levenberg-Marquardt least-squares algorithm to the spectra and measure the fitted properties. 

We focus on the evolution of lines identified as being prevalent tracers of TDE behavior, such as H$\alpha$, H$\beta$, H$\gamma$, \elinesgl{He}{ii}{4686} (here we assume minimal contribution from \eline{N}{iii} and \eline{Fe}{ii} to the broad blended feature), and \elinesgl{He}{i}{5876} \citep{arcavi2014,leloudas2019,velzen2021}. The nature of the spectral lines arising from the disk structure is explored in \ref{sec:disk-modeling}, but for the purposes of investigating the full width at half-maximum (FWHM), offsets, and luminosities, we find fitting the profiles with a single Gaussian is sufficient \citep[e.g.][]{blagorodnova2017,hung2017}. While Gaussian profiles do not account for the asymmetries or double-peaked structure seen in the Balmer profiles of the better resolved spectra, nor the potentially large electron-scattering optical depth, they do provide a straight-forward method for estimating emission widths and centroids.

Due to uncertainties in the blending of the narrow line emission with the TDE emission, we do not remove the narrow line host emission, instead choosing to fit the profiles with the narrow line host contamination in its entirety. Assuming the host contribution does not change over the course of the spectroscopic observations, the relative luminosities are self-consistent at the loss of being more accurate comparisons to other TDEs. The effect of the host emission is taken to be minimal in regards to the line widths and offsets. To account for contamination and blending seen in many of the observed profiles, we fit the collection of emission line models simultaneously with the polynomial scaling of \texttt{Bagpipes} host continuum template. In addition to the seven higher resolution spectra obtained as part of the YSE collaboration, we also include 22 spectra taken with the FLOYDS low resolution spectrograph on the \qty{2}{m} Faulkes Telescope North. While these spectra are too low resolution to disambiguate the double-peaked profiles, they are sufficient for broadly capturing the Gaussian characteristics.

\begin{figure}
    \centering 
    \includegraphics[width=\columnwidth]{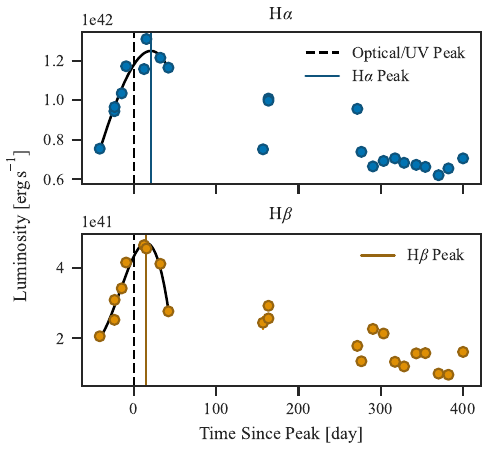}
    \caption{Luminosity evolution of the H$\alpha$ (top) and H$\beta$ (bottom) emission line complexes in the spectra. Fitting is performed with a 4th order polynomial to determine the time of peak luminosity. Time lags (see Section \ref{sec:time-lags}) are calculated as the difference between the optical/UV peak time and the time of peak luminosity as derived from the fitting.}
    \label{fig:balmer_time_lag}
\end{figure}

The velocity widths and offsets of H$\alpha$, H$\beta$, and \eline{He}{ii} are presented in Figure \ref{fig:emission-line-evo}. We omit the luminosities due to contamination from the host narrow lines and from \eline[f]{N}{ii} (for H$\alpha$). The luminosity evolution for H$\alpha$ and H$\beta$ are shown separately in Figure \ref{fig:balmer_time_lag}, demonstrating an initial rise to a time-lagged peak before later declining. We discuss their luminosity behavior in Section \ref{sec:time-lags}). Notably, we observe no significant temporal evolution in the full width at half maximum (FWHM) of H$\alpha$ or H$\beta$, finding a mean velocity widths of \qty{2.98e3}{km.s^{-1}} and \qty{3.97e3}{km.s^{-1}}, respectively, with only a shallow decline at late times. Broad line emission generally exhibits a decrease in width concurrent with the decline in luminosity, a characteristic associated with the presence of an optically thick wind that broadens the lines through electron scattering beyond the photosphere. This is opposite to the behavior observed in AGN studied via reverberation mapping, where line widths tend to increase with decreasing luminosity \citep{peterson2004,denney2009}. Comparing to the TDEs studied by \citet{charalampopoulos2022}, the decline in FWHM for \nov{} is generally shallower, being only marginally consistent with the predicted evolution of line widths in TDEs. One plausible explanation lies in the dependence of line width evolution on the optical depth of the scattering medium. In the unified TDE model proposed by \citet{dai2018}, the optical depth for electron scattering varies with the viewing angle, with higher values observed from the poles to the disk plane. The lack of decline in the FWHM may be due to a viewing angle nearer the pole, or a lack of optically thick wind. The FWHM evolution of the \eline{He}{ii} blended region show some variation at early times near peak, but are consistent with the other TDEs. Interestingly, it does not show an increase with time as in the case of AT~2018hyz. However, the lack of measurable emission between 50 and 150 days after the optical/UV peak makes it difficult to define a trend.Additionally, we do not see any significant change in the behavior of the emission lines during the period of the late-time X-ray brightening.

\subsection{Elliptical Disk Modeling}\label{sec:disk-modeling}

\begin{figure*}[t]
    \centering 
    \includegraphics[width=\textwidth]{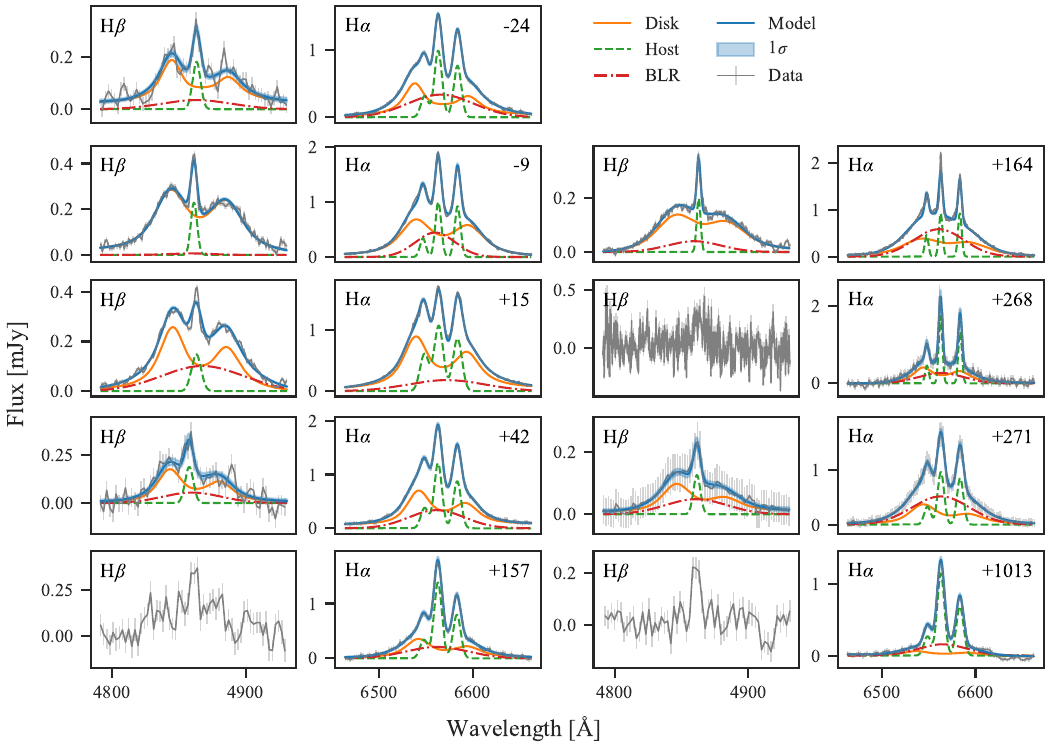}
    \caption{Best-fitting elliptical disk models (orange) to the H$\beta$ and H$\alpha$ line complexes for each spectral observation. Fits required an additional broad Gaussian feature to represent the BLR (red). Narrow host emission was included in the fitting where necessary and are shown in green. The disk model itself is shown in orange. Time since optical/UV peak is shown in the top right of each subplot. Measurable disk emission appears to fade at some time $<$\qty{155}{days}, but shows minimal contribution out to \qty{278}{days}.}
    \label{fig:disk-fits}
\end{figure*}

\begin{table*}
    \centering
\caption{Best-fit Elliptical Disk Model Parameters}
\label{tab:disk-model-params}
\begin{tabular}{ll ccccccc cc}
\toprule
& & \multicolumn{7}{c}{Disk} & \multicolumn{2}{c}{Gaussian} \\
\cmidrule(lr){3-9} \cmidrule(lr){10-11}
Phase & Emission & $\log_{10} \xi_1$ & $\log_{10} \xi_2$ & $i$ & $\log_{10} \sigma_\text{disk}$ & $q$ & $e$ & $\phi$ & $\mu$ & FWHM \\
 & &  [$\mathrm{R_g}$] & [$\mathrm{R_g}$] & [deg] & [km s$^{-1}$] &  &  & [deg] & [km s$^{-1}$] & [km s$^{-1}$] \\
\midrule
-24 & H$\alpha$+H$\beta$ & $3.22_{-0.78}^{+0.48}$ & $4.52_{-0.23}^{+0.18}$ & $47_{-12}^{+16}$ & $2.49_{-0.08}^{+0.07}$ & $1.06_{-0.29}^{+0.25}$ & $0.40_{-0.17}^{+0.23}$ & $223_{-10}^{+12}$ & $125_{-110}^{+110}$ & $3.58_{-0.05}^{+0.04}$ \\
-9 & H$\alpha$+H$\beta$ & $2.79_{-0.47}^{+0.40}$ & $4.73_{-0.14}^{+0.12}$ & $59_{-12}^{+14}$ & $2.70_{-0.03}^{+0.02}$ & $1.21_{-0.07}^{+0.05}$ & $0.45_{-0.16}^{+0.23}$ & $194_{-4}^{+5}$ & $-139_{-43}^{+46}$ & $3.36_{-0.05}^{+0.05}$ \\
15 & H$\alpha$+H$\beta$ & $3.39_{-0.20}^{+0.14}$ & $4.72_{-0.14}^{+0.10}$ & $54_{-10}^{+11}$ & $2.57_{-0.02}^{+0.02}$ & $1.13_{-0.16}^{+0.09}$ & $0.60_{-0.12}^{+0.13}$ & $209_{-4}^{+4}$ & $409_{-101}^{+103}$ & $3.65_{-0.02}^{+0.02}$ \\
42 & H$\alpha$+H$\beta$ & $3.01_{-0.67}^{+0.54}$ & $4.70_{-0.22}^{+0.16}$ & $51_{-13}^{+15}$ & $2.59_{-0.09}^{+0.07}$ & $0.88_{-0.24}^{+0.25}$ & $0.49_{-0.23}^{+0.27}$ & $215_{-11}^{+13}$ & $-69_{-136}^{+124}$ & $3.50_{-0.06}^{+0.08}$ \\
157 & H$\alpha$+H$\beta$ & $3.01_{-0.68}^{+0.59}$ & $4.68_{-0.26}^{+0.19}$ & $50_{-16}^{+17}$ & $2.60_{-0.33}^{+0.19}$ & $1.11_{-0.35}^{+0.32}$ & $0.57_{-0.32}^{+0.30}$ & $222_{-18}^{+21}$ & $-54_{-168}^{+168}$ & $3.55_{-0.10}^{+0.08}$ \\
164 & H$\alpha$ & $3.37_{-0.37}^{+0.20}$ & $4.78_{-0.16}^{+0.13}$ & $51_{-9}^{+12}$ & $2.82_{-0.04}^{+0.03}$ & $1.22_{-0.12}^{+0.11}$ & $0.63_{-0.23}^{+0.21}$ & $201_{-5}^{+6}$ & $-102_{-31}^{+28}$ & $3.46_{-0.01}^{+0.01}$ \\
268 & H$\alpha$ & $3.41_{-0.62}^{+0.33}$ & $4.72_{-0.32}^{+0.20}$ & $38_{-13}^{+14}$ & $2.47_{-0.29}^{+0.25}$ & $1.27_{-0.48}^{+0.40}$ & $0.34_{-0.24}^{+0.35}$ & $210_{-22}^{+23}$ & $-9_{-142}^{+141}$ & $3.41_{-0.15}^{+0.06}$ \\
271 & H$\alpha$+H$\beta$ & $3.02_{-0.64}^{+0.62}$ & $4.72_{-0.25}^{+0.18}$ & $50_{-16}^{+16}$ & $2.67_{-0.16}^{+0.10}$ & $0.95_{-0.29}^{+0.31}$ & $0.62_{-0.26}^{+0.24}$ & $233_{-15}^{+18}$ & $-8_{-90}^{+86}$ & $3.52_{-0.03}^{+0.03}$ \\
1013 & H$\alpha$ & $3.53_{-0.93}^{+0.34}$ & $4.37_{-0.47}^{+0.42}$ & $35_{-14}^{+15}$ & $2.56_{-0.35}^{+0.28}$ & $1.12_{-0.43}^{+0.53}$ & $0.38_{-0.27}^{+0.36}$ & $201_{-31}^{+31}$ & $76_{-109}^{+128}$ & $3.45_{-0.09}^{+0.05}$ \\
\bottomrule
\end{tabular}
\tablecomments{The uncertainties reported represent the 16th and 84th percentiles of the marginal posterior distributions for each parameter.}
\end{table*}

\begin{figure*}[t]
    \centering 
    \includegraphics[width=\textwidth]{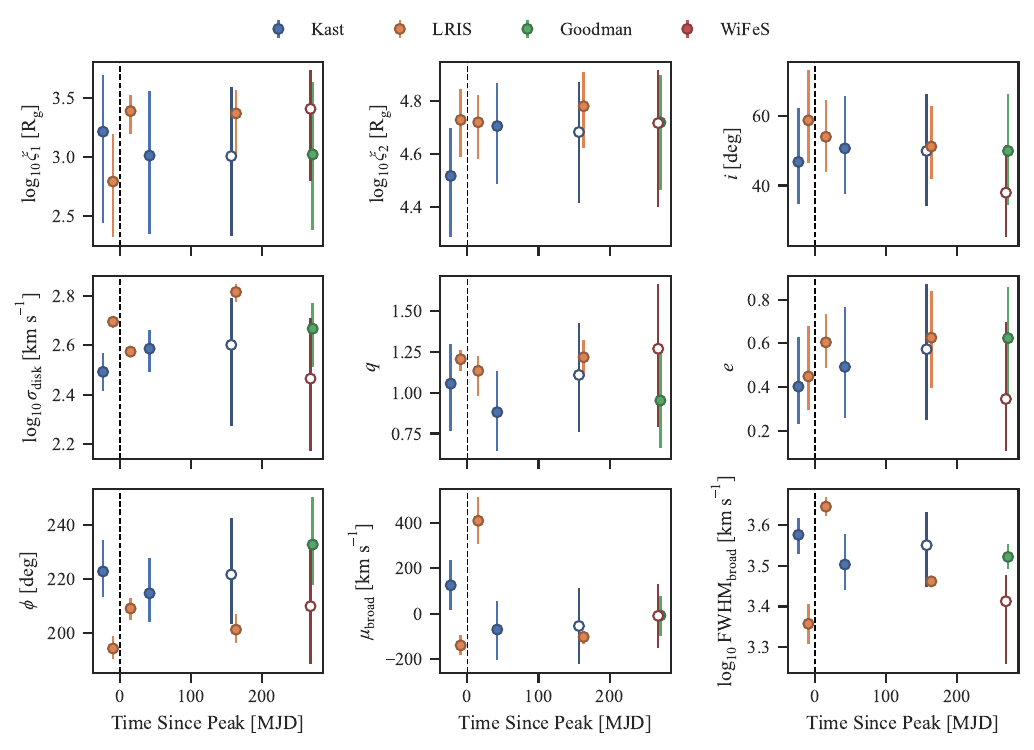}
    \caption{Evolution of the fitted elliptical disk parameters over time. The colors represent the instrument used to obtain the spectrum. The filled circles are the derived parameter determined from fitting both H$\alpha$ and H$\beta$, while the open circles indicate where only the H$\alpha$ profile was fit. The bottom middle plot shows the velocity offset of the broad Gaussian feature. The disk extent remains much higher than other TDEs with observable disk structures. Degeneracies between the parameters result in large uncertainties. Overall, the parameters do not show significant variation over time, suggesting the disk is well-established by the time of the first observation.}
    \label{fig:disk-model-params}
\end{figure*}

The observation of double-peaked emission line features in TDEs have been attributed to the presence of an accretion disk formed from the circularization of the disrupted stellar debris. For our analysis, we use the relativistic elliptical accretion disk model formulated by \citet{eracleous1995} to fit the double-peaked emission line profiles of \nov{}. This disk model, developed to explain the double-peaked emission lines observed in AGN, posits that these lines originate from a relativistic, eccentric accretion disk around a SMBH. In recent years, its use has been extended to TDEs, where the debris from the disrupted star may form highly eccentric accretion disks. The model successfully reproduces the complex and asymmetric line profiles seen in TDEs, such as the double-peaked H$\alpha$ emission lines in PTF09djl, AT~2020zso \citep{wevers2020}, and AT~2018hyz \citep{hung2020, short2020, gomez2020}, as well as the single-peaked but asymmetric lines in ASASSN-14li \citep{liu2017,cao2018}. This approach has proven powerful in probing the structures of accretion disks and coronal X-ray sources in both AGN and TDEs.

The elliptical disk model accounts for the diverse line profiles by considering the orbital motion of the emitting matter within the disk and the orientation of the disk relative to the observer \citep{wevers2022}. It is based on an elliptical geometrically thin, optically thick Keplerian disk described by seven free parameters. The parameters include the emissivity power-law index $q$, which scales as $\xi^{-q}$, where $\xi$ is the radius of the disk in units of gravitational radius $R_g$. Varying this parameter can affect the shape and intensity of the emission profile, particularly at different wavelengths. As pointed out in \citet{hung2020}, changes in the emissivity profile can lead to significant changes in the observed line shapes, particularly when combined with changes in the inclination angle of the disk. The broadening parameter $\sigma$ describes the intrinsic broadening of the line due to Doppler shifts in the disk and affects the apparent shape and intensity of the emission, especially at high inclinations. The azimuthal angle of the elliptical rings $\phi_0$ determines the orientation of the disk with respect to the observer. The inclination angle $i$ is the angle between the disk plane and the observer's line of sight, defined such that $i=0$ for a face-on disk. The eccentricity $e$ describes the deviation of the disk from circularity. Finally, the inner and outer pericenter distances $\xi_1$ and $\xi_2$ define the line-emitting region, expressed in units of $R_g$, and set the radial extent of the disk. 

The model of \citet{eracleous1995} defines the total observed line flux from the disk as
\begin{align}
    F = \int d\nu \int \int d \Omega I_\nu,
\end{align}
where $\nu$, $I_\nu$, and $\Omega$ are the frequency, specific intensity, and solid angle measured in the frame of the observer, respectively. The specific intensity profile of the line is given by \citet{chen1989} as
\begin{align}
    I_{\nu_e} = \frac{1}{4 \pi} \frac{\epsilon_0 \xi^{-q}}{\sqrt{2\pi} \sigma} \exp \left[ - \frac{(\nu_e^2 - \nu_o^2)}{2 \sigma^2} \right],
\end{align}
with $\epsilon(\xi) = \epsilon_0 \xi^{-q}$ describing the line emissivity. In our implementation, we ignore the $\epsilon_0 / 4 \pi$ intensity constant and normalize the peak of the flux profile. Due to the lack of a host spectrum, we prioritize fitting the H$\beta$ emission in cases where the resolution and signal-to-noise of the spectrum are amenable. Otherwise, we simultaneously fit the H$\alpha$ emission with the elliptical disk model and three Gaussian models describing the H$\alpha$, \elinesgl[f]{N}{ii}{6548}, and \elinesgl[f]{N}{ii}{6584} narrow lines originating from the host. In all cases, we follow \citet{hung2020} by including an additional Gaussian representing the broad line region (BLR) which is expected to form from the accreting TDE debris. To compensate for variations in the continuum level due to contribution from the host, we introduce two normalization constants $A$ and $B$, similar to the analysis of AT~2020zso \citep{wevers2022}, such that the disk flux calculation takes the form $A + B \times F$.

We perform the analysis over the spectra from \FirstSpecObsOffset{} to \qty{+1013}{days}, omitting the initial low-resolution classification spectrum and the 22 FLOYDS spectra where the double-peaks features were not resolved. For our spectral modeling, we utilized the nested sampling algorithm implemented in the \texttt{dynesty} \citep{speagle2020} Python package. This approach offers several advantages over traditional Markov Chain Monte Carlo (MCMC) methods, particularly in its ability to simultaneously estimate Bayesian evidence and efficiently sample from complex posterior distributions. The nested sampling framework is well-suited to handling multi-modal likelihoods and wide parameter ranges.

The model simultaneously fits seven physical parameters of the accretion disk to the H$\alpha$ and H$\beta$ profiles, capturing the geometric and dynamical properties of the disk. To accommodate the distinct profiles of H$\alpha$ and H$\beta$, we introduce additional parameters for offset, scaling, and centroid alignment for each line, bringing the initial parameter count to 13. To model the narrow-line components, we include Gaussian profiles for both narrow H$\alpha$ and H$\beta$, as well as two Gaussian profiles for the \elinesgl[f]{N}{ii}{6548} and \elinesgl[f]{N}{ii}{6548} lines. The widths of the \eline[f]{N}{ii} Gaussians are tied in velocity space, while the centroids of the narrow H$\alpha$ and H$\beta$ components are linked to their respective disk model centroids, with other parameters allowed to vary independently. This refinement increases the total parameter count to 20. Finally, we incorporate broad Gaussian components for both H$\alpha$ and H$\beta$ to account for contributions from the BLR. The widths of these broad Gaussians are tied in velocity space, reflecting the shared kinematics of the BLR gas, which results in a final model with 26 free parameters.

The best-fit parameters for our multi-component compound models are presented in Table \ref{tab:disk-model-params}. We discuss the construction of the priors in the Appendix. In Figure \ref{fig:disk-fits} we plot the best-fitting model for the H$\alpha$ and H$\beta$ profiles in the nine spectroscopic epochs. The blue line shows the disk model fit to the data with 1-$\sigma$ uncertainties shown as the blue shaded region, while the narrow line flux is shown as the green dashed line and the broad line flux as the red dash-dotted line. The parameters are determined by both H$\alpha$ and H$\beta$ simultaneously except in cases where the noise in the H$\beta$ complex introduced complications.

We find that despite broad degeneracies between the parameters of the relativistic elliptical disk model, the results indicate that the data are best fit using a two component model consisting of a broad Gaussian component centered on the systemic velocity of the host and a double-peaked elliptical disk component. We show the evolution of the fitted disk parameters in Figure \ref{fig:disk-model-params}. Due to the degeneracies, the parameter uncertainties are quite large. However, while the fits indicate perhaps some minor impact at the time of the optical/UV peak, there appears to be little change in behavior over time. We consider the velocity separation between the peaks ($V_\text{obs}$) of the H$\beta$ line in our $\Delta t = \qty{+22}{d}$ spectrum and calculate the radius of the emitting material assuming Keplerian motion. The result is consistent with our disk fits, yielding a radius of $R/R_g = (2c \sin i / V_\text{obs})^2 \approx \num{5e4}$. Overall, the elliptical disk fitting with an included BLR component provides good fits to the spectral data.

\subsection{Spectral Line Time Lags} \label{sec:time-lags}

In Figure \ref{fig:balmer_time_lag} we present the evolution of the H$\alpha$ and H$\beta$ luminosities, fitted using a 4th-order polynomial around their luminosity peaks. Our modeling of the Balmer emission lines shows that both H$\alpha$ and H$\beta$ continue to rise, reaching their respective peaks well after the optical/UV peak in the light curve. The broader wavelength coverage of the FLOYDS-N spectra allows us to capture a measurable lag between the Balmer lines and the continuum luminosity. \citet{charalampopoulos2022} attributed these line lags in TDEs to light echoes, where ionizing radiation from the flare illuminates surrounding material at different times. They found that the time lags correspond to distances ranging from \qtyrange{2e16}{12e16}{cm}, significantly larger than typical blackbody radii of TDEs. Moreover, \citet{charalampopoulos2022} suggest that lag times are a common feature of TDEs, with seven out of their nine spectroscopic TDEs showing measurable emission line evolution, while only two did not. They also report that the \eline{He}{ii} / H$\alpha$ ratio becomes larger as the photospheric radius of the TDEs recedes, implying the photosphere is stratified with different elements located at different depths. Similar behavior is also observed in AGN, with lag times on the order of days to weeks and a stratification that favors longer lag time times for longer-wavelength emission \citep{clavel1991,peterson1999}. However, type 1 AGN have shown the opposite behavior, with time lags for H$\beta$ larger than those derived from H$\alpha$ suggesting a larger radius for the H$\beta$ emission in type 1 AGN \citep{kovacevic2014}.

We extend our emission line modeling to include \eline[f]{O}{ii}, \eline[f]{O}{iii}, \eline[f]{S}{ii}, and \eline{He}{i}, which, in contrast to the Balmer lines, show minimal response to continuum variations, suggesting an origin in the host. Interestingly, while \eline{He}{ii} exhibits elevated luminosity at the optical/UV peak, it lacks a clear rise and has a much shallower decline over a longer period than the Balmer lines. Interpreting the behavior of the Balmer line lags as light echoes, we use a technique similar to AGN reverberation mapping to determine the structure of the broad line regions \citep{peterson1993}. In a simple model where a ring of gas orbits the black hole at a distance $r$, the average photon travel time to the ring is $\langle \tau \rangle = r/c$, regardless of the system’s inclination angle to the observer \citep{peterson1993,charalampopoulos2022}. Assuming the observed spectral line photon is emitted toward the observer from the instantaneous ionization and recombination of the gas, we measure the time difference between the spectral line peak and the optical light curve peak, finding offsets of \qty{21}{d} for H$\alpha$ and \qty{15}{d} for H$\beta$, corresponding to travel distances of \qty{5.33e16}{cm} and \qty{3.75e16}{cm}, respectively.

The FWHM of the Balmer lines suggests typical velocities on the order of $0.01c$, corresponding to a radius of approximately $10^4 R_{\rm g}$ if the velocity spread is interpreted as Keplerian motion. Consequently, the time lags inferred from our model estimate a black hole mass between \qty{3.4e7}{\msun} and \qty{4.5e7}{\msun}, which aligns reasonably well with mass estimates obtained by other methods (see Section \ref{sec:BH_mass}). It is possible to measure the black hole mass more precisely with a detailed model using our reverberation framework, but we leave this refinement to future studies.

\begin{figure}[t]
    \centering 
    \includegraphics[width=\columnwidth]{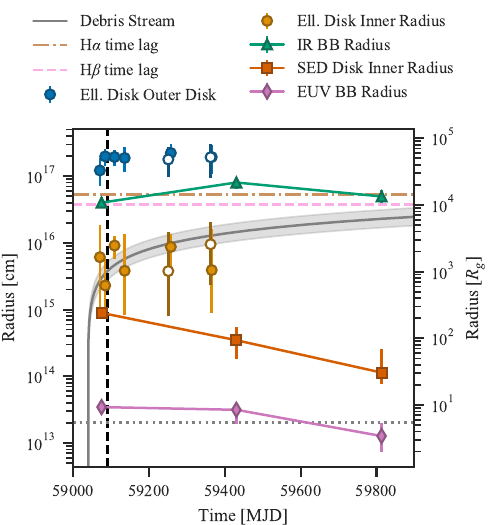}
    \caption{Evolution of various radii for \nov{}. The circles show the behavior of the inner (orange) and outer (blue) radii from the elliptical disk model fitting. The pink diamonds, red squares, and green triangles represent the mid-IR black body radius, inner disk radius estimate, and EUV black body radius from the SED fitting, respectively. The semi-major axis of the orbits of the returning stellar debris, shown by the solid grey line, is determined by approximating Kepler's third law. The approximation assumes that the energy distribution of the gas remains frozen in at the moment of disruption \citep{guillochon2014}. The yellow dash-dotted and pink dashed lines shown the emission radius for the H$\alpha$ and H$\beta$ lines as determined by the line lag analysis. The dotted grey line shows the expected tidal radius for a \qty{1}{\msun} star given our black hole mass of \qty{2.5e7}{\msun}.}
    \label{fig:photospheric-radius}
\end{figure}

In Figure \ref{fig:photospheric-radius}, we plot the radius of the emission line-emitting material, with the dash-dotted purple line representing H$\alpha$ and the dashed magenta line representing H$\beta$, alongside the inner (yellow circles) and outer (pink circles) radii derived from our disk fits. The grey region indicates the estimated semi-major axis of the disrupted debris based on the black hole mass estimate. Notably, the emitting radius of the hydrogen emission lines aligns well with the outer radius of our elliptical disk, illustrating a stratification that is consistent with observations of both TDEs and AGN. This stratification shows that H$\beta$ is located closer to the black hole than H$\alpha$, with Helium emission occurring even nearer, due to its higher ionization temperature requirements.\citep{guillochon2014,roth2016}.

\subsection{X-ray Spectrum} \label{sec:x-ray-spec}

The X-ray emission component of TDEs is thought to originate from the inner radii of the accretion disk formed during the circularization of the disrupted stellar debris, but optically selected TDEs are generally X-ray faint \citep{guillochon2014,auchettl2017, dai2018,parkinson2022, guolo2024}, with most of their high-energy emission either reprocessed by an intervening distribution of material around the black hole \citep{dai2018,parkinson2022,thomsen2022} or minimally produced in shocks between colliding debris streams \citep{piran2015}. Whether or not the primary emission source is predominantly in the X-ray regime makes the characterization and interpretation of the X-ray emission challenging. A measurable soft X-ray component can be interpreted as direct observation of the Rayleigh-Jeans tail of the accretion disk emission; however, the lack of such a component in X-ray faint TDEs may indicate that a significant portion of the radiated energy is emitted at EUV wavelengths.

The X-ray spectral modeling discussed in Section \ref{sec:obs-x-ray} builds upon the work of \citet{auchettl2017} and \citet{guolo2024}, comparing various models employed in the analysis of TDE X-ray data. TDEs are typically distinguished by extreme X-ray softness \citep[notable exceptions include with the rapidly variable hyper-luminous X-ray sources and super soft AGN; see][]{sacchi2023}, with temperatures ranging from \qtyrange{50}{100}{eV} and a monotonic decline in luminosity, or a power law index $\gtrsim 3$ \citep{auchettl2017,komossa2023}. The absorbed power law model (\texttt{powerlaw}) is frequently employed to describe the hot coronae in AGN, where emission is due to the inverse Compton scattering of seed photons \citep{titarchuk1995}. However, this model is inadequate for representing the thermal spectrum of TDEs, particularly in cases where corona formation is not evident (i.e., $\Gamma > 4$). To better model the cooler nature of TDE accretion disks ($kT \le \qty{0.3}{keV}$), \citet{mummery2021a} developed a specialized model called \texttt{tdediscspec}. We tested this model for \nov{} and found that a significant hard X-ray excess remains even after applying the TDE disk model, prompting us to incorporate the \texttt{simPL} \citep{steiner2009} power law model into our analysis. This model accounts for the Compton upscattering of soft photons by the corona near the accretion disk, with parameters that include the fraction of soft component photons upscattered to form the power law ($f$) and the photon index of the resultant power law ($\Gamma$) (for more details, see \citet{guolo2024}, Appendix A). However, the \texttt{simPL} component dominated the fitting, resulting in a worse fit when considering the additional parameters introduced by the \texttt{tdediscspec} model. Therefore, we opted to use the \texttt{powerlaw} model as implemented in the \texttt{XSPEC} package for our X-ray spectral analysis. This yielded particularly AGN-like power law indices, but the lack of a significant soft X-ray supports our interpretation of the primary emission of the TDE occurring predominantly in the EUV.

\section{Discussion} \label{sec:discussion}

The photometric and spectroscopic characteristics of \nov{} are unique in the context of nuclear transients, demonstrating a range of behavior indicative of a TDE, though with some caveats. Primarily, the evolution of the spectra show a strong blue continuum that fades over the course of $\sim$\qty{3}{years} of observation. The event is marked by prominent double-peaked Balmer emission lines in its optical spectra, indicating the presence of a disk-like structure around the black hole. While double-peaked emission line profiles are not a new phenomenon in either TDEs or AGN, \nov{} shows behavior that is inconsistent with both a freshly formed accretion disk composed of the disrupted stellar material, as well as the general characteristics of an active AGN accretion disk. Both our modeling of the spectra and the SED suggest that this extended disk-like structure is relatively quiescent, only being illuminated by energy released from the tidal disruption.

\begin{figure*}[t]
    \centering 
    \includegraphics[width=\textwidth]{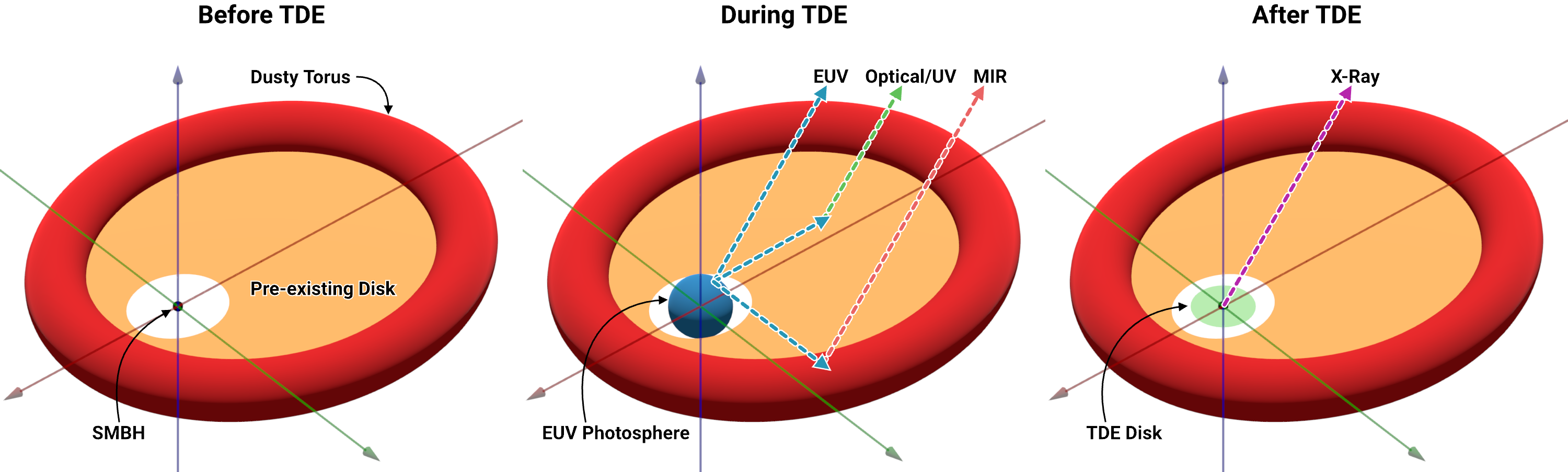}
    \caption{Illustration of \nov{}'s evolution and its associated emission. While we do not render the sizes of each component to scale (for ease of visualization, though see Figure \ref{fig:photospheric-radius}), the eccentricity, inclination, and apocenter of the pre-existing disk are shown from the perspective of the observer based on our best-fit model parameters. \textit{Left:} Before the TDE, the pre-existing disk (orange) remains quiescent. \textit{Middle:} Following the disruption, an EUV photosphere (blue) forms, with its emission reprocessed by the pre-existing disk and the surrounding dusty torus (red), producing the observed optical/UV and MIR emission. \textit{Right:} At later times, as the EUV photosphere recedes, the TDE-driven accretion disk (green) becomes visible, allowing X-ray emission from the accretion of disrupted material onto the SMBH (black) to emerge. At this time, we no longer see the optical/UV emission or double-peaked emission lines from the pre-existing disk.}
    \label{fig:scheme}
\end{figure*}

In \nov{}, the presence of an elliptical disk structure can be established at the time of earliest spectrum, \qty{24}{days} before peak light, implying that this disk was not formed as a direct result of the stellar disruption. The disk we infer from the double peak structure is likely distinct from the accretion disk formed by the disrupted material powering the transient emission. The disrupted material, however, may still rapidly form an accretion disk, aided by interactions with the pre-existing disk at larger radii. In fact, it may be necessary that rapid accretion disk formation occurs in the inner region of the SMBH, as the illumination of the double-peaked emission lines in the extended disk structure would require that larger amounts of energy be released at early times. Figure \ref{fig:scheme} illustrates our interpretation of \nov{}’s evolutionary phases. Before the TDE, the pre-existing disk remains quiescent, with minimal activity. During the disruption, an EUV photosphere forms, with its emission reprocessed into optical/UV by the pre-existing disk and into MIR by the surrounding dusty torus. At later times, as the EUV photosphere recedes, unobstructed X-ray emission emerges. Meanwhile, the optical/UV emission from the reprocessed EUV photosphere fades, and the double-peaked emission lines in the spectra disappear.

Our spectroscopic modeling of the extended disk material reveals an inner radius of \SpecDiskInnerRadius{}, consistent with the expected formation radius of a TDE accretion disk, within the uncertainties. Likewise, the SED disk modeling finds an inner radius of approximately \uqty{240}{24}{27}{R_g}, also inline with theoretical expectations. This difference could be the result of changes in the ionization state with radius, or contamination from the TDE itself. The inner disk radius as measured from the SED shows a decline at late times, diverging from the spectroscopic modeling. This might suggest that the accretion disk formed by the TDE extends out to the pre-existing disk of material. In such a case, the orientation of the TDE accretion disk relative to the pre-existing disk could be misaligned, resulting in an inability to capture the rotation of the material in the emission lines. We note that the elliptical accretion disk model of \citet{eracleous1995} approximates the photon geodesics only in the weak-field regime, and therefore is not valid for $\xi_1 \lesssim \qty{100}{R_g}$, which may prevent the spectroscopic modeling from fully capturing the inner disk behavior at late times.

\subsection{Stream-stream Collisions and Late-time Disk Formation}

Stream-stream collisions offer a possible framework for explaining the observed optical emission and late-time x-ray flare in \nov{}. Hydrodynamic simulations of TDEs indicate that the process of circularization is inefficient when the star's orbital pericenter radius is greater than the Schwarzschild radius of the black hole \citep{shiokawa2015}. This inefficiency is mainly due to the relatively weak relativistic apsidal precession, which significantly influences the path and behavior of the disrupted stellar material. This implies a timescale for the bound debris streams to reconfigure into a circular disk that is significantly longer than the timescale on which the disrupted stellar material falls back onto the black hole. Moreover, due to the minimal apsidal precession, the streams of stellar debris self-intersect, with the point of intersection predominantly occurring near the apocenter. The spatial orientation of the self-intersection further extends the timescale for circularization. Following this insight, \citet{piran2015} proposed that the observed radiation from TDEs is powered by the shocks that occur at the point of self-intersection of the stellar debris streams instead of from accretion onto the SMBH. Under the assumption that the initial energy of the fallback material is purely kinetic and that the thermal energy generated at the shock point is efficiently radiated away without significantly boosting the kinetic energy of the debris streams, it is possible to reproduce the luminosities, low temperatures, and larger radiation radii in optical/UV TDEs \citep{ryu2020b}. Under this model, the accretion of material onto the SMBH is delayed, but can explain how late-time X-ray and possibly optical/UV emission may be observed as a consequence of disk accretion at late times. Stream collisions and delayed-accretion have been invoked to explain the photometric behavior observed in several TDEs including ASASSN-15oi \citep{gezari2017}, AT~2019azh \citep{liu2022}, and AT~2019avd \citep{chen2022}.

However, there remain several features that this scenario fails to explain in regards to the properties of \nov{}. Primarily, this does not account for the early-time disk structure seen in the $\Delta t = \FirstSpecObsOffset{}$ spectrum. Considering the minimal evolution of the emission line complexes shown in Figure \ref{fig:emission-line-evo}, it is not enough that a disk forms by the time of peak optical emission, or even at the time of the first spectrum. Indeed, the constancy of the emission line behavior suggests that the disk is relatively stable.

\subsection{Rapid Disk Formation with Reprocessed Accretion Emission}

We also consider the traditional approach of disrupted stellar material undergoing circularization and rapidly forming an accretion disk, with the high temperatures producing emission mainly in the X-ray. This process of circularization is predicted to occur relatively fast as a consequence of the luminosity following the debris fallback rate \citep{guillochon2015, mockler2019}. In this paradigm, the observed optical/UV emission comes from an atmosphere around the accretion disk which reprocesses high energy radiation from the inner accretion flow and re-radiates it at lower energies, providing an explanation for the characteristically lower temperatures observed in TDEs \citep{loeb1997,strubbe2009,coughlin2014,guillochon2014,roth2016, lu2020}. Once the inner accretion disk fully ionizes the reprocessing layer, the winds surrounding the SMBH become transparent to X-ray emission and the observed X-ray flux rises with the decline in reprocessing efficiency \citep{metzger2017}. This would suggest that the late-time X-ray brightening in \nov{} is the consequence of radiation escaping without reprocessing. optical/UV emission is still expected to be produced by the inner accretion disk even after the reprocessing layer has been fully ionized \citep{strubbe2009,lodato2011}. Late-time observations of TDEs demonstrate a near constant luminosity in their UV light curves, consistent with a viscously spreading, unobscured accretion disk \citep{velzen2019a,mummery2020,mummery2024}. This late-time excess is not as prevalent for TDEs around high mass black holes, but \citet{velzen2019a} demonstrate that the early-time power laws to the UV light curves in their high mass subsample are comparatively shallower, similar to the behavior found in \nov{}.

Recent simulations of TDE accretion disks performed by \citet{thomsen2022} over a range of inclination angles and at different stellar evolution stages found that the viewing angle between the observer and the disk is largely responsible for determining whether an X-ray or optical/UV bright TDE is observed, supporting the unified model of \citet{dai2018}. Additionally, the simulations indicate that the timescale of X-ray emission is dependent on the temporal evolution of the accretion rate. As the rate declines post-peak, the amount of obscuring material available to reprocess the X-ray radiation drops, leading to a corresponding decrease in the optical/UV to X-ray ratio ($L_O/L_X$). At intermediate viewing angles, which appears to be the case for \nov{} (see Table \ref{tab:disk-model-params}), $L_O/L_X$ is expected to reach unity in a few hundred days, consistent with the observations. Figure \ref{fig:sed-fits} illustrates the behavior of the X-ray luminosity compared to the UV/optical. The SED fitting performed with our model (Section \ref{sec:sed-modeling}) does result in $L_O/L_X \approx 0.71$ during the period of the X-ray flaring. Given \nov{}'s super-Eddington nature, it appears to be consistent with the simulations showing super-Eddington-accretion-rate-induced optically and geometrically thick disk structures form around the SMBH and, at late-times, ``open up'' to allow direct viewing of the X-ray accretion disk.

A concern with this interpretation, however, is the lack of distinguishable Bowen fluorescence features in the spectra of \nov{}. If a reprocessing of accretion emission by static or outflowing material is playing a significant role in the optical profile of \nov{}, then it is reasonable to expect the presence of Bowen fluorescence lines as a consequence of a strong X-ray source \citep{leloudas2016}. The interpretation of the \eline{He}{ii} emission feature in Section \ref{sec:spec-analysis} does not support this conclusion, but the ambiguity of the complex may suggest that some X-ray reprocessing could be at work, with the majority of the optical/UV coming from radiation in the EUV.

\subsection{Rapid Circularization by Pre-existing Disk Material}

Notwithstanding the systematic differences in stellar orbit distributions between AGN hosts and inactive galaxies, to first order, a similar number of TDEs should occur in both. Recent theoretical studies suggest that the TDE rate should be higher in AGN \citep{wang2024,kaur2024}, triggered by interactions with the disk instead of as a consequence of relaxation processes.

The spectroscopic analysis of \nov{} reveals that the double-peaked Balmer emission can be well-fit by an extended ($\bar{\xi}_1 = \SpecDiskOuterRadius{}$), elliptical ($\bar{e} = \SpecDiskEccentricity{}$) disk that does not evolve significantly over the course of the tidal disruption. Independent modeling of the optical/UV SEDs demonstrate that the photometric behavior cannot be characterized by a single black body. Proposing that the double-peaked profiles in the spectra do not come from the TDE itself, but instead from pre-existing disk-like material around the SMBH, we can resolve the optical excess as arising from disk-reprocessed EUV emission. The lack of significant pre-flare variability in the optical/UV light curves and tell-tale narrow line region (NLR) emission lines indicate that the structure is relatively quiescent.

The early appearance of double-peaked emission features in the spectra of \nov{}--evident \qty{24}{days} before the peak in the optical/UV light curve--indicates that significant emission is produced at early times. This suggests a mechanism by which material rapidly reaches the SMBH. One such mechanism is the interaction with pre-existing AGN material around the disk which can facilitate rapid circularization of the disrupted stellar material through stream-disk interaction. \citet{chan2019} conducted a series of hydrodynamic simulations to investigate the interaction between the debris stream of a TDE and the pre-existing accretion disk of an AGN. In these simulations, a parabolic debris stream representing the returning stellar debris strikes the accretion disk perpendicularly at pericenter. They found that the shocks generated by the interaction with the disk dissipate the disk gas's kinetic energy and accelerates its inward flow toward the black hole. The debris stream, being significantly more massive than the disk, obstructs the disk's rotation, causing the disk gas to lose angular momentum and rapidly spiral inward. These shocks convert the debris's orbital kinetic energy into heat, which can be radiated away if the cooling time is sufficiently short, thereby reducing the debris orbits' eccentricity and promoting rapid circularization. 

\citet{chan2020} also found that efficient energy dissipation allows the debris to form a compact accretion disk, with the rapid inflow and energy dissipation rates potentially exceeding the Eddington limit. The energy dissipation rate is initially super-Eddington due to the high surface density of the radiatively efficient disk. However, because the inflow time is shorter than the cooling time, much of the dissipated energy is advected into the black hole rather than being radiated away, which regulates the bolometric luminosity to an Eddington-level plateau. This plateau persists as long as the disk surface density remains high enough to sustain the super-Eddington dissipation rate. The duration of the plateau, typically tens of days, is determined by the time it takes for the disk to deplete its mass through shock-driven inflows and the resupply of mass from the debris stream. As the disk mass decreases, the cooling time shortens, eventually leading to a decline in luminosity once the inflow time becomes comparable to the cooling time. This Eddington-level plateau may be driving the prolonged peak of \nov{}.

Stream-disk interaction can also contribute to late-time hard X-ray emission through the formation of strong shocks when the returning tidal debris stream collides with the AGN accretion disk \citep{chan2021}. These shocks dissipate a significant amount of kinetic energy, heating the gas to high temperatures. As the stream material, which is much more dilute than the disk, is stopped by the denser disk, it undergoes Compton cooling, producing hard X-rays and even soft $\gamma$-rays. The energy dissipated at this second impact powers another flare, with most of the energy emitted between approximately \qty{10}{keV} and \qty{1}{MeV} \citep{chan2021}. While the X-ray observations of \nov{} are limited to energies $<$\qty{10}{keV}, collisions between the disrupted material and the pre-existing AGN disk may explain the observed late-time X-ray brightening and harder X-ray spectrum.

\subsection{Primary Emission in the EUV} \label{sec:euv-primary}

As mentioned previously, the theoretical understanding of TDEs predicts that half the disrupted star's mass is bound and accreted by the SMBH with an expected energy release of $\sim$\qtyrange{1e52}{1e53}{erg}. However, for the majority of TDE candidates identified in recent surveys, the observed radiation energy in the optical/UV bands is only $\sim$\qty{1e51}{erg} or less. If we consider that the optical/UV spectra of TDEs are generally well-described by the Rayleigh-Jeans tail of a hot black body, then the peak frequency of the energy emission could be in the unobserved EUV band \citep[e.g.][]{lu2018}. Indeed, the integrated energy released in the EUV from our multi-component SED model reaches $\EradTDE$, in line with theoretical predictions. Constraining the emission in this region from the optical/UV is difficult due to the strong dependence on the line-of-sight extinction of the host galaxy. Nonetheless, the proposed existence of a passive, extended disk-like structure around \nov{} allows for the capture and re-emission of the EUV photons into more easily observed optical wavelengths. To support this model, we can leverage the mid-IR observations, which benefit from negligible extinction from the host galaxy, to perform dust reverberation mapping as an independent measure of the EUV emitted near the SMBH.

The mid-infrared radiation is a consequence of some fraction of UV photons being absorbed and re-radiated by the dusty medium around the nucleus of the galaxy. \citet{lu2018} demonstrates that one can constrain the total emitted UV-optical luminosity by considering the sublimation radius for the dust particles in the optically thin limit. The light-crossing timescale for the radiated dust shell can be derived from the sublimation radius and used to infer the total optical/UV luminosity of the TDE:
\begin{align}
    t_\text{IR} \approx \qty{0.6}{yr} \; \left( \frac{L_\text{UV}}{\qty{1e45}{erg.s^{-1}}} \right)^{1/2}&\left( \frac{T_\text{sub}}{\qty{1800}{K}} \right)^{-2.5} \left( \frac{a}{\qty{0.1}{\micron}} \right)^{-1/2},
\end{align}
where $L_\text{UV}$ is the total UV-optical luminosity of the TDE, $T_\text{sub}$ is the sublimation temperature of the dust, and $a$ is the maximum size of dust grains (the biggest dust grains are the last ones to sublimate and hence they dominate the absorption cross-section for EUV photons).

The mid-infrared emission from surviving dust particles at a radius of approximately $R \sim R_\text{sub}$ persists over the light-crossing timescale of the radiating dust shell. As seen in Figure \ref{fig:ir_lc}, the emission from the hottest dust grains endures for a duration of at least $\qty{0.5}{yr}$, constrained by the temporal sampling of the WISE observations. By applying our predicted IR black body temperature from Table \ref{tab:bb-params} as the sublimation temperature, we can estimate a lower limit on the EUV luminosity of \nov{} at $L_\text{UV} \gtrsim \qty{1e44}{erg.s^{-1}}$, assuming the dust grains are around \qty{0.1}{\micron} in size. If the mid-infrared emission extends to $\sim$\qty{1}{yr} or if the dust grains are larger (e.g. $\sim$\qty{1}{\micron}, as may occur in dense molecular gas), we estimate an EUV luminosity on the order of \qty{1e45}{erg.s^{-1}}. Given that the lower limit of the EUV luminosity is still well above the optical/UV and X-ray emission, along with the uncertainties in dust composition and geometry, we interpret this result as evidence that the primary emission of \nov{} likely occurs in the EUV. This is consistent with our EUV luminosity estimate from the SED fitting.

\subsection{Outflows as Origin for the Double-peaked Lines}

Outflows are an alternative to an elliptical disk for producing double-peaked emission lines, and have been both theoretically predicted \citep{strubbe2009,roth2018,dai2018} and observed in TDEs \citep{alexander2016,kara2018,hung2019,nicholl2020}. These outflows are thought to be driven by the intense radiation pressure generated when stellar debris accretes onto the SMBH at super-Eddington rates \citep{strubbe2011}. Typical, spherical outflows are expected to produce single-peaked, broad emission lines with a blue-shifted asymmetry. However, \citet{parkinson2022} showed that biconical disc winds can produce double-peaked emission lines, particularly when the kinematics of the line-forming regions are dominated by rotation.

As discussed in Section \ref{sec:emission-line-profiles}, the strong double-peaked line profiles of \nov{} maintain a relatively constant velocity separation over time before disappearing after $\sim$\qty{1000}{days}. The persistence of the velocity separation argues against an expanding outflow, which would typically show decreasing velocities as the fallback rate declines and the material decelerates \citep{strubbe2009}. Likewise, outflows in TDEs are often associated with the super-Eddington phase of accretion where intense radiation pressure or winds can drive disk material outward \citep{cao2022,dai2021}. This phase is expected to be relatively short-lived; as the accretion rate decreases, the outflow weakens and eventually ceases. The long-lasting nature of the double-peaked profiles in \nov{} suggest a more stable emission mechanism than what is typically associated with transient outflows.

The double-peaked emission lines in \nov{} show a velocity separation of approximately \qty{2.5e3}{km.s^{-1}}, while the H$\alpha$ and H$\beta$ emitting regions are located at distances of about \qty{5e16}{cm} and \qty{3e16}{cm}, respectively, based on our lag time analysis. Given the outflow velocity inferred from the peak separation, it would take roughly 5 to 6 years for material in the outflow to reach these distances, assuming constant velocity. This timescale is significantly longer than the measured lag times of \qty{20}{days} for H$\alpha$ and \qty{15}{days} for H$\beta$, indicating a discrepancy between the inferred outflow velocity and the rapid response of the emission lines. This inconsistency suggests that the double-peaked profiles are unlikely to be generated by an expanding outflow, as the emission regions appear to be much closer to the central source than could be achieved by an outflow on this timescale.

Super-Eddington accretion in TDEs are generally associated with geometrically thick disks launching wide-angle, optically thick fast outflows \citep{wevers2020}. Likewise, at super-Eddington rates, the X-ray emission is expected to be thermalized and reprocessed by the optically thick outflows \citep{dai2018}. However, neither the emission lines nor the hard X-ray emission appear consistent with these expectations. Furthermore, while the hard X-ray emission observed in \nov{} could be indicative of very hot outflows or jets, the lack of significant radio emission makes this interpretation less plausible. In cases where relativistic jets or energetic outflows are present, radio emission is commonly observed due to synchrotron radiation from shocks generated by the outflow’s interaction with the surrounding medium \citep{decolle2012,auchettl2017,lu2018,alexander2020}. The lack of radio detection suggests that, if an outflow is contributing to the observed X-ray emission, it is likely not energetic enough to produce detectable radio waves. Instead, the hard X-ray emission could plausibly originate from mechanisms near the black hole, such as Compton upscattering in a hot corona or shock heating within the inner regions of the accretion flow \citep{crumley2016,zanazzi2020,mummery2021b}. This scenario would involve compact, high-energy processes rather than a large-scale, radio-bright jet or extended hot outflow.

Collisionally-induced outflows (CIOs) have also been invoked as the source of the emergence of emission lines in TDEs. In this scenario, the shock of the self-crossing of the debris streams caused by relativistic precession can result in unbound debris which acts to reprocess emission and produce time-lagged spectral lines \citep{lu2020}. For a mass traveling at $\sim$$0.1c$, it is possible to reach distances of $\sim$\qty{3e16}{cm} in around \qty{100}{days} (roughly the time of the first double-peaked spectral observation predicted by our light curve analysis). However, while the partial sky coverage of the CIO can approach half the sky, therefore making it likely to act as the obscuring material between the observer and the accretion disk, it is not clear that a similarly sized mass can be ejected at the same time in the opposite direction to produce the double-peaked emission lines seen in the spectra or even that a covering factor of 50\% could be oriented such that the emission lines show such little variation in their FWHM.

\subsection{Comparison to Other TDEs} \label{sec:comp-tdes}

\begin{figure*}[ht]
    \centering 
    \includegraphics[width=\textwidth]{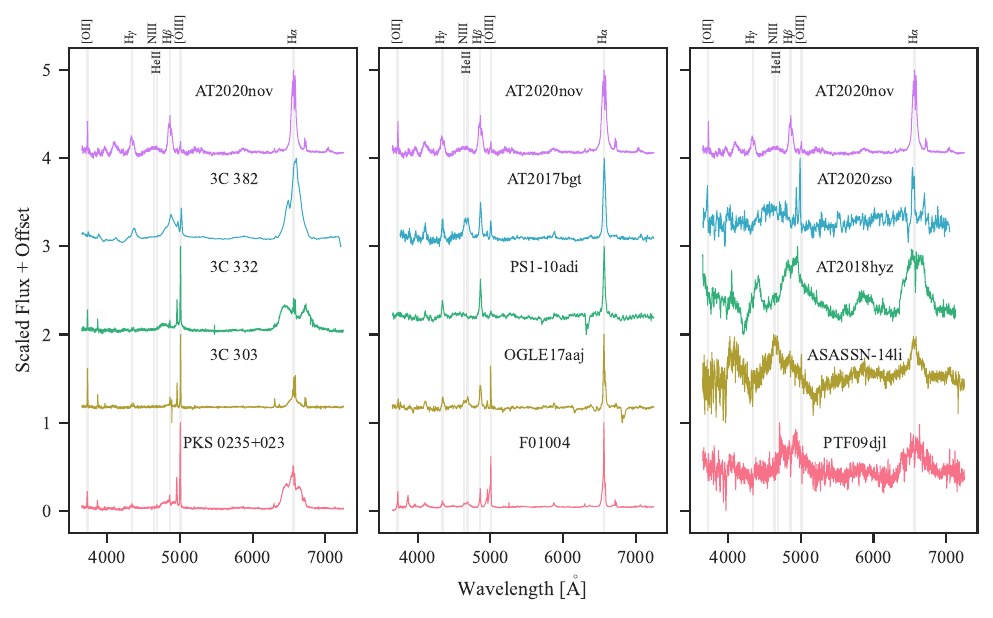}
    \caption{Comparison of the spectra of \nov{} at +11 days with those of other nuclear transients. \textit{Left:} Various AGN with double-peaked structure. \textit{Middle:} Ambiguous nuclear transients, TDE candidates with disk-winds or that have occurred in active AGN hosts. \textit{Right:} Other TDEs that demonstrate double-peaked spectral features, or ``typical'' formation processes. Further discussion of these comparisons is found in Sections \ref{sec:comp-tdes}, \ref{sec:comp-agn}, \& \ref{sec:comp-nuclear-transients}.}
    \label{fig:spec-compare}
\end{figure*}

In the right column of Figure \ref{fig:spec-compare}, we compare the LRIS spectrum near optical/UV peak in \nov{} to a selection of other TDEs in the literature including AT~2020zso, AT~2018hyz, and PTF09djl, which have been suggested to host elliptical accretion disks. AT~2018hyz presented the first unambiguous case of resolved double-peaked Balmer emission in a TDE \citep{hung2020,short2020}. The distinct line profile was well-modeled by a low eccentricity ($e \approx 0.1$) accretion disk extending out to $\sim$\qty{3e3}{R_\mathrm{g}}. This disk was found to have a moderate-to-high inclination angle of $\sim$\qtyrange{50}{70}{\degree}. The model included both a disk component and a Gaussian component, with the latter potentially originating from non-disk clouds or a bipolar outflow.

In contrast, AT~2020zso exhibited a much higher eccentricity and inclination. The accretion disk model for this event yielded an eccentricity of $e \approx 0.97$ and an inclination angle of $i = \qty{85(5)}{\degree}$, indicating a nearly edge-on configuration \citep{wevers2022}. The disk was found to be relatively compact, with inner and outer radii of several \qty{100}{R_\mathrm{g}} and several \qty{1000}{R_\mathrm{g}} respectively. Like AT~2018hyz, the model for AT~2020zso also incorporated both disk and Gaussian components to fully explain the observed line profiles.

PTF09djl represents an extreme case in terms of both eccentricity and inclination. The model for this TDE suggested an accretion disk with eccentricity $e \approx 0.966$ and an inclination angle of \qty{88}{\degree}, even closer to edge-on than AT~2020zso \citep{liu2017}. A semi-major axis of $\sim$\qty{680}{R_\mathrm{g}} was determined from the disk modeling. However, it's worth noting that the spectral data for PTF09djl was more limited and of lower quality compared to the other two events.

\nov{} shows distinctly moderate eccentricity and inclination with average values of \SpecDiskEccentricity{} and \SpecDiskInclination{}, respectively. While AT~2018hyz showed a nearly circular disk, both AT~2020zso and PTF09djl exhibited highly eccentric configurations, with \nov{} occupying a middle ground. This variation could potentially be explained by differences in the circularization efficiency of the stellar debris or the time elapsed since the TDE occurred. The high eccentricities in AT~2020zso and PTF09djl are more consistent with theoretical expectations of inefficient circularization. The inclination angles also varied significantly, from moderate in AT~2018hyz and \nov{} to nearly edge-on in AT~2020zso and PTF09djl. These differences in inclination can have a strong effect on the observed line profiles and may explain some of the diversity seen in TDE spectra. The nearly edge-on configurations of AT~2020zso and PTF09djl may have made the double-peaked features more prominent in these events.

The most divergent characteristic for \nov{} from both the selection of TDEs as well as theoretical expectations are the outer disk radius ($\log_{10} \bar{\xi}_{1} = \LogSpecDiskOuterRadius{}$), which is considerably larger and more extended than that of the other events. The primary factor in determining the extent of the disk is the velocity profiles of the double-peaked structure, which in the case of \nov{} are much narrower than other disk-modeled TDEs. The lack of a distinct trend in the evolution of the radii, and the prevalence of these double-peaks in the early spectra suggest that the disk is formed well before the tidal disruption. 

Another differentiating feature revealed in our spectral modeling is the comparatively flat emissivity power law index of $\bar{q} = \unum{1.37}{0.02}{0.03}$ compared to the other TDEs. For instance, modeling of AT~2018hyz and AT~2020zso has yielded emissivity indices in the range of $q = 2$ to $3$. These higher values are so far typical for TDEs and lie between the indices observed in cataclysmic variables \citep[$q \approx 1.5$,][]{horne1991} and those expected for disks illuminated by isotropic ionizing sources \citep[$q \approx 3$,][]{chen1989,wilkins2012}. The low index for \nov{} suggests a more gradual decline in radiative power with radial distance, possibly hinting that energy from the disruption is being reflected and distributed across an extended, pre-existing disk structure rather than being emitted directly from a compact, newly formed accretion disk.

\subsection{Comparison to AGN} \label{sec:comp-agn}

Analysis of the host galaxy of the \nov{} does not strongly favor the presence of an AGN. When the narrow-line emission from the host is placed on a BPT diagram (see Figure \ref{fig:bpt_diagram}), the weak \eline[f]{O}{iii} emission places it in the star-forming or composite region, with \eline[f]{O}{iii}/H$\beta$ and \eline[f]{N}{ii}/H$\alpha$ indicating that any ionizing radiation is more consistent with stellar processes or a mix of star formation and lower-level nuclear activity, rather than the high-energy output characteristic of AGN.

However, the X-ray properties of \nov{} are quite AGN-like, with power law indices between $\Gamma \sim \numrange{0.93}{2.17}$, placing the X-ray spectrum behavior in-line with type I and type II AGN \citep{auchettl2018,guolo2024}. The lack of soft X-ray emission, resulting in high hardness ratios, is also consistent with AGN behavior. However, the HR appears to display a harder-when-brighter trend or at the very least is constant within uncertainties, both of which are characteristic of TDEs \citep{shemmer2008,auchettl2018}. Likewise, if the primary emission of a \nov{} lies in the EUV range, we would expect a lack of a significant thermal component in the X-ray spectra, with the hard X-ray emission instead sourced from Compton up-scattering of the EUV photons by a hot corona or non-thermal processes in the disk or jet.

The radio observations of \nov{}, conducted with the NRAO Very Large Array (VLA program 20A-372, PI: Alexander) do not significantly support an AGN or AGN flare origin. Initial radio emission was detected 111 days post-discovery, with subsequent low-level emission observed consistently at 228, 246, and 869 days \citep{cendes2023}. This steady luminosity profile contrasts sharply with the transient, elevated radio behavior typical of AGN flares, which generally last around \qty{2.5}{years} and display energy release patterns largely independent of duration \citep{hovatta2008}. Therefore, \citet{cendes2023} concluded that the radio emission in \nov{} likely originates from ongoing star formation within the host galaxy, rather than from any processes related to the TDE or AGN activity. Additionally, while we observe \eline[f]{O}{iii} emission, the luminosity of the line is below $\sim 76\%$ of type-1 and type-2 AGN from \citet{heckman2005}, further implying that the host galaxy does not host a significant AGN.

Pre-flare variability analysis provides further support for this conclusion. In the optical bands, there is no significant variability indicative of AGN activity prior to the TDE, which would typically present as periodic or quasi-periodic fluctuations on a range of timescales (see Figure \ref{fig:pre-flare-lc} in Section \ref{sec:ir-discussion}). In the mid-infrared, we observe only minor variability in the few years leading up to the TDE flare, with behavior that is not convincingly AGN-like and lacks the persistent, structured variability typically associated with AGN. This limited mid-infrared variability could potentially stem from other processes, such as dust heating by star formation or minor fluctuations in the galactic nucleus, rather than robust AGN activity. Furthermore, the variability exhibited by \nov{} contrasts sharply with the continuous optical variability observed in the double-peaked AGN sample studied by \citet{ward2024}. The AGN light curves in their study are characterized by power spectra with amplitudes and power-law indices similar to those of other broad-line AGN. In contrast, nuclear transients like \nov{} display large, single-flare light curves, distinctly different from the more gradual and persistent variability seen in AGN. TDEs also show significant variability amplitudes, often around \qty{1}{mag} over periods $\lesssim$\qty{100}{days}, while AGN generally exhibit less dramatic variability of $\lesssim$\qty{0.1}{mag} on similar timescales \citep{velzen2011a,macleod2012,caplar2017}. Only about 10\% of AGN display variability exceeding \qty{1}{mag}, however, this variability occurs on timescales of approximately \qty{15}{years} \citep{rumbaugh2018}. It is therefore very unlikely that the flare of \nov{} comes from standard AGN variability.

TDE accretion disks tend to be more compact than those in AGN. Modeling of AT~2020zso found inner and outer disk radii of several hundred to a few thousand gravitational radii, while AGN disks typically extend to tens of thousands of gravitational radii \citep{eracleous1995,strateva2003,storchi-bergmann2017,hung2020,wevers2022}. This compact nature of TDE disks is in line with theoretical predictions that the disk should form with a size of about twice the fatal orbit pericenter with broader emission lines compared to AGN. The modeled outer extent of the disk in \nov{} suggests structure more inline with expectations of AGN disks. \citet{zhang2022} show that some broad line AGN are capable of extending out to distances of \qtyrange{40000}{50000}{R_g}. However, TDE disks show evidence of rapid formation, with disk signatures appearing within about a month of disruption, in contrast to the long-lived stable disks observed in AGN. The first column of Figure \ref{fig:spec-compare} compares \nov{} to a sample of broad-line AGN that are known to host disk-like features \citep{eracleous1995,strateva2003}. The minimal \eline{O}{iii} narrow line emission is apparent compared to that of the AGN, and the narrower widths of the Balmer double-peaked profiles stand in contrast to the wider features of the double-peaked AGN sample.

Taken together, these observations suggest that the host of \nov{} is unlikely to harbor a significant AGN. The absence of AGN-like narrow-line emission on the BPT diagram, the low predicted AGN luminosity relative to the TDE peak, and the lack of significant pre-flare variability all point to a relatively quiescent galaxy environment. This context reinforces the classification of \nov{} as a TDE occurring in a non-AGN host, with the flare’s EUV emission and light curve likely dominated by the tidal disruption process rather than pre-existing AGN activity.

\subsection{Comparison to Other Nuclear Transients} \label{sec:comp-nuclear-transients}

The second column of Figure \ref{fig:spec-compare} compares the spectral features of \nov{} with several types of ANTs. AT~2017bgt exemplifies the Bowen fluorescence flare (BFF) subset of ANTs, exhibiting strong \eline{He}{ii} and \eline{N}{iii} emission lines \citep{trakhtenbrot2019}. OGLE17aaj similarly displays a narrow \eline{He}{ii} line width ($\sim \qty{5e3}{km.s^{-1}}$), contrasting with the broad \eline{He}{ii} component observed in \nov{}, which aligns more closely with typical TDE profiles \citep{arcavi2014}. Likewise, F01004--a transient in an Ultra-luminous Infrared Galaxy (ULIRG) \citep{tadhunter2017}--initially showed TDE-like characteristics, though subsequent spectra revealed two distinct emission lines, ultimately suggesting an ANT classification \citep{trakhtenbrot2019}. F01004 has recently undergone another flaring event, indicating the presence of an extreme environment with a potentially high TDE rate, where residual material from previous events may still be impacting the dynamics of the nuclear region \citep{sun2024}. In \nov{}, however, we find little evidence for Bowen fluorescence, and the evolution of the \eline{He}{ii} complex lacks a BFF-like profile. Given that Bowen fluorescence is often a marker for an EUV source like an accretion disk, its absence in \nov{} is intriguing. Although strong EUV emission commonly facilitates Bowen fluorescence, factors such as the absence or low abundance of \eline{O}{iii} or \eline{He}{ii} ions, or high optical depth in the emitting region, can inhibit the required resonance cascade \citep{kastner1990}.

ANTs also exhibit diverse light curve behaviors. While some follow smooth rises and declines \citep[e.g., PS10adi,][]{zhuang2021}, others, like OGLE17aaj and AT~2017bgt \citep{gromadzki2019}, display rapid rises, irregular declines, or plateaus, complicating the use of photometric evolution as a distinguishing feature. Additionally, variations in MIR luminosity have emerged as a common characteristic among ANTs \citep{hinkle2024,wiseman2024}, with these flares often interpreted as dust echoes from a torus-like structure--a feature also noted in \nov{}. Interestingly, the ANT sample studied by \citet{wiseman2024} showed no pre-flare MIR variability, whereas \nov{} presents evidence of some low-level MIR activity prior to the optical/UV flare. However, unlike ANTs, \nov{} does not display the high, AGN-like dust covering factor that defines this class of transients.

\section{Conclusions} \label{sec:conclusion}

We have presented a comprehensive analysis of \nov{}, a transient event that exhibits characteristics of both TDEs and AGN. Our multi-wavelength study, encompassing X-ray, UV, optical, and mid-infrared observations, reveals a complex interplay between the disruption event and its host environment. The transient nature of the double-peaked emission lines and evidence of an extended disk structure suggest that \nov{} represent a TDE occurring in a galaxy with a pre-existing, quiescent accretion disk around its SMBH.

\begin{enumerate}[noitemsep]
    \item \nov{} shows prominent double-peaked Balmer emission lines in its optical spectra, indicating the presence of an elliptical disk-like structure around the SMBH. These double-peaked profiles are evident as early as \qty{24}{days} before the optical/UV peak, much earlier compared to other TDEs. The double-peaked features are transient, persisting with relatively little temporal variation in the fitted parameters before disappearing entirely by \qty{1012}{days}.
    
    \item Modeling of the double-peaked emission lines suggest an elliptical disk with moderate eccentricity ($\bar{e} = \SpecDiskEccentricity{}$) and inclination ($\bar{i} = \SpecDiskInclination{}$). The disk has an inner radius (\SpecDiskInnerRadius{}) consistent with TDE accretion disks, while the outer radius (\SpecDiskOuterRadius{}) is much more extended compared to other TDEs, but is consistent with AGN disk sizes.
    
    \item \nov{}'s light curve shows a broad peak and decline that roughly follows a $t^{-5/3}$ power law, consistent with TDE predictions. The SED analysis over several epochs reveals the need for a passive disk component, reprocessing high-energy EUV photons into optical/near-UV wavelengths as a single black body is insufficient to capture the SED behavior. Modeling the SED with the included passive disk component reveals a high-energy source with a peak luminosity in the EUV of \LpeakTDE. This supports the presence of an extended disk re-radiating energy from the TDE.
    
    \item \nov{} exhibits AGN-like X-ray properties, with power law indices between $\Gamma \sim \numrange{0.93}{2.17}$, inconsistent with the generally thermal emission from TDEs, but may be consistent with a primary emission source primarily in the EUV. A late-time X-ray brightening occurs about \qty{300}{days} after the optical/UV peak similar to late-time flares observed in many other optically bright TDEs.

    \item We infer a black hole mass of $\log_{10} M_\text{bh}/\msun = \num{7.4(0.4)}$ from the host galaxy absorption lines and scaling relations. \nov{} exhibits an Eddington ratio of approximately 1.25, placing it in the super-Eddington regime, but does not demonstrate strong evidence for outflows as is expected for TDEs around high mass black holes.
    
    \item Mid-infrared data indicates dust echoes, a common feature in TDEs, but without the high dust covering factors seen in ANTs or AGN. The dust covering factor is estimated to be about 1.1\%, consistent with typical optically selected TDEs. The mid-infrared emission allows for dust reverberation mapping, which provided an independent estimate of the EUV luminosity. By modeling the dust echoes, we constrained the EUV luminosity to at least \qty{1e44}{erg.s^{-1}}, with an upper estimate of $\sim \qty{1e45}{erg.s^{-1}}$ depending on the size of the dust grains. These findings confirm significant EUV emission, consistent with the SED model.
    
    \item Low-level MIR activity was detected prior to the flare, yet no strong pre-flare optical or MIR variability, pointing to a relatively quiescent host environment without ongoing AGN activity. Likewise, the host galaxy lacks significant AGN-like narrow-line emission on the BPT diagram at late times, with optical/UV spectra further ruling out active AGN processes.
\end{enumerate}

Based on these characteristics, we conclude that \nov{} is likely a TDE occurring in a galaxy with a pre-existing, quiescent accretion disk around its SMBH. The event’s unique features can be attributed to interactions between the TDE and this established disk structure. The early presence of double-peaked emission lines suggests that the TDE illuminates an existing disk rather than forming a new one. The large disk radii and low emissivity index further support the existence of an extended, pre-existing disk. The multi-component SED model, with a significant EUV component, indicates that the primary emission from the TDE is likely reprocessed by this extended disk structure. Additionally, the late-time X-ray brightening may result from TDE debris interacting with the disk, creating shocks and producing hard X-ray emission. The absence of strong AGN signatures in the host galaxy, coupled with the presence of a disk-like structure, implies that this disk was likely quiescent before the TDE occurred.

\section*{Data Availability}

All photometric and spectroscopic data used in this study will be made publicly available through the Weizmann Interactive Supernova Data Repository (WISeREP)\footnote{\url{www.wiserep.org}} upon publication.

\section*{Facilities}

Our analysis of \nov{} makes extensive use of the transient survey management platform YSE-PZ \citep{jones2021,coulter2022,coulter2023}. YSE-PZ was developed by the UC Santa Cruz Transients Team with support from The UCSC team and is supported in part by NASA grants NNG17PX03C, 80NSSC19K1386, and 80NSSC20K0953; NSF grants AST-1518052, AST-1815935, and AST-1911206; the Gordon \& Betty Moore Foundation; the Heising-Simons Foundation; a fellowship from the David and Lucile Packard Foundation to R. J. Foley; Gordon and Betty Moore Foundation postdoctoral fellowships and a NASA Einstein fellowship, as administered through the NASA Hubble Fellowship program and grant HST-HF2-51462.001, to D. O. Jones; and a National Science Foundation Graduate Research Fellowship, administered through grant No. DGE-1339067, to D. A. Coulter.

\section*{Acknowledgements}
N.E., K.D.F., M.E.V., and M.S.\ acknowledge support from NSF grant AST 22-06164.
E.R.-R.\ thanks the Heising-Simons Foundation, NSF (AST-2150255 and AST-2307710), Swift (80NSSC21K1409, 80NSSC19K1391), and Chandra (22-0142) for support.
S.I.R.\ acknowledges support by the Science and Technology Facilities Council (STFC) of the UK Research and Innovation via grant reference ST/Y002644/1.
The UCSC team is supported in part by NASA grant 80NSSC20K0953, NSF grant AST--1815935, the Gordon \& Betty Moore Foundation, the Heising-Simons Foundation, and by a fellowship from the David and Lucile Packard Foundation to R.J.F.
I.A.\ acknowledges support from the European Research Council (ERC) under the European Union’s Horizon 2020 research and innovation program (grant agreement number 852097), from the Israel Science Foundation (grant number 2752/19), from the United States - Israel Binational Science Foundation (BSF; grant number 2018166), and from the Pazy foundation (grant number 216312).
M.R.D.\ acknowledges support from the NSERC through grant RGPIN-2019-06186, the Canada Research Chairs Program, and the Dunlap Institute at the University of Toronto.
C.G.\ is supported by a VILLUM FONDEN Young Investigator Grant (project number 25501).
D.L.\ was supported by research grants VIL16599 and VIL54489 from VILLUM FONDEN.
M.S.\ acknowledges support from the Illinois Space Grant Consortium.
M.E.V.\ acknowledges support from the Illinois Space Grant Consortium and the Center for Astrophysical Surveys Graduate Fellowship.
This research was supported in part by grant NSF PHY-2309135 to the Kavli Institute for Theoretical Physics (KITP).
Parts of this research were supported by the Australian Research Council Discovery Early Career Researcher Award (DECRA) through project number DE230101069.
This material is based upon work supported by the National Science Foundation under Grant No.\ 2206165. 

The Young Supernova Experiment (YSE) and its research infrastructure is supported by the European Research Council under the European Union's Horizon 2020 research and innovation programme (ERC Grant Agreement 101002652, PI K.\ Mandel), the Heising-Simons Foundation (2018-0913, PI R.\ Foley; 2018-0911, PI R.\ Margutti), NASA (NNG17PX03C, PI R.\ Foley), NSF (AST--1720756, AST--1815935, PI R.\ Foley; AST--1909796, AST-1944985, PI R.\ Margutti), the David \& Lucille Packard Foundation (PI R.\ Foley), VILLUM FONDEN (project 16599, PI J.\ Hjorth), and the Center for AstroPhysical Surveys (CAPS) at the National Center for Supercomputing Applications (NCSA) and the University of Illinois Urbana-Champaign.

Pan-STARRS is a project of the Institute for Astronomy of the University of Hawaii, and is supported by the NASA SSO Near Earth Observation Program under grants 80NSSC18K0971, NNX14AM74G, NNX12AR65G, NNX13AQ47G, NNX08AR22G, 80NSSC21K1572, and by the State of Hawaii.  The Pan-STARRS1 Surveys (PS1) and the PS1 public science archive have been made possible through contributions by the Institute for Astronomy, the University of Hawaii, the Pan-STARRS Project Office, the Max-Planck Society and its participating institutes, the Max Planck Institute for Astronomy, Heidelberg and the Max Planck Institute for Extraterrestrial Physics, Garching, The Johns Hopkins University, Durham University, the University of Edinburgh, the Queen's University Belfast, the Harvard-Smithsonian Center for Astrophysics, the Las Cumbres Observatory Global Telescope Network Incorporated, the National Central University of Taiwan, STScI, NASA under grant NNX08AR22G issued through the Planetary Science Division of the NASA Science Mission Directorate, NSF grant AST-1238877, the University of Maryland, Eotvos Lorand University (ELTE), the Los Alamos National Laboratory, and the Gordon and Betty Moore Foundation.

A subset of the data presented herein were obtained at the W.\ M.\ Keck Observatory. NASA Keck time is administered by the NASA Exoplanet Science Institute. The Observatory was made possible by the generous financial support of the W.\ M.\ Keck Foundation. The authors wish to recognize and acknowledge the very significant cultural role and reverence that the summit of Maunakea has always had within the indigenous Hawaiian community.  We are most fortunate to have the opportunity to conduct observations from this mountain.
A major upgrade of the Kast spectrograph on the Shane 3~m telescope at Lick Observatory was made possible through generous gifts from the Heising-Simons Foundation as well as William and Marina Kast. Research at Lick Observatory is partially supported by a generous gift from Google.
Based in part on observations obtained at the Southern Astrophysical Research (SOAR) telescope, which is a joint project of the Minist\'{e}rio da Ci\^{e}ncia, Tecnologia e Inova\c{c}\~{o}es (MCTI/LNA) do Brasil, the US NSF's NOIRLab, the University of North Carolina at Chapel Hill (UNC), and Michigan State University (MSU).
This work makes use of observations from the Las Cumbres Observatory global telescope network.
We acknowledge the use of public data from the Swift data archive.
Based on observations obtained with XMM-Newton, an ESA science mission with instruments and contributions directly funded by ESA Member States and NASA.
Based on observations obtained with the Samuel Oschin Telescope 48-inch and the 60-inch Telescope at the Palomar Observatory as part of the Zwicky Transient Facility project. ZTF is supported by the National Science Foundation under Grant Nos. AST-1440341, AST-2034437, and a collaboration including Caltech, IPAC, the Weizmann Institute for Science, the Oskar Klein Center at Stockholm University, the University of Maryland, the University of Washington, Deutsches Elektronen-Synchrotron and Humboldt University, the TANGO Consortium of Taiwan, the University of Wisconsin at Milwaukee, Trinity College Dublin, Lawrence Livermore National Laboratories, and IN2P3, France. Operations are conducted by COO, IPAC, and UW.
This research is based on observations made with the Galaxy Evolution Explorer, obtained from the MAST data archive at the Space Telescope Science Institute, which is operated by the Association of Universities for Research in Astronomy, Inc., under NASA contract NAS 5--26555.
This publication makes use of data products from the Two Micron All Sky Survey, which is a joint project of the University of Massachusetts and the Infrared Processing and Analysis Center/California Institute of Technology, funded by the National Aeronautics and Space Administration and the National Science Foundation.
This publication makes use of data products from the Wide-field Infrared Survey Explorer, which is a joint project of the University of California, Los Angeles, and the Jet Propulsion Laboratory/California Institute of Technology, funded by the National Aeronautics and Space Administration.

YSE-PZ was developed by the UC Santa Cruz Transients Team. The UCSC team is supported in part by NASA grants NNG17PX03C, 80NSSC19K1386, and 80NSSC20K0953; NSF grants AST--1518052, AST--1815935, and AST--1911206; the Gordon \& Betty Moore Foundation; the Heising-Simons Foundation; a fellowship from the David and Lucile Packard Foundation to R.J.\ Foley; Gordon and Betty Moore Foundation postdoctoral fellowships and a NASA Einstein Fellowship, as administered through the NASA Hubble Fellowship program and grant HST-HF2-51462.001, to D.O.\ Jones; and an NSF Graduate Research Fellowship, administered through grant DGE--1339067, to D.A.\ Coulter.

\appendix

Table \ref{tab:unif-priors} provides a summary of the priors applied to the nine parameters of the elliptical accretion disk model and the three parameters describing the Gaussian components used in our spectral fitting. An initial exploratory \texttt{dynesty} run was conducted with a low number of samples and uniform priors across all parameters to probe the parameter space and assess potential degeneracies. This preliminary analysis revealed significant degeneracies among certain accretion disk parameters, specifically the inclination, eccentricity, apocenter, and inner radius. To address these degeneracies and improve the fit, we subsequently adopted normal priors centered near the average of the posterior distributions obtained from the exploratory run for these parameters for the inclination and apocenter. Uniform priors were retained for the remaining accretion disk parameters to allow flexibility in exploring the broader parameter space.

\begin{table*}
    \centering
    \caption{Priors for Elliptical Disk Modeling}
    \label{tab:unif-priors}
    \begin{tabular}{lclccccc} %L{1.75cm} c P{1.5cm} c c}
    \toprule
        Parameter & Symbol & Unit & Distribution & Min & Max & Location & Scale \\ 
    \midrule
    \multicolumn{8}{c}{Elliptical-disk component} \\
    \midrule
        Inner radius & $\xi_1$ & $R_g$ & Log Uniform & \num{1e2} & \num{1e4} & & \\
        Outer radius & $\xi_2$ & $R_g$ & Log Uniform & \num{1e3} & \num{1e5} & & \\
        Inclination & $i$ & rad & Normal & 0 & $\pi / 2$ & $\pi / 4$ & $\pi / 8$ \\
        Intrinsic broadening parameter & $\sigma_\mathrm{disk}$ & \unit{km.s^{-1}} & Log Uniform & \num{1e2} & \num{1e3} & & \\
        Eccentricity & $e$ & ~ & Uniform & 0 & 1 & &  \\
        Emissivity power-law exponent & $q$ & ~ & Uniform & 0.5 & 2 & & \\
        Orientation angle & $\phi$ & rad & Normal & 0 & $2 \pi$ & $7 \pi / 6$ & $\pi / 6$ \\
        Scale factor & $A$ & ~ & Uniform & 0 & 1 & & \\
        Offset & $B$ & Normalized flux & Uniform & 0 & 0.1 & & \\
    \midrule
        \multicolumn{8}{c}{Gaussian component} \\
    \midrule
        Amplitude & $A$ & Normalized flux & Uniform & 0 & 1 & & \\
        Offset (narrow) & $\mu_{\rm narrow}$ & \unit{\AA} & Uniform & \num{-5} & \num{5} &  &  \\
        Offset (broad) & $\mu_{\rm broad}$ & \unit{km.s^{-1}} & Normal & \num{-1e3} & {1e3} & 0 & 250 \\
        Standard deviation (narrow) & $\sigma_\mathrm{narrow}$ & \unit{km.s^{-1}} & Log Uniform & \num{10} & \num{1e3} &  \\
        Standard deviation (broad) & $\sigma_\mathrm{broad}$ & \unit{km.s^{-1}} & Log Uniform & \num{1e3} & \num{1e4} &  \\
    \bottomrule
    \end{tabular}
\end{table*}

\begin{figure}
    \centering 
    \includegraphics[width=\columnwidth]{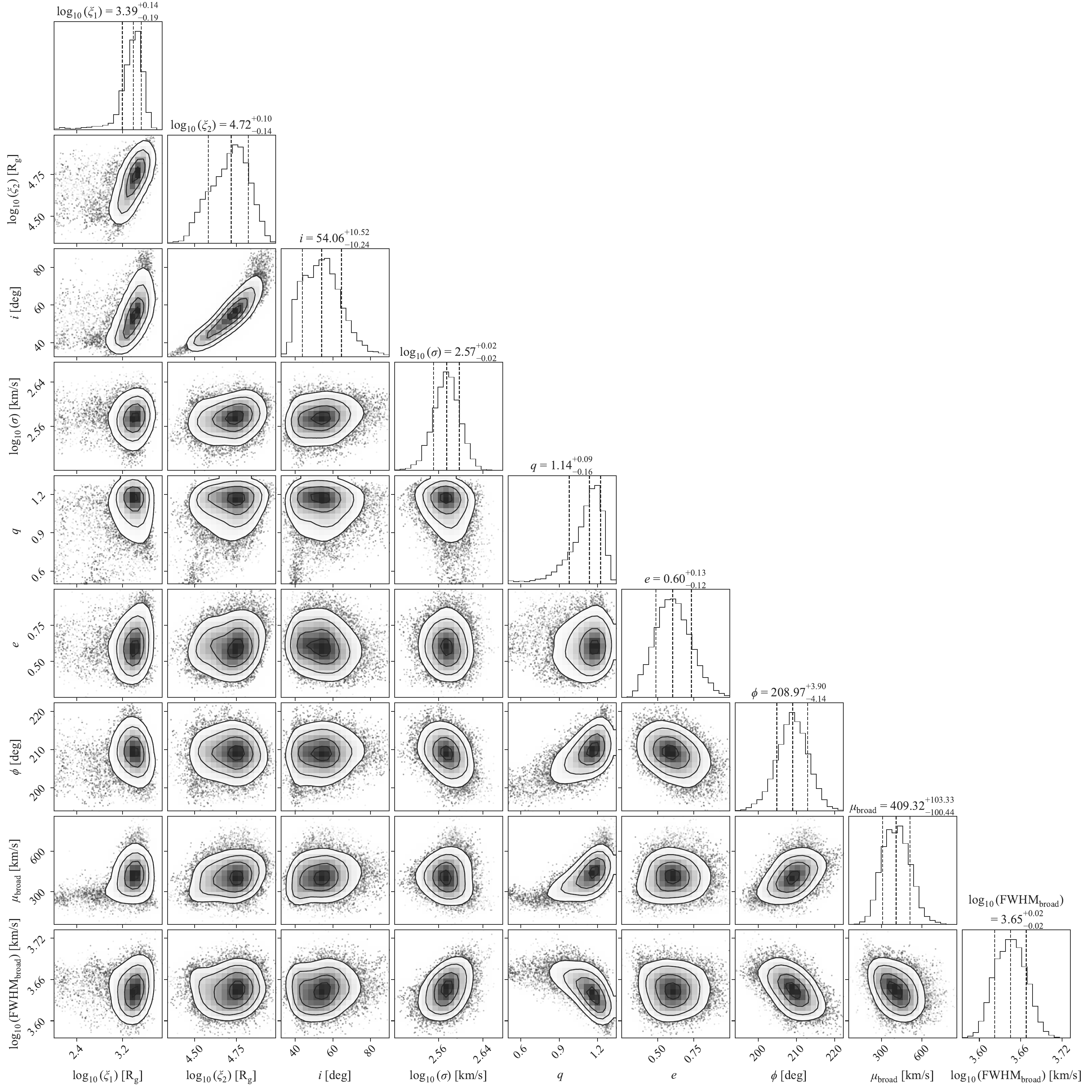}
    \caption{Posterior probability density functions for the elliptical accretion disk and Gaussian models used in the spectroscopic fitting of the H$\alpha$ and H$\beta$ emission line complexes of the LRIS $+15$ day spectrum.
    }
    \label{fig:ap-posteriors}
\end{figure}

For the Gaussian components representing the narrow and broad line features, uniform priors were employed for the amplitude, centroid (represented as a velocity offset), and width parameters of the narrow lines. In contrast, the velocity offset used in the broad Gaussian features was assigned a normal prior centered at zero to reflect physical expectations and further constrain the fit. We demonstrate the resulting posterior probability density functions for the parameters in the LRIS $+15$ day model fit in Figure \ref{fig:ap-posteriors}.

\bibliography{references}{}

\begin{thebibliography}{}
\expandafter\ifx\csname natexlab\endcsname\relax\def\natexlab#1{#1}\fi
\providecommand{\url}[1]{\href{#1}{#1}}
\providecommand{\dodoi}[1]{doi:~\href{http://doi.org/#1}{\nolinkurl{#1}}}
\providecommand{\doeprint}[1]{\href{http://ascl.net/#1}{\nolinkurl{http://ascl.net/#1}}}
\providecommand{\doarXiv}[1]{\href{https://arxiv.org/abs/#1}{\nolinkurl{https://arxiv.org/abs/#1}}}

\bibitem[{Akaike(1974)}]{akaike1974}
Akaike, H. 1974, IEEE Transactions on Automatic Control, 19, 716,
  \dodoi{10.1109/TAC.1974.1100705}

\bibitem[{{Alexander} {et~al.}(2016){Alexander}, {Berger}, {Guillochon},
  {Zauderer}, \& {Williams}}]{alexander2016}
{Alexander}, K.~D., {Berger}, E., {Guillochon}, J., {Zauderer}, B.~A., \&
  {Williams}, P.~K.~G. 2016, \apjl, 819, L25,
  \dodoi{10.3847/2041-8205/819/2/L25}

\bibitem[{{Alexander} {et~al.}(2020){Alexander}, {van Velzen}, {Horesh}, \&
  {Zauderer}}]{alexander2020}
{Alexander}, K.~D., {van Velzen}, S., {Horesh}, A., \& {Zauderer}, B.~A. 2020,
  \ssr, 216, 81, \dodoi{10.1007/s11214-020-00702-w}

\bibitem[{{Aranzana} {et~al.}(2018){Aranzana}, {K{\"o}rding}, {Uttley},
  {Scaringi}, \& {Bloemen}}]{aranzana2018}
{Aranzana}, E., {K{\"o}rding}, E., {Uttley}, P., {Scaringi}, S., \& {Bloemen},
  S. 2018, \mnras, 476, 2501, \dodoi{10.1093/mnras/sty413}

\bibitem[{{Arcavi} {et~al.}(2014){Arcavi}, {Gal-Yam}, {Sullivan}, {Pan},
  {Cenko}, {Horesh}, {Ofek}, {De Cia}, {Yan}, {Yang}, {Howell}, {Tal},
  {Kulkarni}, {Tendulkar}, {Tang}, {Xu}, {Sternberg}, {Cohen}, {Bloom},
  {Nugent}, {Kasliwal}, {Perley}, {Quimby}, {Miller}, {Theissen}, \&
  {Laher}}]{arcavi2014}
{Arcavi}, I., {Gal-Yam}, A., {Sullivan}, M., {et~al.} 2014, \apj, 793, 38,
  \dodoi{10.1088/0004-637X/793/1/38}

\bibitem[{{Arnaud}(1996)}]{arnaud1996}
{Arnaud}, K.~A. 1996, in Astronomical Society of the Pacific Conference Series,
  Vol. 101, Astronomical Data Analysis Software and Systems V, ed. G.~H.
  {Jacoby} \& J.~{Barnes}, 17

\bibitem[{{Auchettl} {et~al.}(2017){Auchettl}, {Guillochon}, \&
  {Ramirez-Ruiz}}]{auchettl2017}
{Auchettl}, K., {Guillochon}, J., \& {Ramirez-Ruiz}, E. 2017, \apj, 838, 149,
  \dodoi{10.3847/1538-4357/aa633b}

\bibitem[{{Auchettl} {et~al.}(2018){Auchettl}, {Ramirez-Ruiz}, \&
  {Guillochon}}]{auchettl2018}
{Auchettl}, K., {Ramirez-Ruiz}, E., \& {Guillochon}, J. 2018, \apj, 852, 37,
  \dodoi{10.3847/1538-4357/aa9b7c}

\bibitem[{{Baldwin} {et~al.}(1981){Baldwin}, {Phillips}, \&
  {Terlevich}}]{baldwin1981}
{Baldwin}, J.~A., {Phillips}, M.~M., \& {Terlevich}, R. 1981, \pasp, 93, 5,
  \dodoi{10.1086/130766}

\bibitem[{{Becker}(2015)}]{becker2015}
{Becker}, A. 2015, {HOTPANTS: High Order Transform of PSF ANd Template
  Subtraction}, Astrophysics Source Code Library, record ascl:1504.004.
\newblock \doeprint{1504.004}

\bibitem[{{Bell} \& {de Jong}(2001)}]{bell2001}
{Bell}, E.~F., \& {de Jong}, R.~S. 2001, \apj, 550, 212, \dodoi{10.1086/319728}

\bibitem[{{Bellm} {et~al.}(2019){Bellm}, {Kulkarni}, {Graham}, {Dekany},
  {Smith}, {Riddle}, {Masci}, {Helou}, {Prince}, {Adams}, {Barbarino},
  {Barlow}, {Bauer}, {Beck}, {Belicki}, {Biswas}, {Blagorodnova}, {Bodewits},
  {Bolin}, {Brinnel}, {Brooke}, {Bue}, {Bulla}, {Burruss}, {Cenko}, {Chang},
  {Connolly}, {Coughlin}, {Cromer}, {Cunningham}, {De}, {Delacroix}, {Desai},
  {Duev}, {Eadie}, {Farnham}, {Feeney}, {Feindt}, {Flynn}, {Franckowiak},
  {Frederick}, {Fremling}, {Gal-Yam}, {Gezari}, {Giomi}, {Goldstein},
  {Golkhou}, {Goobar}, {Groom}, {Hacopians}, {Hale}, {Henning}, {Ho}, {Hover},
  {Howell}, {Hung}, {Huppenkothen}, {Imel}, {Ip}, {Ivezi{\'c}}, {Jackson},
  {Jones}, {Juric}, {Kasliwal}, {Kaspi}, {Kaye}, {Kelley}, {Kowalski},
  {Kramer}, {Kupfer}, {Landry}, {Laher}, {Lee}, {Lin}, {Lin}, {Lunnan},
  {Giomi}, {Mahabal}, {Mao}, {Miller}, {Monkewitz}, {Murphy}, {Ngeow},
  {Nordin}, {Nugent}, {Ofek}, {Patterson}, {Penprase}, {Porter}, {Rauch},
  {Rebbapragada}, {Reiley}, {Rigault}, {Rodriguez}, {van Roestel}, {Rusholme},
  {van Santen}, {Schulze}, {Shupe}, {Singer}, {Soumagnac}, {Stein}, {Surace},
  {Sollerman}, {Szkody}, {Taddia}, {Terek}, {Van Sistine}, {van Velzen},
  {Vestrand}, {Walters}, {Ward}, {Ye}, {Yu}, {Yan}, \& {Zolkower}}]{bellm2019}
{Bellm}, E.~C., {Kulkarni}, S.~R., {Graham}, M.~J., {et~al.} 2019, \pasp, 131,
  018002, \dodoi{10.1088/1538-3873/aaecbe}

\bibitem[{{Blagorodnova} {et~al.}(2017){Blagorodnova}, {Gezari}, {Hung},
  {Kulkarni}, {Cenko}, {Pasham}, {Yan}, {Arcavi}, {Ben-Ami}, {Bue}, {Cantwell},
  {Cao}, {Castro-Tirado}, {Fender}, {Fremling}, {Gal-Yam}, {Ho}, {Horesh},
  {Hosseinzadeh}, {Kasliwal}, {Kong}, {Laher}, {Leloudas}, {Lunnan}, {Masci},
  {Mooley}, {Neill}, {Nugent}, {Powell}, {Valeev}, {Vreeswijk}, {Walters}, \&
  {Wozniak}}]{blagorodnova2017}
{Blagorodnova}, N., {Gezari}, S., {Hung}, T., {et~al.} 2017, \apj, 844, 46,
  \dodoi{10.3847/1538-4357/aa7579}

\bibitem[{{Blagorodnova} {et~al.}(2018){Blagorodnova}, {Neill}, {Walters},
  {Kulkarni}, {Fremling}, {Ben-Ami}, {Dekany}, {Fucik}, {Konidaris}, {Nash},
  {Ngeow}, {Ofek}, {O' Sullivan}, {Quimby}, {Ritter}, \&
  {Vyhmeister}}]{blagorodnova2018}
{Blagorodnova}, N., {Neill}, J.~D., {Walters}, R., {et~al.} 2018, \pasp, 130,
  035003, \dodoi{10.1088/1538-3873/aaa53f}

\bibitem[{{Blagorodnova} {et~al.}(2019){Blagorodnova}, {Cenko}, {Kulkarni},
  {Arcavi}, {Bloom}, {Duggan}, {Filippenko}, {Fremling}, {Horesh},
  {Hosseinzadeh}, {Karamehmetoglu}, {Levan}, {Masci}, {Nugent}, {Pasham},
  {Veilleux}, {Walters}, {Yan}, \& {Zheng}}]{blagorodnova2019}
{Blagorodnova}, N., {Cenko}, S.~B., {Kulkarni}, S.~R., {et~al.} 2019, \apj,
  873, 92, \dodoi{10.3847/1538-4357/ab04b0}

\bibitem[{{Blanchard} {et~al.}(2017){Blanchard}, {Nicholl}, {Berger},
  {Guillochon}, {Margutti}, {Chornock}, {Alexander}, {Leja}, \&
  {Drout}}]{blanchard2017}
{Blanchard}, P.~K., {Nicholl}, M., {Berger}, E., {et~al.} 2017, \apj, 843, 106,
  \dodoi{10.3847/1538-4357/aa77f7}

\bibitem[{{Bogdanovi{\'c}} {et~al.}(2004){Bogdanovi{\'c}}, {Eracleous},
  {Mahadevan}, {Sigurdsson}, \& {Laguna}}]{bogdanovic2004}
{Bogdanovi{\'c}}, T., {Eracleous}, M., {Mahadevan}, S., {Sigurdsson}, S., \&
  {Laguna}, P. 2004, \apj, 610, 707, \dodoi{10.1086/421758}

\bibitem[{{Bonnerot} \& {Lu}(2020)}]{bonnerot2020}
{Bonnerot}, C., \& {Lu}, W. 2020, \mnras, 495, 1374,
  \dodoi{10.1093/mnras/staa1246}

\bibitem[{{Bonnerot} {et~al.}(2017){Bonnerot}, {Rossi}, \&
  {Lodato}}]{bonnerot2017}
{Bonnerot}, C., {Rossi}, E.~M., \& {Lodato}, G. 2017, \mnras, 464, 2816,
  \dodoi{10.1093/mnras/stw2547}

\bibitem[{{Bowen}(1935)}]{bowen1935}
{Bowen}, I.~S. 1935, \apj, 81, 1, \dodoi{10.1086/143613}

\bibitem[{{Brown} {et~al.}(2013){Brown}, {Baliber}, {Bianco}, {Bowman},
  {Burleson}, {Conway}, {Crellin}, {Depagne}, {De Vera}, {Dilday}, {Dragomir},
  {Dubberley}, {Eastman}, {Elphick}, {Falarski}, {Foale}, {Ford}, {Fulton},
  {Garza}, {Gomez}, {Graham}, {Greene}, {Haldeman}, {Hawkins}, {Haworth},
  {Haynes}, {Hidas}, {Hjelstrom}, {Howell}, {Hygelund}, {Lister}, {Lobdill},
  {Martinez}, {Mullins}, {Norbury}, {Parrent}, {Paulson}, {Petry}, {Pickles},
  {Posner}, {Rosing}, {Ross}, {Sand}, {Saunders}, {Shobbrook}, {Shporer},
  {Street}, {Thomas}, {Tsapras}, {Tufts}, {Valenti}, {Vander Horst}, {Walker},
  {White}, \& {Willis}}]{brown2013}
{Brown}, T.~M., {Baliber}, N., {Bianco}, F.~B., {et~al.} 2013, \pasp, 125,
  1031, \dodoi{10.1086/673168}

\bibitem[{{Bu} {et~al.}(2022){Bu}, {Qiao}, {Yang}, {Liu}, {Chen}, \&
  {Wu}}]{bu2022}
{Bu}, D.-F., {Qiao}, E., {Yang}, X.-H., {et~al.} 2022, \mnras, 516, 2833,
  \dodoi{10.1093/mnras/stac2399}

\bibitem[{{Burrows} {et~al.}(2005){Burrows}, {Hill}, {Nousek}, {Kennea},
  {Wells}, {Osborne}, {Abbey}, {Beardmore}, {Mukerjee}, {Short}, {Chincarini},
  {Campana}, {Citterio}, {Moretti}, {Pagani}, {Tagliaferri}, {Giommi},
  {Capalbi}, {Tamburelli}, {Angelini}, {Cusumano}, {Br{\"a}uninger}, {Burkert},
  \& {Hartner}}]{burrows2005}
{Burrows}, D.~N., {Hill}, J.~E., {Nousek}, J.~A., {et~al.} 2005, \ssr, 120,
  165, \dodoi{10.1007/s11214-005-5097-2}

\bibitem[{{Butler} \& {Bloom}(2011)}]{butler2011}
{Butler}, N.~R., \& {Bloom}, J.~S. 2011, \aj, 141, 93,
  \dodoi{10.1088/0004-6256/141/3/93}

\bibitem[{{Calzetti} {et~al.}(2000){Calzetti}, {Armus}, {Bohlin}, {Kinney},
  {Koornneef}, \& {Storchi-Bergmann}}]{calzetti2000}
{Calzetti}, D., {Armus}, L., {Bohlin}, R.~C., {et~al.} 2000, \apj, 533, 682,
  \dodoi{10.1086/308692}

\bibitem[{{Cao} {et~al.}(2018){Cao}, {Liu}, {Zhou}, {Komossa}, \&
  {Ho}}]{cao2018}
{Cao}, R., {Liu}, F.~K., {Zhou}, Z.~Q., {Komossa}, S., \& {Ho}, L.~C. 2018,
  \mnras, 480, 2929, \dodoi{10.1093/mnras/sty1997}

\bibitem[{{Cao} \& {Gu}(2022)}]{cao2022}
{Cao}, X., \& {Gu}, W.-M. 2022, \apj, 936, 141,
  \dodoi{10.3847/1538-4357/ac8980}

\bibitem[{{Caplar} {et~al.}(2017){Caplar}, {Lilly}, \&
  {Trakhtenbrot}}]{caplar2017}
{Caplar}, N., {Lilly}, S.~J., \& {Trakhtenbrot}, B. 2017, \apj, 834, 111,
  \dodoi{10.3847/1538-4357/834/2/111}

\bibitem[{{Cappellari}(2017)}]{cappellari2017}
{Cappellari}, M. 2017, \mnras, 466, 798, \dodoi{10.1093/mnras/stw3020}

\bibitem[{{Carnall} {et~al.}(2018){Carnall}, {McLure}, {Dunlop}, \&
  {Dav{\'e}}}]{carnall2018}
{Carnall}, A.~C., {McLure}, R.~J., {Dunlop}, J.~S., \& {Dav{\'e}}, R. 2018,
  \mnras, 480, 4379, \dodoi{10.1093/mnras/sty2169}

\bibitem[{{Carr} {et~al.}(2024){Carr}, {Davis}, {Camilleri}, {Lidman},
  {Freeman}, \& {Scolnic}}]{carr2024}
{Carr}, A., {Davis}, T.~M., {Camilleri}, R., {et~al.} 2024, arXiv e-prints,
  arXiv:2402.13484, \dodoi{10.48550/arXiv.2402.13484}

\bibitem[{{Cendes} {et~al.}(2023){Cendes}, {Berger}, {Alexander}, {Chornock},
  {Margutti}, {Metzger}, {Wieringa}, {Bietenholz}, {Hajela}, {Laskar}, {Stroh},
  \& {Terreran}}]{cendes2023}
{Cendes}, Y., {Berger}, E., {Alexander}, K.~D., {et~al.} 2023, arXiv e-prints,
  arXiv:2308.13595, \dodoi{10.48550/arXiv.2308.13595}

\bibitem[{{Chambers} \& {Pan-STARRS Team}(2017)}]{chambers2017}
{Chambers}, K.~C., \& {Pan-STARRS Team}. 2017, in American Astronomical Society
  Meeting Abstracts, Vol. 229, American Astronomical Society Meeting Abstracts
  \#229, 223.03

\bibitem[{{Chan} {et~al.}(2020){Chan}, {Piran}, \& {Krolik}}]{chan2020}
{Chan}, C.-H., {Piran}, T., \& {Krolik}, J.~H. 2020, \apj, 903, 17,
  \dodoi{10.3847/1538-4357/abb776}

\bibitem[{{Chan} {et~al.}(2021){Chan}, {Piran}, \& {Krolik}}]{chan2021}
---. 2021, \apj, 914, 107, \dodoi{10.3847/1538-4357/abf0a7}

\bibitem[{{Chan} {et~al.}(2019){Chan}, {Piran}, {Krolik}, \&
  {Saban}}]{chan2019}
{Chan}, C.-H., {Piran}, T., {Krolik}, J.~H., \& {Saban}, D. 2019, \apj, 881,
  113, \dodoi{10.3847/1538-4357/ab2b40}

\bibitem[{{Charalampopoulos} {et~al.}(2022){Charalampopoulos}, {Leloudas},
  {Malesani}, {Wevers}, {Arcavi}, {Nicholl}, {Pursiainen}, {Lawrence},
  {Anderson}, {Benetti}, {Cannizzaro}, {Chen}, {Galbany}, {Gromadzki},
  {Guti{\'e}rrez}, {Inserra}, {Jonker}, {M{\"u}ller-Bravo}, {Onori}, {Short},
  {Sollerman}, \& {Young}}]{charalampopoulos2022}
{Charalampopoulos}, P., {Leloudas}, G., {Malesani}, D.~B., {et~al.} 2022, \aap,
  659, A34, \dodoi{10.1051/0004-6361/202142122}

\bibitem[{{Chen} {et~al.}(2022){Chen}, {Dou}, \& {Shen}}]{chen2022}
{Chen}, J.-H., {Dou}, L.-M., \& {Shen}, R.-F. 2022, \apj, 928, 63,
  \dodoi{10.3847/1538-4357/ac558d}

\bibitem[{{Chen} {et~al.}(1989){Chen}, {Halpern}, \& {Filippenko}}]{chen1989}
{Chen}, K., {Halpern}, J.~P., \& {Filippenko}, A.~V. 1989, \apj, 339, 742,
  \dodoi{10.1086/167332}

\bibitem[{{Chiang} \& {Goldreich}(1997)}]{chiang1997}
{Chiang}, E.~I., \& {Goldreich}, P. 1997, \apj, 490, 368,
  \dodoi{10.1086/304869}

\bibitem[{{Childress} {et~al.}(2014){Childress}, {Vogt}, {Nielsen}, \&
  {Sharp}}]{childress2014}
{Childress}, M.~J., {Vogt}, F. P.~A., {Nielsen}, J., \& {Sharp}, R.~G. 2014,
  \apss, 349, 617, \dodoi{10.1007/s10509-013-1682-0}

\bibitem[{{Clavel} {et~al.}(1991){Clavel}, {Reichert}, {Alloin}, {Crenshaw},
  {Kriss}, {Krolik}, {Malkan}, {Netzer}, {Peterson}, {Wamsteker}, {Altamore},
  {Baribaud}, {Barr}, {Beck}, {Binette}, {Bromage}, {Brosch}, {Diaz},
  {Filippenko}, {Fricke}, {Gaskell}, {Giommi}, {Glass}, {Gondhalekar},
  {Hackney}, {Halpern}, {Hutter}, {Joersaeter}, {Kinney}, {Kollatschny},
  {Koratkar}, {Korista}, {Laor}, {Lasota}, {Leibowitz}, {Maoz}, {Martin},
  {Mazeh}, {Meurs}, {Nair}, {O'Brien}, {Pelat}, {Perez}, {Perola}, {Ptak},
  {Rodriguez-Pascual}, {Rosenblatt}, {Sadun}, {Santos-Lleo}, {Shaw}, {Smith},
  {Stirpe}, {Stoner}, {Sun}, {Ulrich}, {van Groningen}, \&
  {Zheng}}]{clavel1991}
{Clavel}, J., {Reichert}, G.~A., {Alloin}, D., {et~al.} 1991, \apj, 366, 64,
  \dodoi{10.1086/169540}

\bibitem[{{Clemens} {et~al.}(2004){Clemens}, {Crain}, \&
  {Anderson}}]{clemens2004}
{Clemens}, J.~C., {Crain}, J.~A., \& {Anderson}, R. 2004, in Society of
  Photo-Optical Instrumentation Engineers (SPIE) Conference Series, Vol. 5492,
  Ground-based Instrumentation for Astronomy, ed. A.~F.~M. {Moorwood} \&
  M.~{Iye}, 331--340, \dodoi{10.1117/12.550069}

\bibitem[{{Coughlin} \& {Begelman}(2014)}]{coughlin2014}
{Coughlin}, E.~R., \& {Begelman}, M.~C. 2014, \apj, 781, 82,
  \dodoi{10.1088/0004-637X/781/2/82}

\bibitem[{{Coughlin} \& {Nixon}(2022)}]{coughlin2022}
{Coughlin}, E.~R., \& {Nixon}, C.~J. 2022, \mnras, 517, L26,
  \dodoi{10.1093/mnrasl/slac106}

\bibitem[{{Coulter} {et~al.}(2022){Coulter}, {Jones}, {McGill}, {Foley},
  {Aleo}, {Bustamante-Rosell}, {Chatterjee}, {Davis}, {Engel}, {Gagliano},
  {Jacobson-Gal{\'a}n}, {Kilpatrick}, {Pan}, {Rojas-Bravo}, {Siebert},
  {Taggart}, {Tinyanont}, \& {Wang}}]{coulter2022}
{Coulter}, D.~A., {Jones}, D.~O., {McGill}, P., {et~al.} 2022, {YSE-PZ: An
  Open-source Target and Observation Management System}, v0.3.0,  Zenodo,
  \dodoi{10.5281/zenodo.7278430}

\bibitem[{{Coulter} {et~al.}(2023){Coulter}, {Jones}, {McGill}, {Foley},
  {Aleo}, {Bustamante-Rosell}, {Chatterjee}, {Davis}, {Dickinson}, {Engel},
  {Gagliano}, {Jacobson-Gal{\'a}n}, {Kilpatrick}, {Kutcka}, {Le Saux},
  {Malanchev}, {Pan}, {Qui{\~n}onez}, {Rojas-Bravo}, {Siebert}, {Taggart},
  {Tinyanont}, \& {Wang}}]{coulter2023}
---. 2023, \pasp, 135, 064501, \dodoi{10.1088/1538-3873/acd662}

\bibitem[{{Crumley} {et~al.}(2016){Crumley}, {Lu}, {Santana}, {Hern{\'a}ndez},
  {Kumar}, \& {Markoff}}]{crumley2016}
{Crumley}, P., {Lu}, W., {Santana}, R., {et~al.} 2016, \mnras, 460, 396,
  \dodoi{10.1093/mnras/stw967}

\bibitem[{{Cutri} {et~al.}(2003){Cutri}, {Skrutskie}, {van Dyk}, {Beichman},
  {Carpenter}, {Chester}, {Cambresy}, {Evans}, {Fowler}, {Gizis}, {Howard},
  {Huchra}, {Jarrett}, {Kopan}, {Kirkpatrick}, {Light}, {Marsh}, {McCallon},
  {Schneider}, {Stiening}, {Sykes}, {Weinberg}, {Wheaton}, {Wheelock}, \&
  {Zacarias}}]{cutri2003}
{Cutri}, R.~M., {Skrutskie}, M.~F., {van Dyk}, S., {et~al.} 2003, {VizieR
  Online Data Catalog: 2MASS All-Sky Catalog of Point Sources (Cutri+ 2003)},
  VizieR On-line Data Catalog: II/246. Originally published in: University of
  Massachusetts and Infrared Processing and Analysis Center, (IPAC/California
  Institute of Technology) (2003)

\bibitem[{{Dahiwale} \& {Fremling}(2020)}]{dahiwale2020}
{Dahiwale}, A., \& {Fremling}, C. 2020, Transient Name Server Classification
  Report, 2020-3800, 1

\bibitem[{{Dai} {et~al.}(2021){Dai}, {Lodato}, \& {Cheng}}]{dai2021}
{Dai}, J.~L., {Lodato}, G., \& {Cheng}, R. 2021, \ssr, 217, 12,
  \dodoi{10.1007/s11214-020-00747-x}

\bibitem[{{Dai} {et~al.}(2015){Dai}, {McKinney}, \& {Miller}}]{dai2015}
{Dai}, L., {McKinney}, J.~C., \& {Miller}, M.~C. 2015, \apjl, 812, L39,
  \dodoi{10.1088/2041-8205/812/2/L39}

\bibitem[{{Dai} {et~al.}(2018){Dai}, {McKinney}, {Roth}, {Ramirez-Ruiz}, \&
  {Miller}}]{dai2018}
{Dai}, L., {McKinney}, J.~C., {Roth}, N., {Ramirez-Ruiz}, E., \& {Miller},
  M.~C. 2018, \apjl, 859, L20, \dodoi{10.3847/2041-8213/aab429}

\bibitem[{{De} {et~al.}(2020){De}, {Kasliwal}, {Tzanidakis}, {Fremling},
  {Adams}, {Aloisi}, {Andreoni}, {Bagdasaryan}, {Bellm}, {Bildsten},
  {Cannella}, {Cook}, {Delacroix}, {Drake}, {Duev}, {Dugas}, {Frederick},
  {Gal-Yam}, {Goldstein}, {Golkhou}, {Graham}, {Hale}, {Hankins}, {Helou},
  {Ho}, {Irani}, {Jencson}, {Kaplan}, {Kaye}, {Kulkarni}, {Kupfer}, {Laher},
  {Leadbeater}, {Lunnan}, {Masci}, {Miller}, {Neill}, {Ofek}, {Perley},
  {Polin}, {Prince}, {Quataert}, {Reiley}, {Riddle}, {Rusholme}, {Sharma},
  {Shupe}, {Sollerman}, {Tartaglia}, {Walters}, {Yan}, \& {Yao}}]{de2020}
{De}, K., {Kasliwal}, M.~M., {Tzanidakis}, A., {et~al.} 2020, \apj, 905, 58,
  \dodoi{10.3847/1538-4357/abb45c}

\bibitem[{{De Colle} {et~al.}(2012){De Colle}, {Guillochon}, {Naiman}, \&
  {Ramirez-Ruiz}}]{decolle2012}
{De Colle}, F., {Guillochon}, J., {Naiman}, J., \& {Ramirez-Ruiz}, E. 2012,
  \apj, 760, 103, \dodoi{10.1088/0004-637X/760/2/103}

\bibitem[{{Denney} {et~al.}(2009){Denney}, {Peterson}, {Pogge}, {Adair},
  {Atlee}, {Au-Yong}, {Bentz}, {Bird}, {Brokofsky}, {Chisholm}, {Comins},
  {Dietrich}, {Doroshenko}, {Eastman}, {Efimov}, {Ewald}, {Ferbey}, {Gaskell},
  {Hedrick}, {Jackson}, {Klimanov}, {Klimek}, {Kruse}, {Lad{\'e}route}, {Lamb},
  {Leighly}, {Minezaki}, {Nazarov}, {Onken}, {Petersen}, {Peterson},
  {Poindexter}, {Sakata}, {Schlesinger}, {Sergeev}, {Skolski}, {Stieglitz},
  {Tobin}, {Unterborn}, {Vestergaard}, {Watkins}, {Watson}, \&
  {Yoshii}}]{denney2009}
{Denney}, K.~D., {Peterson}, B.~M., {Pogge}, R.~W., {et~al.} 2009, \apjl, 704,
  L80, \dodoi{10.1088/0004-637X/704/2/L80}

\bibitem[{{Dodd} {et~al.}(2021){Dodd}, {Law-Smith}, {Auchettl}, {Ramirez-Ruiz},
  \& {Foley}}]{dodd2021}
{Dodd}, S.~A., {Law-Smith}, J. A.~P., {Auchettl}, K., {Ramirez-Ruiz}, E., \&
  {Foley}, R.~J. 2021, \apjl, 907, L21, \dodoi{10.3847/2041-8213/abd852}

\bibitem[{{Dodd} {et~al.}(2023){Dodd}, {Nukala}, {Connor}, {Auchettl},
  {French}, {Law-Smith}, {Hammerstein}, \& {Ramirez-Ruiz}}]{dodd2023}
{Dodd}, S.~A., {Nukala}, A., {Connor}, I., {et~al.} 2023, \apjl, 959, L19,
  \dodoi{10.3847/2041-8213/ad1112}

\bibitem[{{Dopita} {et~al.}(2007){Dopita}, {Hart}, {McGregor}, {Oates},
  {Bloxham}, \& {Jones}}]{dopita2007}
{Dopita}, M., {Hart}, J., {McGregor}, P., {et~al.} 2007, \apss, 310, 255,
  \dodoi{10.1007/s10509-007-9510-z}

\bibitem[{{Dopita} {et~al.}(2010){Dopita}, {Rhee}, {Farage}, {McGregor},
  {Bloxham}, {Green}, {Roberts}, {Neilson}, {Wilson}, {Young}, {Firth},
  {Busarello}, \& {Merluzzi}}]{dopita2010}
{Dopita}, M., {Rhee}, J., {Farage}, C., {et~al.} 2010, \apss, 327, 245,
  \dodoi{10.1007/s10509-010-0335-9}

\bibitem[{{Eracleous} {et~al.}(1995){Eracleous}, {Livio}, {Halpern}, \&
  {Storchi-Bergmann}}]{eracleous1995}
{Eracleous}, M., {Livio}, M., {Halpern}, J.~P., \& {Storchi-Bergmann}, T. 1995,
  \apj, 438, 610, \dodoi{10.1086/175104}

\bibitem[{{Evans} {et~al.}(2009){Evans}, {Beardmore}, {Page}, {Osborne},
  {O'Brien}, {Willingale}, {Starling}, {Burrows}, {Godet}, {Vetere}, {Racusin},
  {Goad}, {Wiersema}, {Angelini}, {Capalbi}, {Chincarini}, {Gehrels}, {Kennea},
  {Margutti}, {Morris}, {Mountford}, {Pagani}, {Perri}, {Romano}, \&
  {Tanvir}}]{evans2009}
{Evans}, P.~A., {Beardmore}, A.~P., {Page}, K.~L., {et~al.} 2009, \mnras, 397,
  1177, \dodoi{10.1111/j.1365-2966.2009.14913.x}

\bibitem[{{Fitzpatrick}(1999)}]{fitzpatrick1999}
{Fitzpatrick}, E.~L. 1999, \pasp, 111, 63, \dodoi{10.1086/316293}

\bibitem[{{Foley} {et~al.}(2003){Foley}, {Papenkova}, {Swift}, {Filippenko},
  {Li}, {Mazzali}, {Chornock}, {Leonard}, \& {Van Dyk}}]{foley2003}
{Foley}, R.~J., {Papenkova}, M.~S., {Swift}, B.~J., {et~al.} 2003, \pasp, 115,
  1220, \dodoi{10.1086/378242}

\bibitem[{{French} {et~al.}(2020){French}, {Wevers}, {Law-Smith}, {Graur}, \&
  {Zabludoff}}]{french2020b}
{French}, K.~D., {Wevers}, T., {Law-Smith}, J., {Graur}, O., \& {Zabludoff},
  A.~I. 2020, \ssr, 216, 32, \dodoi{10.1007/s11214-020-00657-y}

\bibitem[{{Gehrels} {et~al.}(2004){Gehrels}, {Chincarini}, {Giommi}, {Mason},
  {Nousek}, {Wells}, {White}, {Barthelmy}, {Burrows}, {Cominsky}, {Hurley},
  {Marshall}, {M{\'e}sz{\'a}ros}, {Roming}, {Angelini}, {Barbier}, {Belloni},
  {Campana}, {Caraveo}, {Chester}, {Citterio}, {Cline}, {Cropper}, {Cummings},
  {Dean}, {Feigelson}, {Fenimore}, {Frail}, {Fruchter}, {Garmire}, {Gendreau},
  {Ghisellini}, {Greiner}, {Hill}, {Hunsberger}, {Krimm}, {Kulkarni}, {Kumar},
  {Lebrun}, {Lloyd-Ronning}, {Markwardt}, {Mattson}, {Mushotzky}, {Norris},
  {Osborne}, {Paczynski}, {Palmer}, {Park}, {Parsons}, {Paul}, {Rees},
  {Reynolds}, {Rhoads}, {Sasseen}, {Schaefer}, {Short}, {Smale}, {Smith},
  {Stella}, {Tagliaferri}, {Takahashi}, {Tashiro}, {Townsley}, {Tueller},
  {Turner}, {Vietri}, {Voges}, {Ward}, {Willingale}, {Zerbi}, \&
  {Zhang}}]{gehrels2004}
{Gehrels}, N., {Chincarini}, G., {Giommi}, P., {et~al.} 2004, \apj, 611, 1005,
  \dodoi{10.1086/422091}

\bibitem[{Gezari(2014)}]{gezari2014}
Gezari, S. 2014, Physics Today, 67, 37, \dodoi{10.1063/PT.3.2382}

\bibitem[{{Gezari}(2021)}]{gezari2021}
{Gezari}, S. 2021, \araa, 59, 21, \dodoi{10.1146/annurev-astro-111720-030029}

\bibitem[{{Gezari} {et~al.}(2017){Gezari}, {Cenko}, \& {Arcavi}}]{gezari2017}
{Gezari}, S., {Cenko}, S.~B., \& {Arcavi}, I. 2017, \apjl, 851, L47,
  \dodoi{10.3847/2041-8213/aaa0c2}

\bibitem[{{Gezari} {et~al.}(2012){Gezari}, {Chornock}, {Rest}, {Huber},
  {Forster}, {Berger}, {Challis}, {Neill}, {Martin}, {Heckman}, {Lawrence},
  {Norman}, {Narayan}, {Foley}, {Marion}, {Scolnic}, {Chomiuk}, {Soderberg},
  {Smith}, {Kirshner}, {Riess}, {Smartt}, {Stubbs}, {Tonry}, {Wood-Vasey},
  {Burgett}, {Chambers}, {Grav}, {Heasley}, {Kaiser}, {Kudritzki}, {Magnier},
  {Morgan}, \& {Price}}]{gezari2012}
{Gezari}, S., {Chornock}, R., {Rest}, A., {et~al.} 2012, \nat, 485, 217,
  \dodoi{10.1038/nature10990}

\bibitem[{{Golightly} {et~al.}(2019){Golightly}, {Coughlin}, \&
  {Nixon}}]{golightly2019}
{Golightly}, E. C.~A., {Coughlin}, E.~R., \& {Nixon}, C.~J. 2019, \apj, 872,
  163, \dodoi{10.3847/1538-4357/aafd2f}

\bibitem[{{Gomez} {et~al.}(2020){Gomez}, {Nicholl}, {Short}, {Margutti},
  {Alexander}, {Blanchard}, {Berger}, {Eftekhari}, {Schulze}, {Anderson},
  {Arcavi}, {Chornock}, {Cowperthwaite}, {Galbany}, {Herzog}, {Hiramatsu},
  {Hosseinzadeh}, {Laskar}, {M{\"u}ller Bravo}, {Patton}, \&
  {Terreran}}]{gomez2020}
{Gomez}, S., {Nicholl}, M., {Short}, P., {et~al.} 2020, \mnras, 497, 1925,
  \dodoi{10.1093/mnras/staa2099}

\bibitem[{{Green} {et~al.}(1995){Green}, {Schartel}, {Anderson}, {Hewett},
  {Foltz}, {Brinkmann}, {Fink}, {Truemper}, \& {Margon}}]{green1995}
{Green}, P.~J., {Schartel}, N., {Anderson}, S.~F., {et~al.} 1995, \apj, 450,
  51, \dodoi{10.1086/176118}

\bibitem[{{Greene} {et~al.}(2020){Greene}, {Strader}, \& {Ho}}]{greene2020}
{Greene}, J.~E., {Strader}, J., \& {Ho}, L.~C. 2020, \araa, 58, 257,
  \dodoi{10.1146/annurev-astro-032620-021835}

\bibitem[{{Gromadzki} {et~al.}(2019){Gromadzki}, {Hamanowicz}, {Wyrzykowski},
  {Sokolovsky}, {Fraser}, {Koz{\l}owski}, {Guillochon}, {Arcavi},
  {Trakhtenbrot}, {Jonker}, {Mattila}, {Udalski}, {Szyma{\'n}ski},
  {Soszy{\'n}ski}, {Poleski}, {Pietrukowicz}, {Skowron}, {Mr{\'o}z}, {Ulaczyk},
  {Pawlak}, {Rybicki}, {Sollerman}, {Taddia}, {Kostrzewa-Rutkowska}, {Onori},
  {Young}, {Maguire}, {Smartt}, {Inserra}, {Gal-Yam}, {Rau}, {Chen}, {Angus},
  \& {Buckley}}]{gromadzki2019}
{Gromadzki}, M., {Hamanowicz}, A., {Wyrzykowski}, L., {et~al.} 2019, \aap, 622,
  L2, \dodoi{10.1051/0004-6361/201833682}

\bibitem[{{Guillochon} {et~al.}(2014){Guillochon}, {Manukian}, \&
  {Ramirez-Ruiz}}]{guillochon2014}
{Guillochon}, J., {Manukian}, H., \& {Ramirez-Ruiz}, E. 2014, \apj, 783, 23,
  \dodoi{10.1088/0004-637X/783/1/23}

\bibitem[{{Guillochon} \& {Ramirez-Ruiz}(2013)}]{guillochon2013}
{Guillochon}, J., \& {Ramirez-Ruiz}, E. 2013, \apj, 767, 25,
  \dodoi{10.1088/0004-637X/767/1/25}

\bibitem[{{Guillochon} \& {Ramirez-Ruiz}(2015)}]{guillochon2015}
---. 2015, \apj, 809, 166, \dodoi{10.1088/0004-637X/809/2/166}

\bibitem[{{Guolo} {et~al.}(2024){Guolo}, {Gezari}, {Yao}, {van Velzen},
  {Hammerstein}, {Cenko}, \& {Tokayer}}]{guolo2024}
{Guolo}, M., {Gezari}, S., {Yao}, Y., {et~al.} 2024, \apj, 966, 160,
  \dodoi{10.3847/1538-4357/ad2f9f}

\bibitem[{{Hajela} {et~al.}(2024){Hajela}, {Alexander}, {Margutti}, {Chornock},
  {Bietenholz}, {Christy}, {Stroh}, {Terreran}, {Saxton}, {Komossa}, {Bright},
  {Ramirez-Ruiz}, {Coppejans}, {Leung}, {Cendes}, {Wiston}, {Laskar}, {Horesh},
  {Schroeder}, {Nayana A.}, {Wieringa}, {Velez}, {Berger}, {Blanchard},
  {Eftekhari}, {Gomez}, {Nicholl}, {Sears}, \& {Zauderer}}]{hajela2024}
{Hajela}, A., {Alexander}, K.~D., {Margutti}, R., {et~al.} 2024, arXiv
  e-prints, arXiv:2407.19019, \dodoi{10.48550/arXiv.2407.19019}

\bibitem[{{Hammerstein} {et~al.}(2023){Hammerstein}, {van Velzen}, {Gezari},
  {Cenko}, {Yao}, {Ward}, {Frederick}, {Villanueva}, {Somalwar}, {Graham},
  {Kulkarni}, {Stern}, {Andreoni}, {Bellm}, {Dekany}, {Dhawan}, {Drake},
  {Fremling}, {Gatkine}, {Groom}, {Ho}, {Kasliwal}, {Karambelkar}, {Kool},
  {Masci}, {Medford}, {Perley}, {Purdum}, {van Roestel}, {Sharma}, {Sollerman},
  {Taggart}, \& {Yan}}]{hammerstein2023}
{Hammerstein}, E., {van Velzen}, S., {Gezari}, S., {et~al.} 2023, \apj, 942, 9,
  \dodoi{10.3847/1538-4357/aca283}

\bibitem[{{Hayasaki} \& {Loeb}(2016)}]{hayasaki2016a}
{Hayasaki}, K., \& {Loeb}, A. 2016, Scientific Reports, 6, 35629,
  \dodoi{10.1038/srep35629}

\bibitem[{{Hayasaki} {et~al.}(2013){Hayasaki}, {Stone}, \&
  {Loeb}}]{hayasaki2013a}
{Hayasaki}, K., {Stone}, N., \& {Loeb}, A. 2013, \mnras, 434, 909,
  \dodoi{10.1093/mnras/stt871}

\bibitem[{{Heckman} {et~al.}(2004){Heckman}, {Kauffmann}, {Brinchmann},
  {Charlot}, {Tremonti}, \& {White}}]{heckman2004}
{Heckman}, T.~M., {Kauffmann}, G., {Brinchmann}, J., {et~al.} 2004, \apj, 613,
  109, \dodoi{10.1086/422872}

\bibitem[{{Heckman} {et~al.}(2005){Heckman}, {Ptak}, {Hornschemeier}, \&
  {Kauffmann}}]{heckman2005}
{Heckman}, T.~M., {Ptak}, A., {Hornschemeier}, A., \& {Kauffmann}, G. 2005,
  \apj, 634, 161, \dodoi{10.1086/491665}

\bibitem[{{Hern{\'a}ndez-Garc{\'\i}a}
  {et~al.}(2016){Hern{\'a}ndez-Garc{\'\i}a}, {Masegosa},
  {Gonz{\'a}lez-Mart{\'\i}n}, {M{\'a}rquez}, \& {Perea}}]{hernandez-garcia2016}
{Hern{\'a}ndez-Garc{\'\i}a}, L., {Masegosa}, J., {Gonz{\'a}lez-Mart{\'\i}n},
  O., {M{\'a}rquez}, I., \& {Perea}, J. 2016, \apj, 824, 7,
  \dodoi{10.3847/0004-637X/824/1/7}

\bibitem[{{HI4PI Collaboration} {et~al.}(2016){HI4PI Collaboration}, {Ben
  Bekhti}, {Fl{\"o}er}, {Keller}, {Kerp}, {Lenz}, {Winkel}, {Bailin},
  {Calabretta}, {Dedes}, {Ford}, {Gibson}, {Haud}, {Janowiecki}, {Kalberla},
  {Lockman}, {McClure-Griffiths}, {Murphy}, {Nakanishi}, {Pisano}, \&
  {Staveley-Smith}}]{hi4pi2016}
{HI4PI Collaboration}, {Ben Bekhti}, N., {Fl{\"o}er}, L., {et~al.} 2016, \aap,
  594, A116, \dodoi{10.1051/0004-6361/201629178}

\bibitem[{{Hills}(1975)}]{hills1975}
{Hills}, J.~G. 1975, \nat, 254, 295, \dodoi{10.1038/254295a0}

\bibitem[{{Hinkle}(2024)}]{hinkle2024}
{Hinkle}, J.~T. 2024, \mnras, 531, 2603, \dodoi{10.1093/mnras/stae1229}

\bibitem[{{Hinkle} {et~al.}(2021{\natexlab{a}}){Hinkle}, {Holoien}, {Shappee},
  \& {Auchettl}}]{hinkle2021b}
{Hinkle}, J.~T., {Holoien}, T. W.~S., {Shappee}, B.~J., \& {Auchettl}, K.
  2021{\natexlab{a}}, \apj, 910, 83, \dodoi{10.3847/1538-4357/abe4d8}

\bibitem[{{Hinkle} {et~al.}(2021{\natexlab{b}}){Hinkle}, {Holoien}, {Auchettl},
  {Shappee}, {Neustadt}, {Payne}, {Brown}, {Kochanek}, {Stanek}, {Graham},
  {Tucker}, {Do}, {Anderson}, {Bose}, {Chen}, {Coulter}, {Dimitriadis}, {Dong},
  {Foley}, {Huber}, {Hung}, {Kilpatrick}, {Pignata}, {Piro}, {Rojas-Bravo},
  {Siebert}, {Stalder}, {Thompson}, {Tonry}, {Vallely}, \&
  {Wisniewski}}]{hinkle2021a}
{Hinkle}, J.~T., {Holoien}, T.~W.~S., {Auchettl}, K., {et~al.}
  2021{\natexlab{b}}, \mnras, 500, 1673, \dodoi{10.1093/mnras/staa3170}

\bibitem[{{Hodgkin} {et~al.}(2013){Hodgkin}, {Wyrzykowski}, {Blagorodnova}, \&
  {Koposov}}]{hodgkin2013}
{Hodgkin}, S.~T., {Wyrzykowski}, L., {Blagorodnova}, N., \& {Koposov}, S. 2013,
  Philosophical Transactions of the Royal Society of London Series A, 371,
  20120239, \dodoi{10.1098/rsta.2012.0239}

\bibitem[{{Holoien} {et~al.}(2019){Holoien}, {Vallely}, {Auchettl}, {Stanek},
  {Kochanek}, {French}, {Prieto}, {Shappee}, {Brown}, {Fausnaugh}, {Dong},
  {Thompson}, {Bose}, {Neustadt}, {Cacella}, {Brimacombe}, {Kendurkar},
  {Beaton}, {Boutsia}, {Chomiuk}, {Connor}, {Morrell}, {Newman}, {Rudie},
  {Shishkovksy}, \& {Strader}}]{holoien2019a}
{Holoien}, T. W.~S., {Vallely}, P.~J., {Auchettl}, K., {et~al.} 2019, \apj,
  883, 111, \dodoi{10.3847/1538-4357/ab3c66}

\bibitem[{{Holoien} {et~al.}(2020){Holoien}, {Auchettl}, {Tucker}, {Shappee},
  {Patel}, {Miller-Jones}, {Mockler}, {Groenewald}, {Hinkle}, {Brown},
  {Kochanek}, {Stanek}, {Chen}, {Dong}, {Prieto}, {Thompson}, {Beaton},
  {Connor}, {Cowperthwaite}, {Dahmen}, {French}, {Morrell}, {Buckley},
  {Gromadzki}, {Roy}, {Coulter}, {Dimitriadis}, {Foley}, {Kilpatrick}, {Piro},
  {Rojas-Bravo}, {Siebert}, \& {van Velzen}}]{holoien2020}
{Holoien}, T. W.~S., {Auchettl}, K., {Tucker}, M.~A., {et~al.} 2020, \apj, 898,
  161, \dodoi{10.3847/1538-4357/ab9f3d}

\bibitem[{{Holoien} {et~al.}(2022){Holoien}, {Neustadt}, {Vallely}, {Auchettl},
  {Hinkle}, {Romero-Ca{\~n}izales}, {Shappee}, {Kochanek}, {Stanek}, {Chen},
  {Dong}, {Prieto}, {Thompson}, {Brink}, {Filippenko}, {Zheng}, {Bersier},
  {Bose}, {Burgasser}, {Channa}, {de Jaeger}, {Hestenes}, {Im}, {Jeffers},
  {Jun}, {Lansbury}, {Post}, {Ross}, {Stern}, {Tang}, {Tucker}, {Valenti},
  {Yunus}, \& {Zhang}}]{holoien2022}
{Holoien}, T. W.~S., {Neustadt}, J. M.~M., {Vallely}, P.~J., {et~al.} 2022,
  \apj, 933, 196, \dodoi{10.3847/1538-4357/ac74b9}

\bibitem[{{Horne}(1986)}]{horne1986}
{Horne}, K. 1986, \pasp, 98, 609, \dodoi{10.1086/131801}

\bibitem[{{Horne} \& {Saar}(1991)}]{horne1991}
{Horne}, K., \& {Saar}, S.~H. 1991, \apjl, 374, L55, \dodoi{10.1086/186070}

\bibitem[{{Hovatta} {et~al.}(2008){Hovatta}, {Nieppola}, {Tornikoski},
  {Valtaoja}, {Aller}, \& {Aller}}]{hovatta2008}
{Hovatta}, T., {Nieppola}, E., {Tornikoski}, M., {et~al.} 2008, \aap, 485, 51,
  \dodoi{10.1051/0004-6361:200809806}

\bibitem[{{Hung} {et~al.}(2017){Hung}, {Gezari}, {Blagorodnova}, {Roth},
  {Cenko}, {Kulkarni}, {Horesh}, {Arcavi}, {McCully}, {Yan}, {Lunnan},
  {Fremling}, {Cao}, {Nugent}, \& {Wozniak}}]{hung2017}
{Hung}, T., {Gezari}, S., {Blagorodnova}, N., {et~al.} 2017, \apj, 842, 29,
  \dodoi{10.3847/1538-4357/aa7337}

\bibitem[{{Hung} {et~al.}(2019){Hung}, {Cenko}, {Roth}, {Gezari}, {Veilleux},
  {van Velzen}, {Gaskell}, {Foley}, {Blagorodnova}, {Yan}, {Graham}, {Brown},
  {Siebert}, {Frederick}, {Ward}, {Gatkine}, {Gal-Yam}, {Yang}, {Schulze},
  {Dimitriadis}, {Kupfer}, {Shupe}, {Rusholme}, {Masci}, {Riddle}, {Soumagnac},
  {van Roestel}, \& {Dekany}}]{hung2019}
{Hung}, T., {Cenko}, S.~B., {Roth}, N., {et~al.} 2019, \apj, 879, 119,
  \dodoi{10.3847/1538-4357/ab24de}

\bibitem[{{Hung} {et~al.}(2020){Hung}, {Foley}, {Ramirez-Ruiz}, {Dai},
  {Auchettl}, {Kilpatrick}, {Mockler}, {Brown}, {Coulter}, {Dimitriadis},
  {Holoien}, {Law-Smith}, {Piro}, {Rest}, {Rojas-Bravo}, \&
  {Siebert}}]{hung2020}
{Hung}, T., {Foley}, R.~J., {Ramirez-Ruiz}, E., {et~al.} 2020, \apj, 903, 31,
  \dodoi{10.3847/1538-4357/abb606}

\bibitem[{{Jiang} {et~al.}(2021{\natexlab{a}}){Jiang}, {Wang}, {Hu}, {Sun},
  {Dou}, \& {Xiao}}]{jiang2021b}
{Jiang}, N., {Wang}, T., {Hu}, X., {et~al.} 2021{\natexlab{a}}, \apj, 911, 31,
  \dodoi{10.3847/1538-4357/abe772}

\bibitem[{{Jiang} {et~al.}(2019){Jiang}, {Wang}, {Mou}, {Liu}, {Dou}, {Sheng},
  \& {Wang}}]{jiang2019a}
{Jiang}, N., {Wang}, T., {Mou}, G., {et~al.} 2019, \apj, 871, 15,
  \dodoi{10.3847/1538-4357/aaf6b2}

\bibitem[{{Jiang} {et~al.}(2017){Jiang}, {Wang}, {Yan}, {Xiao}, {Yang}, {Dou},
  {Wang}, {Cutri}, \& {Mainzer}}]{jiang2017}
{Jiang}, N., {Wang}, T., {Yan}, L., {et~al.} 2017, \apj, 850, 63,
  \dodoi{10.3847/1538-4357/aa93f5}

\bibitem[{{Jiang} {et~al.}(2021{\natexlab{b}}){Jiang}, {Wang}, {Dou}, {Shu},
  {Hu}, {Liu}, {Wang}, {Yan}, {Sheng}, {Yang}, {Sun}, \& {Zhou}}]{jiang2021a}
{Jiang}, N., {Wang}, T., {Dou}, L., {et~al.} 2021{\natexlab{b}}, \apjs, 252,
  32, \dodoi{10.3847/1538-4365/abd1dc}

\bibitem[{{Jones} {et~al.}(2019){Jones}, {French}, {Agnello}, {Angus},
  {Ansari}, {Arendse}, {Gall}, {Grillo}, {Bruun}, {Hede}, {Hjorth}, {Izzo},
  {Korhonen}, {Raimundo}, {Ramanah}, {Sarangi}, {Wojtak}, {Chambers}, {Huber},
  {Magnier}, {Boer}, {Fairlamb}, {Lin}, {Wainscoat}, {Lowe}, {Willman},
  {Bulger}, {Schultz}, {Engel}, {Gagliano}, {Narayan}, {Soraisam}, {Wang},
  {Rest}, {Smartt}, {Smith}, {Alexander}, {Baldeschi}, {Blanchard},
  {Coppejans}, {DeMarchi}, {Hajela}, {Jacobson-Galan}, {Margutti}, {Matthews},
  {Stauffer}, {Stroh}, {Terreran}, {Drout}, {Coulter}, {Dimitriadis}, {Foley},
  {Hung}, {Kilpatrick}, {Rojas-Bravo}, {Siebert}, {Auchettl}, \&
  {Ramirez-Ruiz}}]{jones2019}
{Jones}, D.~O., {French}, K.~D., {Agnello}, A., {et~al.} 2019, Transient Name
  Server AstroNote, 148, 1

\bibitem[{{Jones} {et~al.}(2021){Jones}, {Foley}, {Narayan}, {Hjorth}, {Huber},
  {Aleo}, {Alexander}, {Angus}, {Auchettl}, {Baldassare}, {Bruun}, {Chambers},
  {Chatterjee}, {Coppejans}, {Coulter}, {DeMarchi}, {Dimitriadis}, {Drout},
  {Engel}, {French}, {Gagliano}, {Gall}, {Hung}, {Izzo}, {Jacobson-Gal{\'a}n},
  {Kilpatrick}, {Korhonen}, {Margutti}, {Raimundo}, {Ramirez-Ruiz}, {Rest},
  {Rojas-Bravo}, {Siebert}, {Smartt}, {Smith}, {Terreran}, {Wang}, {Wojtak},
  {Agnello}, {Ansari}, {Arendse}, {Baldeschi}, {Blanchard}, {Brethauer},
  {Bright}, {Brown}, {de Boer}, {Dodd}, {Fairlamb}, {Grillo}, {Hajela}, {Hede},
  {Kolborg}, {Law-Smith}, {Lin}, {Magnier}, {Malanchev}, {Matthews}, {Mockler},
  {Muthukrishna}, {Pan}, {Pfister}, {Ramanah}, {Rest}, {Sarangi},
  {Schr{\o}der}, {Stauffer}, {Stroh}, {Taggart}, {Tinyanont}, {Wainscoat}, \&
  {Young Supernova Experiment}}]{jones2021}
{Jones}, D.~O., {Foley}, R.~J., {Narayan}, G., {et~al.} 2021, \apj, 908, 143,
  \dodoi{10.3847/1538-4357/abd7f5}

\bibitem[{{Jonker} {et~al.}(2020){Jonker}, {Stone}, {Generozov}, {van Velzen},
  \& {Metzger}}]{jonker2020}
{Jonker}, P.~G., {Stone}, N.~C., {Generozov}, A., {van Velzen}, S., \&
  {Metzger}, B. 2020, \apj, 889, 166, \dodoi{10.3847/1538-4357/ab659c}

\bibitem[{{Kaiser} {et~al.}(2002){Kaiser}, {Aussel}, {Burke}, {Boesgaard},
  {Chambers}, {Chun}, {Heasley}, {Hodapp}, {Hunt}, {Jedicke}, {Jewitt},
  {Kudritzki}, {Luppino}, {Maberry}, {Magnier}, {Monet}, {Onaka}, {Pickles},
  {Rhoads}, {Simon}, {Szalay}, {Szapudi}, {Tholen}, {Tonry}, {Waterson}, \&
  {Wick}}]{kaiser2002}
{Kaiser}, N., {Aussel}, H., {Burke}, B.~E., {et~al.} 2002, in Society of
  Photo-Optical Instrumentation Engineers (SPIE) Conference Series, Vol. 4836,
  Survey and Other Telescope Technologies and Discoveries, ed. J.~A. {Tyson} \&
  S.~{Wolff}, 154--164, \dodoi{10.1117/12.457365}

\bibitem[{{Kankare} {et~al.}(2017){Kankare}, {Kotak}, {Mattila}, {Lundqvist},
  {Ward}, {Fraser}, {Lawrence}, {Smartt}, {Meikle}, {Bruce}, {Harmanen},
  {Hutton}, {Inserra}, {Kangas}, {Pastorello}, {Reynolds},
  {Romero-Ca{\~n}izales}, {Smith}, {Valenti}, {Chambers}, {Hodapp}, {Huber},
  {Kaiser}, {Kudritzki}, {Magnier}, {Tonry}, {Wainscoat}, \&
  {Waters}}]{kankare2017}
{Kankare}, E., {Kotak}, R., {Mattila}, S., {et~al.} 2017, Nature Astronomy, 1,
  865, \dodoi{10.1038/s41550-017-0290-2}

\bibitem[{{Kara} {et~al.}(2018){Kara}, {Dai}, {Reynolds}, \&
  {Kallman}}]{kara2018}
{Kara}, E., {Dai}, L., {Reynolds}, C.~S., \& {Kallman}, T. 2018, \mnras, 474,
  3593, \dodoi{10.1093/mnras/stx3004}

\bibitem[{{Karas} \& {{\v{S}}ubr}(2007)}]{karas2007}
{Karas}, V., \& {{\v{S}}ubr}, L. 2007, \aap, 470, 11,
  \dodoi{10.1051/0004-6361:20066068}

\bibitem[{{Kastner} \& {Bhatia}(1990)}]{kastner1990}
{Kastner}, S.~O., \& {Bhatia}, A.~K. 1990, \apj, 362, 745,
  \dodoi{10.1086/169312}

\bibitem[{{Kaur} \& {Stone}(2024)}]{kaur2024}
{Kaur}, K., \& {Stone}, N.~C. 2024, arXiv e-prints, arXiv:2405.18500,
  \dodoi{10.48550/arXiv.2405.18500}

\bibitem[{{Kewley} {et~al.}(2006){Kewley}, {Groves}, {Kauffmann}, \&
  {Heckman}}]{kewley2006}
{Kewley}, L.~J., {Groves}, B., {Kauffmann}, G., \& {Heckman}, T. 2006, \mnras,
  372, 961, \dodoi{10.1111/j.1365-2966.2006.10859.x}

\bibitem[{{Kochanek}(1994)}]{kochanek1994}
{Kochanek}, C.~S. 1994, \apj, 422, 508, \dodoi{10.1086/173745}

\bibitem[{{Komossa} \& {Grupe}(2023)}]{komossa2023}
{Komossa}, S., \& {Grupe}, D. 2023, Astronomische Nachrichten, 344, e20230015,
  \dodoi{10.1002/asna.20230015}

\bibitem[{{Kormendy} \& {Ho}(2013)}]{kormendy2013}
{Kormendy}, J., \& {Ho}, L.~C. 2013, \araa, 51, 511,
  \dodoi{10.1146/annurev-astro-082708-101811}

\bibitem[{{Kova{\v{c}}evi{\'c}} {et~al.}(2014){Kova{\v{c}}evi{\'c}},
  {Popovi{\'c}}, {Shapovalova}, {Ili{\'c}}, {Burenkov}, \&
  {Chavushyan}}]{kovacevic2014}
{Kova{\v{c}}evi{\'c}}, A., {Popovi{\'c}}, L.~{\v{C}}., {Shapovalova}, A.~I.,
  {et~al.} 2014, Advances in Space Research, 54, 1414,
  \dodoi{10.1016/j.asr.2014.06.025}

\bibitem[{{Koz{\l}owski} {et~al.}(2016){Koz{\l}owski}, {Kochanek}, {Ashby},
  {Assef}, {Brodwin}, {Eisenhardt}, {Jannuzi}, \& {Stern}}]{kozlowski2016}
{Koz{\l}owski}, S., {Kochanek}, C.~S., {Ashby}, M. L.~N., {et~al.} 2016, \apj,
  817, 119, \dodoi{10.3847/0004-637X/817/2/119}

\bibitem[{{Lang}(2014)}]{lang2014}
{Lang}, D. 2014, \aj, 147, 108, \dodoi{10.1088/0004-6256/147/5/108}

\bibitem[{{Law-Smith} {et~al.}(2017){Law-Smith}, {Ramirez-Ruiz}, {Ellison}, \&
  {Foley}}]{law-smith2017}
{Law-Smith}, J., {Ramirez-Ruiz}, E., {Ellison}, S.~L., \& {Foley}, R.~J. 2017,
  \apj, 850, 22, \dodoi{10.3847/1538-4357/aa94c7}

\bibitem[{{Law-Smith} {et~al.}(2020){Law-Smith}, {Coulter}, {Guillochon},
  {Mockler}, \& {Ramirez-Ruiz}}]{law-smith2020}
{Law-Smith}, J. A.~P., {Coulter}, D.~A., {Guillochon}, J., {Mockler}, B., \&
  {Ramirez-Ruiz}, E. 2020, \apj, 905, 141, \dodoi{10.3847/1538-4357/abc489}

\bibitem[{{Lawrence} \& {Papadakis}(1993)}]{lawrence1993}
{Lawrence}, A., \& {Papadakis}, I. 1993, \apjl, 414, L85,
  \dodoi{10.1086/187002}

\bibitem[{{Lawrence} {et~al.}(1987){Lawrence}, {Watson}, {Pounds}, \&
  {Elvis}}]{lawrence1987}
{Lawrence}, A., {Watson}, M.~G., {Pounds}, K.~A., \& {Elvis}, M. 1987, \nat,
  325, 694, \dodoi{10.1038/325694a0}

\bibitem[{{Leloudas} {et~al.}(2016){Leloudas}, {Fraser}, {Stone}, {van Velzen},
  {Jonker}, {Arcavi}, {Fremling}, {Maund}, {Smartt}, {Kr{\`\i}hler},
  {Miller-Jones}, {Vreeswijk}, {Gal-Yam}, {Mazzali}, {De Cia}, {Howell},
  {Inserra}, {Patat}, {de Ugarte Postigo}, {Yaron}, {Ashall}, {Bar},
  {Campbell}, {Chen}, {Childress}, {Elias-Rosa}, {Harmanen}, {Hosseinzadeh},
  {Johansson}, {Kangas}, {Kankare}, {Kim}, {Kuncarayakti}, {Lyman}, {Magee},
  {Maguire}, {Malesani}, {Mattila}, {McCully}, {Nicholl}, {Prentice},
  {Romero-Ca{\~n}izales}, {Schulze}, {Smith}, {Sollerman}, {Sullivan},
  {Tucker}, {Valenti}, {Wheeler}, \& {Young}}]{leloudas2016}
{Leloudas}, G., {Fraser}, M., {Stone}, N.~C., {et~al.} 2016, Nature Astronomy,
  1, 0002, \dodoi{10.1038/s41550-016-0002}

\bibitem[{{Leloudas} {et~al.}(2019){Leloudas}, {Dai}, {Arcavi}, {Vreeswijk},
  {Mockler}, {Roy}, {Malesani}, {Schulze}, {Wevers}, {Fraser}, {Ramirez-Ruiz},
  {Auchettl}, {Burke}, {Cannizzaro}, {Charalampopoulos}, {Chen}, {Cikota},
  {Della Valle}, {Galbany}, {Gromadzki}, {Heintz}, {Hiramatsu}, {Jonker},
  {Kostrzewa-Rutkowska}, {Maguire}, {Mandel}, {Nicholl}, {Onori}, {Roth},
  {Smartt}, {Wyrzykowski}, \& {Young}}]{leloudas2019}
{Leloudas}, G., {Dai}, L., {Arcavi}, I., {et~al.} 2019, \apj, 887, 218,
  \dodoi{10.3847/1538-4357/ab5792}

\bibitem[{{Liu} {et~al.}(2017){Liu}, {Zhou}, {Cao}, {Ho}, \&
  {Komossa}}]{liu2017}
{Liu}, F.~K., {Zhou}, Z.~Q., {Cao}, R., {Ho}, L.~C., \& {Komossa}, S. 2017,
  \mnras, 472, L99, \dodoi{10.1093/mnrasl/slx147}

\bibitem[{{Liu} {et~al.}(2022){Liu}, {Dou}, {Chen}, \& {Shen}}]{liu2022}
{Liu}, X.-L., {Dou}, L.-M., {Chen}, J.-H., \& {Shen}, R.-F. 2022, \apj, 925,
  67, \dodoi{10.3847/1538-4357/ac33a9}

\bibitem[{{Liu} {et~al.}(2020){Liu}, {Li}, {Liu}, {Lu}, {Yuan}, {Dou}, \&
  {Shen}}]{liu2020}
{Liu}, Z., {Li}, D., {Liu}, H.-Y., {et~al.} 2020, \apj, 894, 93,
  \dodoi{10.3847/1538-4357/ab880f}

\bibitem[{{Lodato}(2012)}]{lodato2012}
{Lodato}, G. 2012, in European Physical Journal Web of Conferences, Vol.~39,
  European Physical Journal Web of Conferences, 01001,
  \dodoi{10.1051/epjconf/20123901001}

\bibitem[{{Lodato} {et~al.}(2009){Lodato}, {King}, \& {Pringle}}]{lodato2009}
{Lodato}, G., {King}, A.~R., \& {Pringle}, J.~E. 2009, \mnras, 392, 332,
  \dodoi{10.1111/j.1365-2966.2008.14049.x}

\bibitem[{{Lodato} \& {Rossi}(2011)}]{lodato2011}
{Lodato}, G., \& {Rossi}, E.~M. 2011, \mnras, 410, 359,
  \dodoi{10.1111/j.1365-2966.2010.17448.x}

\bibitem[{{Loeb} \& {Ulmer}(1997)}]{loeb1997}
{Loeb}, A., \& {Ulmer}, A. 1997, \apj, 489, 573, \dodoi{10.1086/304814}

\bibitem[{{Lu} \& {Bonnerot}(2020)}]{lu2020}
{Lu}, W., \& {Bonnerot}, C. 2020, \mnras, 492, 686,
  \dodoi{10.1093/mnras/stz3405}

\bibitem[{{Lu} \& {Kumar}(2018)}]{lu2018}
{Lu}, W., \& {Kumar}, P. 2018, \apj, 865, 128, \dodoi{10.3847/1538-4357/aad54a}

\bibitem[{{MacLeod} {et~al.}(2012){MacLeod}, {Ivezi{\'c}}, {Sesar}, {de Vries},
  {Kochanek}, {Kelly}, {Becker}, {Lupton}, {Hall}, {Richards}, {Anderson}, \&
  {Schneider}}]{macleod2012}
{MacLeod}, C.~L., {Ivezi{\'c}}, {\v{Z}}., {Sesar}, B., {et~al.} 2012, \apj,
  753, 106, \dodoi{10.1088/0004-637X/753/2/106}

\bibitem[{{Mainzer} {et~al.}(2014){Mainzer}, {Bauer}, {Cutri}, {Grav},
  {Masiero}, {Beck}, {Clarkson}, {Conrow}, {Dailey}, {Eisenhardt}, {Fabinsky},
  {Fajardo-Acosta}, {Fowler}, {Gelino}, {Grillmair}, {Heinrichsen}, {Kendall},
  {Kirkpatrick}, {Liu}, {Masci}, {McCallon}, {Nugent}, {Papin}, {Rice},
  {Royer}, {Ryan}, {Sevilla}, {Sonnett}, {Stevenson}, {Thompson}, {Wheelock},
  {Wiemer}, {Wittman}, {Wright}, \& {Yan}}]{mainzer2014}
{Mainzer}, A., {Bauer}, J., {Cutri}, R.~M., {et~al.} 2014, \apj, 792, 30,
  \dodoi{10.1088/0004-637X/792/1/30}

\bibitem[{{Martin} {et~al.}(2005){Martin}, {Fanson}, {Schiminovich},
  {Morrissey}, {Friedman}, {Barlow}, {Conrow}, {Grange}, {Jelinsky},
  {Milliard}, {Siegmund}, {Bianchi}, {Byun}, {Donas}, {Forster}, {Heckman},
  {Lee}, {Madore}, {Malina}, {Neff}, {Rich}, {Small}, {Surber}, {Szalay},
  {Welsh}, \& {Wyder}}]{martin2005}
{Martin}, D.~C., {Fanson}, J., {Schiminovich}, D., {et~al.} 2005, \apjl, 619,
  L1, \dodoi{10.1086/426387}

\bibitem[{{Masci} {et~al.}(2019){Masci}, {Laher}, {Rusholme}, {Shupe}, {Groom},
  {Surace}, {Jackson}, {Monkewitz}, {Beck}, {Flynn}, {Terek}, {Landry},
  {Hacopians}, {Desai}, {Howell}, {Brooke}, {Imel}, {Wachter}, {Ye}, {Lin},
  {Cenko}, {Cunningham}, {Rebbapragada}, {Bue}, {Miller}, {Mahabal}, {Bellm},
  {Patterson}, {Juri{\'c}}, {Golkhou}, {Ofek}, {Walters}, {Graham}, {Kasliwal},
  {Dekany}, {Kupfer}, {Burdge}, {Cannella}, {Barlow}, {Van Sistine}, {Giomi},
  {Fremling}, {Blagorodnova}, {Levitan}, {Riddle}, {Smith}, {Helou}, {Prince},
  \& {Kulkarni}}]{masci2019}
{Masci}, F.~J., {Laher}, R.~R., {Rusholme}, B., {et~al.} 2019, \pasp, 131,
  018003, \dodoi{10.1088/1538-3873/aae8ac}

\bibitem[{{Masterson} {et~al.}(2024){Masterson}, {De}, {Panagiotou}, {Kara},
  {Arcavi}, {Eilers}, {Frostig}, {Gezari}, {Grotova}, {Liu}, {Malyali},
  {Meisner}, {Merloni}, {Newsome}, {Rau}, {Simcoe}, \& {van
  Velzen}}]{masterson2024}
{Masterson}, M., {De}, K., {Panagiotou}, C., {et~al.} 2024, \apj, 961, 211,
  \dodoi{10.3847/1538-4357/ad18bb}

\bibitem[{Mattila {et~al.}(2017)Mattila, Graham, Kankare, Kool, Moriya,
  Perez-Torres, \& Wyrzykowski}]{mattila2017}
Mattila, S., Graham, M.~J., Kankare, E., {et~al.} 2017, Proceedings of the
  International Astronomical Union, 14, 263–268,
  \dodoi{10.1017/S1743921318002727}

\bibitem[{{McHardy} \& {Czerny}(1987)}]{mchardy1987}
{McHardy}, I., \& {Czerny}, B. 1987, \nat, 325, 696, \dodoi{10.1038/325696a0}

\bibitem[{{Meisner} {et~al.}(2018){Meisner}, {Lang}, \&
  {Schlegel}}]{meisner2018}
{Meisner}, A.~M., {Lang}, D.~A., \& {Schlegel}, D.~J. 2018, Research Notes of
  the American Astronomical Society, 2, 202, \dodoi{10.3847/2515-5172/aaecd5}

\bibitem[{{Merloni} {et~al.}(2015){Merloni}, {Dwelly}, {Salvato},
  {Georgakakis}, {Greiner}, {Krumpe}, {Nandra}, {Ponti}, \&
  {Rau}}]{merloni2015}
{Merloni}, A., {Dwelly}, T., {Salvato}, M., {et~al.} 2015, \mnras, 452, 69,
  \dodoi{10.1093/mnras/stv1095}

\bibitem[{{Metzger}(2022)}]{metzger2022}
{Metzger}, B.~D. 2022, \apjl, 937, L12, \dodoi{10.3847/2041-8213/ac90ba}

\bibitem[{{Metzger} \& {Stone}(2016)}]{metzger2016}
{Metzger}, B.~D., \& {Stone}, N.~C. 2016, \mnras, 461, 948,
  \dodoi{10.1093/mnras/stw1394}

\bibitem[{{Metzger} \& {Stone}(2017)}]{metzger2017}
---. 2017, \apj, 844, 75, \dodoi{10.3847/1538-4357/aa7a16}

\bibitem[{Miller \& Stone(1994)}]{miller1994}
Miller, J., \& Stone, R. 1994, The Kast Double Spectograph, Lick Observatory
  technical reports (University of California Observatories/Lick Observatory).
\newblock \url{https://books.google.com/books?id=QXk2AQAAIAAJ}

\bibitem[{{Mockler} {et~al.}(2019){Mockler}, {Guillochon}, \&
  {Ramirez-Ruiz}}]{mockler2019}
{Mockler}, B., {Guillochon}, J., \& {Ramirez-Ruiz}, E. 2019, \apj, 872, 151,
  \dodoi{10.3847/1538-4357/ab010f}

\bibitem[{{Mockler} \& {Ramirez-Ruiz}(2021)}]{mockler2021}
{Mockler}, B., \& {Ramirez-Ruiz}, E. 2021, \apj, 906, 101,
  \dodoi{10.3847/1538-4357/abc955}

\bibitem[{{Mummery}(2021)}]{mummery2021a}
{Mummery}, A. 2021, \mnras, 507, L24, \dodoi{10.1093/mnrasl/slab088}

\bibitem[{{Mummery} \& {Balbus}(2020)}]{mummery2020}
{Mummery}, A., \& {Balbus}, S.~A. 2020, \mnras, 492, 5655,
  \dodoi{10.1093/mnras/staa192}

\bibitem[{{Mummery} \& {Balbus}(2021)}]{mummery2021b}
---. 2021, \mnras, 504, 4730, \dodoi{10.1093/mnras/stab1184}

\bibitem[{{Mummery} {et~al.}(2024){Mummery}, {van Velzen}, {Nathan}, {Ingram},
  {Hammerstein}, {Fraser-Taliente}, \& {Balbus}}]{mummery2024}
{Mummery}, A., {van Velzen}, S., {Nathan}, E., {et~al.} 2024, \mnras, 527,
  2452, \dodoi{10.1093/mnras/stad3001}

\bibitem[{{Neustadt} {et~al.}(2020){Neustadt}, {Holoien}, {Kochanek},
  {Auchettl}, {Brown}, {Shappee}, {Pogge}, {Dong}, {Stanek}, {Tucker}, {Bose},
  {Chen}, {Ricci}, {Vallely}, {Prieto}, {Thompson}, {Coulter}, {Drout},
  {Foley}, {Kilpatrick}, {Piro}, {Rojas-Bravo}, {Buckley}, {Gromadzki},
  {Dimitriadis}, {Siebert}, {Do}, {Huber}, \& {Payne}}]{neustadt2020}
{Neustadt}, J.~M.~M., {Holoien}, T.~W.~S., {Kochanek}, C.~S., {et~al.} 2020,
  \mnras, 494, 2538, \dodoi{10.1093/mnras/staa859}

\bibitem[{{Nicholl} {et~al.}(2022){Nicholl}, {Lanning}, {Ramsden}, {Mockler},
  {Lawrence}, {Short}, \& {Ridley}}]{nicholl2022}
{Nicholl}, M., {Lanning}, D., {Ramsden}, P., {et~al.} 2022, \mnras, 515, 5604,
  \dodoi{10.1093/mnras/stac2206}

\bibitem[{{Nicholl} {et~al.}(2020){Nicholl}, {Wevers}, {Oates}, {Alexander},
  {Leloudas}, {Onori}, {Jerkstrand}, {Gomez}, {Campana}, {Arcavi},
  {Charalampopoulos}, {Gromadzki}, {Ihanec}, {Jonker}, {Lawrence}, {Mandel},
  {Schulze}, {Short}, {Burke}, {McCully}, {Hiramatsu}, {Howell}, {Pellegrino},
  {Abbot}, {Anderson}, {Berger}, {Blanchard}, {Cannizzaro}, {Chen},
  {Dennefeld}, {Galbany}, {Gonz{\'a}lez-Gait{\'a}n}, {Hosseinzadeh}, {Inserra},
  {Irani}, {Kuin}, {M{\"u}ller-Bravo}, {Pineda}, {Ross}, {Roy}, {Smartt},
  {Smith}, {Tucker}, {Wyrzykowski}, \& {Young}}]{nicholl2020}
{Nicholl}, M., {Wevers}, T., {Oates}, S.~R., {et~al.} 2020, \mnras, 499, 482,
  \dodoi{10.1093/mnras/staa2824}

\bibitem[{{Oke} {et~al.}(1995){Oke}, {Cohen}, {Carr}, {Cromer}, {Dingizian},
  {Harris}, {Labrecque}, {Lucinio}, {Schaal}, {Epps}, \& {Miller}}]{oke1995}
{Oke}, J.~B., {Cohen}, J.~G., {Carr}, M., {et~al.} 1995, \pasp, 107, 375,
  \dodoi{10.1086/133562}

\bibitem[{{Panagiotou} {et~al.}(2023){Panagiotou}, {De}, {Masterson}, {Kara},
  {Calzadilla}, {Eilers}, {Frostig}, {Karambelkar}, {Kasliwal}, {Lourie},
  {Meisner}, {Simcoe}, {Stein}, \& {Zolkower}}]{panagiotou2023}
{Panagiotou}, C., {De}, K., {Masterson}, M., {et~al.} 2023, \apjl, 948, L5,
  \dodoi{10.3847/2041-8213/acc02f}

\bibitem[{{Parkinson} {et~al.}(2022){Parkinson}, {Knigge}, {Matthews}, {Long},
  {Higginbottom}, {Sim}, \& {Mangham}}]{parkinson2022}
{Parkinson}, E.~J., {Knigge}, C., {Matthews}, J.~H., {et~al.} 2022, \mnras,
  510, 5426, \dodoi{10.1093/mnras/stac027}

\bibitem[{{Patterson} {et~al.}(2019){Patterson}, {Bellm}, {Rusholme}, {Masci},
  {Juric}, {Krughoff}, {Golkhou}, {Graham}, {Kulkarni}, {Helou}, \& {Zwicky
  Transient Facility Collaboration}}]{patterson2019}
{Patterson}, M.~T., {Bellm}, E.~C., {Rusholme}, B., {et~al.} 2019, \pasp, 131,
  018001, \dodoi{10.1088/1538-3873/aae904}

\bibitem[{{Payne} {et~al.}(2021){Payne}, {Shappee}, {Hinkle}, {Vallely},
  {Kochanek}, {Holoien}, {Auchettl}, {Stanek}, {Thompson}, {Neustadt},
  {Tucker}, {Armstrong}, {Brimacombe}, {Cacella}, {Cornect}, {Denneau},
  {Fausnaugh}, {Flewelling}, {Grupe}, {Heinze}, {Lopez}, {Monard}, {Prieto},
  {Schneider}, {Sheppard}, {Tonry}, \& {Weiland}}]{payne2021}
{Payne}, A.~V., {Shappee}, B.~J., {Hinkle}, J.~T., {et~al.} 2021, \apj, 910,
  125, \dodoi{10.3847/1538-4357/abe38d}

\bibitem[{{Peterson}(1993)}]{peterson1993}
{Peterson}, B.~M. 1993, \pasp, 105, 247, \dodoi{10.1086/133140}

\bibitem[{{Peterson} \& {Wandel}(1999)}]{peterson1999}
{Peterson}, B.~M., \& {Wandel}, A. 1999, \apjl, 521, L95,
  \dodoi{10.1086/312190}

\bibitem[{{Peterson} {et~al.}(2004){Peterson}, {Ferrarese}, {Gilbert}, {Kaspi},
  {Malkan}, {Maoz}, {Merritt}, {Netzer}, {Onken}, {Pogge}, {Vestergaard}, \&
  {Wandel}}]{peterson2004}
{Peterson}, B.~M., {Ferrarese}, L., {Gilbert}, K.~M., {et~al.} 2004, \apj, 613,
  682, \dodoi{10.1086/423269}

\bibitem[{{Phinney}(1989)}]{phinney1989c}
{Phinney}, E.~S. 1989, in The Center of the Galaxy, ed. M.~{Morris}, Vol. 136,
  543

\bibitem[{{Piran} {et~al.}(2015){Piran}, {Svirski}, {Krolik}, {Cheng}, \&
  {Shiokawa}}]{piran2015}
{Piran}, T., {Svirski}, G., {Krolik}, J., {Cheng}, R.~M., \& {Shiokawa}, H.
  2015, \apj, 806, 164, \dodoi{10.1088/0004-637X/806/2/164}

\bibitem[{{Planck Collaboration} {et~al.}(2020){Planck Collaboration},
  {Aghanim}, {Akrami}, {Ashdown}, {Aumont}, {Baccigalupi}, {Ballardini},
  {Banday}, {Barreiro}, {Bartolo}, {Basak}, {Battye}, {Benabed}, {Bernard},
  {Bersanelli}, {Bielewicz}, {Bock}, {Bond}, {Borrill}, {Bouchet}, {Boulanger},
  {Bucher}, {Burigana}, {Butler}, {Calabrese}, {Cardoso}, {Carron},
  {Challinor}, {Chiang}, {Chluba}, {Colombo}, {Combet}, {Contreras}, {Crill},
  {Cuttaia}, {de Bernardis}, {de Zotti}, {Delabrouille}, {Delouis}, {Di
  Valentino}, {Diego}, {Dor{\'e}}, {Douspis}, {Ducout}, {Dupac}, {Dusini},
  {Efstathiou}, {Elsner}, {En{\ss}lin}, {Eriksen}, {Fantaye}, {Farhang},
  {Fergusson}, {Fernandez-Cobos}, {Finelli}, {Forastieri}, {Frailis},
  {Fraisse}, {Franceschi}, {Frolov}, {Galeotta}, {Galli}, {Ganga},
  {G{\'e}nova-Santos}, {Gerbino}, {Ghosh}, {Gonz{\'a}lez-Nuevo}, {G{\'o}rski},
  {Gratton}, {Gruppuso}, {Gudmundsson}, {Hamann}, {Handley}, {Hansen},
  {Herranz}, {Hildebrandt}, {Hivon}, {Huang}, {Jaffe}, {Jones}, {Karakci},
  {Keih{\"a}nen}, {Keskitalo}, {Kiiveri}, {Kim}, {Kisner}, {Knox},
  {Krachmalnicoff}, {Kunz}, {Kurki-Suonio}, {Lagache}, {Lamarre}, {Lasenby},
  {Lattanzi}, {Lawrence}, {Le Jeune}, {Lemos}, {Lesgourgues}, {Levrier},
  {Lewis}, {Liguori}, {Lilje}, {Lilley}, {Lindholm}, {L{\'o}pez-Caniego},
  {Lubin}, {Ma}, {Mac{\'\i}as-P{\'e}rez}, {Maggio}, {Maino}, {Mandolesi},
  {Mangilli}, {Marcos-Caballero}, {Maris}, {Martin}, {Martinelli},
  {Mart{\'\i}nez-Gonz{\'a}lez}, {Matarrese}, {Mauri}, {McEwen}, {Meinhold},
  {Melchiorri}, {Mennella}, {Migliaccio}, {Millea}, {Mitra},
  {Miville-Desch{\^e}nes}, {Molinari}, {Montier}, {Morgante}, {Moss}, {Natoli},
  {N{\o}rgaard-Nielsen}, {Pagano}, {Paoletti}, {Partridge}, {Patanchon},
  {Peiris}, {Perrotta}, {Pettorino}, {Piacentini}, {Polastri}, {Polenta},
  {Puget}, {Rachen}, {Reinecke}, {Remazeilles}, {Renzi}, {Rocha}, {Rosset},
  {Roudier}, {Rubi{\~n}o-Mart{\'\i}n}, {Ruiz-Granados}, {Salvati}, {Sandri},
  {Savelainen}, {Scott}, {Shellard}, {Sirignano}, {Sirri}, {Spencer},
  {Sunyaev}, {Suur-Uski}, {Tauber}, {Tavagnacco}, {Tenti}, {Toffolatti},
  {Tomasi}, {Trombetti}, {Valenziano}, {Valiviita}, {Van Tent}, {Vibert},
  {Vielva}, {Villa}, {Vittorio}, {Wandelt}, {Wehus}, {White}, {White},
  {Zacchei}, \& {Zonca}}]{planck2020}
{Planck Collaboration}, {Aghanim}, N., {Akrami}, Y., {et~al.} 2020, \aap, 641,
  A6, \dodoi{10.1051/0004-6361/201833910}

\bibitem[{{Ramirez-Ruiz} \& {Rosswog}(2009)}]{ramirez-ruiz2009}
{Ramirez-Ruiz}, E., \& {Rosswog}, S. 2009, \apjl, 697, L77,
  \dodoi{10.1088/0004-637X/697/2/L77}

\bibitem[{{Rees}(1988)}]{rees1988}
{Rees}, M.~J. 1988, \nat, 333, 523, \dodoi{10.1038/333523a0}

\bibitem[{{Reines} \& {Volonteri}(2015)}]{reines2015}
{Reines}, A.~E., \& {Volonteri}, M. 2015, \apj, 813, 82,
  \dodoi{10.1088/0004-637X/813/2/82}

\bibitem[{{Rest} {et~al.}(2014){Rest}, {Scolnic}, {Foley}, {Huber}, {Chornock},
  {Narayan}, {Tonry}, {Berger}, {Soderberg}, {Stubbs}, {Riess}, {Kirshner},
  {Smartt}, {Schlafly}, {Rodney}, {Botticella}, {Brout}, {Challis}, {Czekala},
  {Drout}, {Hudson}, {Kotak}, {Leibler}, {Lunnan}, {Marion}, {McCrum},
  {Milisavljevic}, {Pastorello}, {Sanders}, {Smith}, {Stafford}, {Thilker},
  {Valenti}, {Wood-Vasey}, {Zheng}, {Burgett}, {Chambers}, {Denneau}, {Draper},
  {Flewelling}, {Hodapp}, {Kaiser}, {Kudritzki}, {Magnier}, {Metcalfe},
  {Price}, {Sweeney}, {Wainscoat}, \& {Waters}}]{rest2014}
{Rest}, A., {Scolnic}, D., {Foley}, R.~J., {et~al.} 2014, \apj, 795, 44,
  \dodoi{10.1088/0004-637X/795/1/44}

\bibitem[{{Roberts} \& {Warwick}(2000)}]{roberts2000}
{Roberts}, T.~P., \& {Warwick}, R.~S. 2000, \mnras, 315, 98,
  \dodoi{10.1046/j.1365-8711.2000.03384.x}

\bibitem[{{Roming} {et~al.}(2005){Roming}, {Kennedy}, {Mason}, {Nousek}, {Ahr},
  {Bingham}, {Broos}, {Carter}, {Hancock}, {Huckle}, {Hunsberger}, {Kawakami},
  {Killough}, {Koch}, {McLelland}, {Smith}, {Smith}, {Soto}, {Boyd},
  {Breeveld}, {Holland}, {Ivanushkina}, {Pryzby}, {Still}, \&
  {Stock}}]{roming2005}
{Roming}, P. W.~A., {Kennedy}, T.~E., {Mason}, K.~O., {et~al.} 2005, \ssr, 120,
  95, \dodoi{10.1007/s11214-005-5095-4}

\bibitem[{{Roth} \& {Kasen}(2018)}]{roth2018}
{Roth}, N., \& {Kasen}, D. 2018, \apj, 855, 54,
  \dodoi{10.3847/1538-4357/aaaec6}

\bibitem[{{Roth} {et~al.}(2016){Roth}, {Kasen}, {Guillochon}, \&
  {Ramirez-Ruiz}}]{roth2016}
{Roth}, N., {Kasen}, D., {Guillochon}, J., \& {Ramirez-Ruiz}, E. 2016, \apj,
  827, 3, \dodoi{10.3847/0004-637X/827/1/3}

\bibitem[{{Rumbaugh} {et~al.}(2018){Rumbaugh}, {Shen}, {Morganson}, {Liu},
  {Banerji}, {McMahon}, {Abdalla}, {Benoit-L{\'e}vy}, {Bertin}, {Brooks},
  {Buckley-Geer}, {Capozzi}, {Carnero Rosell}, {Carrasco Kind}, {Carretero},
  {Cunha}, {D'Andrea}, {da Costa}, {DePoy}, {Desai}, {Doel}, {Frieman},
  {Garc{\'\i}a-Bellido}, {Gruen}, {Gruendl}, {Gschwend}, {Gutierrez},
  {Honscheid}, {James}, {Kuehn}, {Kuhlmann}, {Kuropatkin}, {Lima}, {Maia},
  {Marshall}, {Martini}, {Menanteau}, {Plazas}, {Reil}, {Roodman}, {Sanchez},
  {Scarpine}, {Schindler}, {Schubnell}, {Sheldon}, {Smith}, {Soares-Santos},
  {Sobreira}, {Suchyta}, {Swanson}, {Walker}, {Wester}, \& {DES
  Collaboration}}]{rumbaugh2018}
{Rumbaugh}, N., {Shen}, Y., {Morganson}, E., {et~al.} 2018, \apj, 854, 160,
  \dodoi{10.3847/1538-4357/aaa9b6}

\bibitem[{{Ryde}(1999)}]{ryde1999}
{Ryde}, F. 1999, Astrophysical Letters and Communications, 39, 281,
  \dodoi{10.48550/arXiv.astro-ph/9811462}

\bibitem[{{Ryu} {et~al.}(2020){Ryu}, {Krolik}, \& {Piran}}]{ryu2020b}
{Ryu}, T., {Krolik}, J., \& {Piran}, T. 2020, \apj, 904, 73,
  \dodoi{10.3847/1538-4357/abbf4d}

\bibitem[{{Sacchi} {et~al.}(2023){Sacchi}, {Risaliti}, \&
  {Miniutti}}]{sacchi2023}
{Sacchi}, A., {Risaliti}, G., \& {Miniutti}, G. 2023, \aap, 671, A33,
  \dodoi{10.1051/0004-6361/202244983}

\bibitem[{{Sarin} \& {Metzger}(2024)}]{sarin2024}
{Sarin}, N., \& {Metzger}, B.~D. 2024, \apjl, 961, L19,
  \dodoi{10.3847/2041-8213/ad16d8}

\bibitem[{{Schlafly} \& {Finkbeiner}(2011)}]{schlafly2011}
{Schlafly}, E.~F., \& {Finkbeiner}, D.~P. 2011, \apj, 737, 103,
  \dodoi{10.1088/0004-637X/737/2/103}

\bibitem[{{Schulze} {et~al.}(2011){Schulze}, {Klose}, {Bj{\"o}rnsson},
  {Jakobsson}, {Kann}, {Rossi}, {Kr{\"u}hler}, {Greiner}, \&
  {Ferrero}}]{schulze2011}
{Schulze}, S., {Klose}, S., {Bj{\"o}rnsson}, G., {et~al.} 2011, \aap, 526, A23,
  \dodoi{10.1051/0004-6361/201015581}

\bibitem[{{Seibert} {et~al.}(2012){Seibert}, {Wyder}, {Neill}, {Madore},
  {Bianchi}, {Smith}, {Shiao}, {Schiminovich}, {Rich}, {Conrow}, {Martin}, \&
  {GALEX Catalog Team}}]{seibert2012}
{Seibert}, M., {Wyder}, T., {Neill}, J., {et~al.} 2012, in American
  Astronomical Society Meeting Abstracts, Vol. 219, American Astronomical
  Society Meeting Abstracts \#219, 340.01

\bibitem[{{Shemmer} {et~al.}(2008){Shemmer}, {Brandt}, {Netzer}, {Maiolino}, \&
  {Kaspi}}]{shemmer2008}
{Shemmer}, O., {Brandt}, W.~N., {Netzer}, H., {Maiolino}, R., \& {Kaspi}, S.
  2008, \apj, 682, 81, \dodoi{10.1086/588776}

\bibitem[{{Shiokawa} {et~al.}(2015){Shiokawa}, {Krolik}, {Cheng}, {Piran}, \&
  {Noble}}]{shiokawa2015}
{Shiokawa}, H., {Krolik}, J.~H., {Cheng}, R.~M., {Piran}, T., \& {Noble}, S.~C.
  2015, \apj, 804, 85, \dodoi{10.1088/0004-637X/804/2/85}

\bibitem[{{Short} {et~al.}(2020){Short}, {Nicholl}, {Lawrence}, {Gomez},
  {Arcavi}, {Wevers}, {Leloudas}, {Schulze}, {Anderson}, {Berger}, {Blanchard},
  {Burke}, {Castro Segura}, {Charalampopoulos}, {Chornock}, {Galbany},
  {Gromadzki}, {Herzog}, {Hiramatsu}, {Horne}, {Hosseinzadeh}, {Howell},
  {Ihanec}, {Inserra}, {Kankare}, {Maguire}, {McCully}, {M{\"u}ller Bravo},
  {Onori}, {Sollerman}, \& {Young}}]{short2020}
{Short}, P., {Nicholl}, M., {Lawrence}, A., {et~al.} 2020, \mnras, 498, 4119,
  \dodoi{10.1093/mnras/staa2065}

\bibitem[{{Short} {et~al.}(2023){Short}, {Lawrence}, {Nicholl}, {Ward},
  {Reynolds}, {Mattila}, {Yin}, {Arcavi}, {Carnall}, {Charalampopoulos},
  {Gromadzki}, {Jonker}, {Kim}, {Leloudas}, {Mandel}, {Onori}, {Pursiainen},
  {Schulze}, {Villforth}, \& {Wevers}}]{short2023}
{Short}, P., {Lawrence}, A., {Nicholl}, M., {et~al.} 2023, \mnras, 525, 1568,
  \dodoi{10.1093/mnras/stad2270}

\bibitem[{{Shu} {et~al.}(2018){Shu}, {Wang}, {Dou}, {Jiang}, {Wang}, \&
  {Wang}}]{shu2018}
{Shu}, X.~W., {Wang}, S.~S., {Dou}, L.~M., {et~al.} 2018, \apjl, 857, L16,
  \dodoi{10.3847/2041-8213/aaba17}

\bibitem[{{Siebert} {et~al.}(2019){Siebert}, {Foley}, {Jones}, {Angulo},
  {Davis}, {Duarte}, {Strasburger}, {Conlon}, {Kazmi}, {Nishimoto}, {Schubert},
  {Sun}, \& {Tippens}}]{siebert2019b}
{Siebert}, M.~R., {Foley}, R.~J., {Jones}, D.~O., {et~al.} 2019, \mnras, 486,
  5785, \dodoi{10.1093/mnras/stz1209}

\bibitem[{{Silverman} {et~al.}(2012){Silverman}, {Ganeshalingam}, {Li}, \&
  {Filippenko}}]{silverman2012b}
{Silverman}, J.~M., {Ganeshalingam}, M., {Li}, W., \& {Filippenko}, A.~V. 2012,
  \mnras, 425, 1889, \dodoi{10.1111/j.1365-2966.2012.21526.x}

\bibitem[{{Speagle}(2020)}]{speagle2020}
{Speagle}, J.~S. 2020, \mnras, 493, 3132, \dodoi{10.1093/mnras/staa278}

\bibitem[{{Stalevski} {et~al.}(2016){Stalevski}, {Ricci}, {Ueda}, {Lira},
  {Fritz}, \& {Baes}}]{stalevski2016}
{Stalevski}, M., {Ricci}, C., {Ueda}, Y., {et~al.} 2016, \mnras, 458, 2288,
  \dodoi{10.1093/mnras/stw444}

\bibitem[{{Steiner} {et~al.}(2009){Steiner}, {Narayan}, {McClintock}, \&
  {Ebisawa}}]{steiner2009}
{Steiner}, J.~F., {Narayan}, R., {McClintock}, J.~E., \& {Ebisawa}, K. 2009,
  \pasp, 121, 1279, \dodoi{10.1086/648535}

\bibitem[{{Stern} {et~al.}(2012){Stern}, {Assef}, {Benford}, {Blain}, {Cutri},
  {Dey}, {Eisenhardt}, {Griffith}, {Jarrett}, {Lake}, {Masci}, {Petty},
  {Stanford}, {Tsai}, {Wright}, {Yan}, {Harrison}, \& {Madsen}}]{stern2012}
{Stern}, D., {Assef}, R.~J., {Benford}, D.~J., {et~al.} 2012, \apj, 753, 30,
  \dodoi{10.1088/0004-637X/753/1/30}

\bibitem[{{Storchi-Bergmann} {et~al.}(2017){Storchi-Bergmann}, {Schimoia},
  {Peterson}, {Elvis}, {Denney}, {Eracleous}, \&
  {Nemmen}}]{storchi-bergmann2017}
{Storchi-Bergmann}, T., {Schimoia}, J.~S., {Peterson}, B.~M., {et~al.} 2017,
  \apj, 835, 236, \dodoi{10.3847/1538-4357/835/2/236}

\bibitem[{{Strateva} {et~al.}(2003){Strateva}, {Strauss}, {Hao}, {Schlegel},
  {Hall}, {Gunn}, {Li}, {Ivezi{\'c}}, {Richards}, {Zakamska}, {Voges},
  {Anderson}, {Lupton}, {Schneider}, {Brinkmann}, \& {Nichol}}]{strateva2003}
{Strateva}, I.~V., {Strauss}, M.~A., {Hao}, L., {et~al.} 2003, \aj, 126, 1720,
  \dodoi{10.1086/378367}

\bibitem[{{Strubbe} \& {Quataert}(2009)}]{strubbe2009}
{Strubbe}, L.~E., \& {Quataert}, E. 2009, \mnras, 400, 2070,
  \dodoi{10.1111/j.1365-2966.2009.15599.x}

\bibitem[{{Strubbe} \& {Quataert}(2011)}]{strubbe2011}
---. 2011, \mnras, 415, 168, \dodoi{10.1111/j.1365-2966.2011.18686.x}

\bibitem[{{Sun} {et~al.}(2024){Sun}, {Jiang}, {Dou}, {Shu}, {Zhu}, {Dong},
  {Buckley}, {Cenko}, {Fan}, {Gromadzki}, {Liu}, {Wang}, {Wang}, {Wang}, {Wu},
  {Yang}, {Zhang}, {Zhang}, \& {Zhang}}]{sun2024}
{Sun}, L., {Jiang}, N., {Dou}, L., {et~al.} 2024, arXiv e-prints,
  arXiv:2410.09720, \dodoi{10.48550/arXiv.2410.09720}

\bibitem[{{Tadhunter} {et~al.}(2017){Tadhunter}, {Spence}, {Rose}, {Mullaney},
  \& {Crowther}}]{tadhunter2017}
{Tadhunter}, C., {Spence}, R., {Rose}, M., {Mullaney}, J., \& {Crowther}, P.
  2017, Nature Astronomy, 1, 0061, \dodoi{10.1038/s41550-017-0061}

\bibitem[{{Thomsen} {et~al.}(2022){Thomsen}, {Kwan}, {Dai}, {Wu}, {Roth}, \&
  {Ramirez-Ruiz}}]{thomsen2022}
{Thomsen}, L.~L., {Kwan}, T.~M., {Dai}, L., {et~al.} 2022, \apjl, 937, L28,
  \dodoi{10.3847/2041-8213/ac911f}

\bibitem[{{Titarchuk} \& {Hua}(1995)}]{titarchuk1995}
{Titarchuk}, L., \& {Hua}, X.-M. 1995, \apj, 452, 226, \dodoi{10.1086/176293}

\bibitem[{{Tonry} {et~al.}(2018){Tonry}, {Denneau}, {Heinze}, {Stalder},
  {Smith}, {Smartt}, {Stubbs}, {Weiland}, \& {Rest}}]{tonry2018}
{Tonry}, J.~L., {Denneau}, L., {Heinze}, A.~N., {et~al.} 2018, \pasp, 130,
  064505, \dodoi{10.1088/1538-3873/aabadf}

\bibitem[{{Trakhtenbrot} {et~al.}(2019){Trakhtenbrot}, {Arcavi}, {Ricci},
  {Tacchella}, {Stern}, {Netzer}, {Jonker}, {Horesh}, {Mej{\'\i}a-Restrepo},
  {Hosseinzadeh}, {Hallefors}, {Howell}, {McCully}, {Balokovi{\'c}}, {Heida},
  {Kamraj}, {Lansbury}, {Wyrzykowski}, {Gromadzki}, {Hamanowicz}, {Cenko},
  {Sand}, {Hsiao}, {Phillips}, {Diamond}, {Kara}, {Gendreau}, {Arzoumanian}, \&
  {Remillard}}]{trakhtenbrot2019}
{Trakhtenbrot}, B., {Arcavi}, I., {Ricci}, C., {et~al.} 2019, Nature Astronomy,
  3, 242, \dodoi{10.1038/s41550-018-0661-3}

\bibitem[{{Valenti} {et~al.}(2014){Valenti}, {Sand}, {Pastorello}, {Graham},
  {Howell}, {Parrent}, {Tomasella}, {Ochner}, {Fraser}, {Benetti}, {Yuan},
  {Smartt}, {Maund}, {Arcavi}, {Gal-Yam}, {Inserra}, \& {Young}}]{valenti2014}
{Valenti}, S., {Sand}, D., {Pastorello}, A., {et~al.} 2014, \mnras, 438, L101,
  \dodoi{10.1093/mnrasl/slt171}

\bibitem[{{van Velzen} {et~al.}(2020){van Velzen}, {Holoien}, {Onori}, {Hung},
  \& {Arcavi}}]{velzen2020}
{van Velzen}, S., {Holoien}, T. W.~S., {Onori}, F., {Hung}, T., \& {Arcavi}, I.
  2020, \ssr, 216, 124, \dodoi{10.1007/s11214-020-00753-z}

\bibitem[{{van Velzen} {et~al.}(2019{\natexlab{a}}){van Velzen}, {Stone},
  {Metzger}, {Gezari}, {Brown}, \& {Fruchter}}]{velzen2019b}
{van Velzen}, S., {Stone}, N.~C., {Metzger}, B.~D., {et~al.}
  2019{\natexlab{a}}, \apj, 878, 82, \dodoi{10.3847/1538-4357/ab1844}

\bibitem[{{van Velzen} {et~al.}(2011){van Velzen}, {Farrar}, {Gezari},
  {Morrell}, {Zaritsky}, {{\"O}stman}, {Smith}, {Gelfand}, \&
  {Drake}}]{velzen2011a}
{van Velzen}, S., {Farrar}, G.~R., {Gezari}, S., {et~al.} 2011, \apj, 741, 73,
  \dodoi{10.1088/0004-637X/741/2/73}

\bibitem[{{van Velzen} {et~al.}(2019{\natexlab{b}}){van Velzen}, {Gezari},
  {Cenko}, {Kara}, {Miller-Jones}, {Hung}, {Bright}, {Roth}, {Blagorodnova},
  {Huppenkothen}, {Yan}, {Ofek}, {Sollerman}, {Frederick}, {Ward}, {Graham},
  {Fender}, {Kasliwal}, {Canella}, {Stein}, {Giomi}, {Brinnel}, {van Santen},
  {Nordin}, {Bellm}, {Dekany}, {Fremling}, {Golkhou}, {Kupfer}, {Kulkarni},
  {Laher}, {Mahabal}, {Masci}, {Miller}, {Neill}, {Riddle}, {Rigault},
  {Rusholme}, {Soumagnac}, \& {Tachibana}}]{velzen2019a}
{van Velzen}, S., {Gezari}, S., {Cenko}, S.~B., {et~al.} 2019{\natexlab{b}},
  \apj, 872, 198, \dodoi{10.3847/1538-4357/aafe0c}

\bibitem[{{van Velzen} {et~al.}(2021){van Velzen}, {Gezari}, {Hammerstein},
  {Roth}, {Frederick}, {Ward}, {Hung}, {Cenko}, {Stein}, {Perley}, {Taggart},
  {Foley}, {Sollerman}, {Blagorodnova}, {Andreoni}, {Bellm}, {Brinnel}, {De},
  {Dekany}, {Feeney}, {Fremling}, {Giomi}, {Golkhou}, {Graham}, {Ho},
  {Kasliwal}, {Kilpatrick}, {Kulkarni}, {Kupfer}, {Laher}, {Mahabal}, {Masci},
  {Miller}, {Nordin}, {Riddle}, {Rusholme}, {van Santen}, {Sharma}, {Shupe}, \&
  {Soumagnac}}]{velzen2021}
{van Velzen}, S., {Gezari}, S., {Hammerstein}, E., {et~al.} 2021, \apj, 908, 4,
  \dodoi{10.3847/1538-4357/abc258}

\bibitem[{{Verro} {et~al.}(2022){Verro}, {Trager}, {Peletier}, {Lan{\c{c}}on},
  {Gonneau}, {Vazdekis}, {Prugniel}, {Chen}, {Coelho},
  {S{\'a}nchez-Bl{\'a}zquez}, {Martins}, {Arentsen}, {Lyubenova},
  {Falc{\'o}n-Barroso}, \& {Dries}}]{verro2022}
{Verro}, K., {Trager}, S.~C., {Peletier}, R.~F., {et~al.} 2022, \aap, 660, A34,
  \dodoi{10.1051/0004-6361/202142388}

\bibitem[{{Wang} {et~al.}(2024){Wang}, {Lin}, {Zhang}, \& {Zhu}}]{wang2024}
{Wang}, Y., {Lin}, D. N.~C., {Zhang}, B., \& {Zhu}, Z. 2024, \apjl, 962, L7,
  \dodoi{10.3847/2041-8213/ad20e5}

\bibitem[{{Ward} {et~al.}(2024){Ward}, {Gezari}, {Nugent}, {Kerr}, {Eracleous},
  {Frederick}, {Hammerstein}, {Graham}, {van Velzen}, {Kasliwal}, {Laher},
  {Masci}, {Purdum}, {Racine}, \& {Smith}}]{ward2024}
{Ward}, C., {Gezari}, S., {Nugent}, P., {et~al.} 2024, \apj, 961, 172,
  \dodoi{10.3847/1538-4357/ad147d}

\bibitem[{{Wevers}(2020)}]{wevers2020}
{Wevers}, T. 2020, \mnras, 497, L1, \dodoi{10.1093/mnrasl/slaa097}

\bibitem[{{Wevers} {et~al.}(2019){Wevers}, {Pasham}, {van Velzen}, {Leloudas},
  {Schulze}, {Miller-Jones}, {Jonker}, {Gromadzki}, {Kankare}, {Hodgkin},
  {Wyrzykowski}, {Kostrzewa-Rutkowska}, {Moran}, {Berton}, {Maguire}, {Onori},
  {Mattila}, \& {Nicholl}}]{wevers2019b}
{Wevers}, T., {Pasham}, D.~R., {van Velzen}, S., {et~al.} 2019, \mnras, 488,
  4816, \dodoi{10.1093/mnras/stz1976}

\bibitem[{{Wevers} {et~al.}(2022){Wevers}, {Nicholl}, {Guolo},
  {Charalampopoulos}, {Gromadzki}, {Reynolds}, {Kankare}, {Leloudas},
  {Anderson}, {Arcavi}, {Cannizzaro}, {Chen}, {Ihanec}, {Inserra},
  {Guti{\'e}rrez}, {Jonker}, {Lawrence}, {Magee}, {M{\"u}ller-Bravo}, {Onori},
  {Ridley}, {Schulze}, {Short}, {Hiramatsu}, {Newsome}, {Terwel}, {Yang}, \&
  {Young}}]{wevers2022}
{Wevers}, T., {Nicholl}, M., {Guolo}, M., {et~al.} 2022, \aap, 666, A6,
  \dodoi{10.1051/0004-6361/202142616}

\bibitem[{{Wilkins} \& {Fabian}(2012)}]{wilkins2012}
{Wilkins}, D.~R., \& {Fabian}, A.~C. 2012, \mnras, 424, 1284,
  \dodoi{10.1111/j.1365-2966.2012.21308.x}

\bibitem[{{Wilms} {et~al.}(2000){Wilms}, {Allen}, \& {McCray}}]{wilms2000}
{Wilms}, J., {Allen}, A., \& {McCray}, R. 2000, \apj, 542, 914,
  \dodoi{10.1086/317016}

\bibitem[{{Wiseman} {et~al.}(2024){Wiseman}, {Williams}, {Arcavi}, {Galbany},
  {Graham}, {H{\"o}nig}, {Newsome}, {Subrayan}, {Sullivan}, {Wang}, {Ili{\'c}},
  {Nicholl}, {Oates}, {Petrushevska}, \& {Smith}}]{wiseman2024}
{Wiseman}, P., {Williams}, R.~D., {Arcavi}, I., {et~al.} 2024, arXiv e-prints,
  arXiv:2406.11552, \dodoi{10.48550/arXiv.2406.11552}

\bibitem[{{Wright} {et~al.}(2010){Wright}, {Eisenhardt}, {Mainzer}, {Ressler},
  {Cutri}, {Jarrett}, {Kirkpatrick}, {Padgett}, {McMillan}, {Skrutskie},
  {Stanford}, {Cohen}, {Walker}, {Mather}, {Leisawitz}, {Gautier}, {McLean},
  {Benford}, {Lonsdale}, {Blain}, {Mendez}, {Irace}, {Duval}, {Liu}, {Royer},
  {Heinrichsen}, {Howard}, {Shannon}, {Kendall}, {Walsh}, {Larsen}, {Cardon},
  {Schick}, {Schwalm}, {Abid}, {Fabinsky}, {Naes}, \& {Tsai}}]{wright2010}
{Wright}, E.~L., {Eisenhardt}, P. R.~M., {Mainzer}, A.~K., {et~al.} 2010, \aj,
  140, 1868, \dodoi{10.1088/0004-6256/140/6/1868}

\bibitem[{{Yan} \& {Xie}(2018)}]{yan2018}
{Yan}, Z., \& {Xie}, F.-G. 2018, \mnras, 475, 1190,
  \dodoi{10.1093/mnras/stx3259}

\bibitem[{{Yang} {et~al.}(2019){Yang}, {Shen}, {Liu}, {Wu}, {Jiang},
  {Shangguan}, {Graham}, \& {Yao}}]{yang2019}
{Yang}, Q., {Shen}, Y., {Liu}, X., {et~al.} 2019, \apj, 885, 110,
  \dodoi{10.3847/1538-4357/ab481a}

\bibitem[{{Yao} {et~al.}(2023){Yao}, {Ravi}, {Gezari}, {van Velzen}, {Lu},
  {Schulze}, {Somalwar}, {Kulkarni}, {Hammerstein}, {Nicholl}, {Graham},
  {Perley}, {Cenko}, {Stein}, {Ricarte}, {Chadayammuri}, {Quataert}, {Bellm},
  {Bloom}, {Dekany}, {Drake}, {Groom}, {Mahabal}, {Prince}, {Riddle},
  {Rusholme}, {Sharma}, {Sollerman}, \& {Yan}}]{yao2023}
{Yao}, Y., {Ravi}, V., {Gezari}, S., {et~al.} 2023, \apjl, 955, L6,
  \dodoi{10.3847/2041-8213/acf216}

\bibitem[{{Yuan} {et~al.}(1998){Yuan}, {Brinkmann}, {Siebert}, \&
  {Voges}}]{yuan1993}
{Yuan}, W., {Brinkmann}, W., {Siebert}, J., \& {Voges}, W. 1998, \aap, 330,
  108, \dodoi{10.48550/arXiv.astro-ph/9805015}

\bibitem[{{Zabludoff} {et~al.}(2021){Zabludoff}, {Arcavi}, {LaMassa}, {Perets},
  {Trakhtenbrot}, {Zauderer}, {Auchettl}, {Dai}, {French}, {Hung}, {Kara},
  {Lodato}, {Maksym}, {Qin}, {Ramirez-Ruiz}, {Roth}, {Runnoe}, \&
  {Wevers}}]{zabludoff2021}
{Zabludoff}, A., {Arcavi}, I., {LaMassa}, S., {et~al.} 2021, \ssr, 217, 54,
  \dodoi{10.1007/s11214-021-00829-4}

\bibitem[{{Zackay} {et~al.}(2016){Zackay}, {Ofek}, \& {Gal-Yam}}]{zackay2016}
{Zackay}, B., {Ofek}, E.~O., \& {Gal-Yam}, A. 2016, \apj, 830, 27,
  \dodoi{10.3847/0004-637X/830/1/27}

\bibitem[{{Zanazzi} \& {Ogilvie}(2020)}]{zanazzi2020}
{Zanazzi}, J.~J., \& {Ogilvie}, G.~I. 2020, \mnras, 499, 5562,
  \dodoi{10.1093/mnras/staa3127}

\bibitem[{{Zhang}(2022)}]{zhang2022}
{Zhang}, X. 2022, \apjs, 260, 31, \dodoi{10.3847/1538-4365/ac6020}

\bibitem[{{Zhuang} \& {Shen}(2021)}]{zhuang2021}
{Zhuang}, J., \& {Shen}, R.-F. 2021, Journal of High Energy Astrophysics, 32,
  11, \dodoi{10.1016/j.jheap.2021.06.001}

\end{thebibliography}
\bibliographystyle{aasjournal}

\end{document}